\pdfoutput=1
\documentclass[cernpreprint, texlive=2023, txfonts=true, orcidlogo=true, USenglish, texmf, orcidlogo]{atlasdoc}

\usepackage[backend=biber,subcaption]{atlaspackage}
\usepackage{atlasbiblatex}
\usepackage{atlasphysics}
 
\graphicspath{{logos/}{figures/}}

\usepackage{anyfontsize}
\usepackage{esvect}
\usepackage{mathrsfs}
 
\AtlasTitle{Measurement of the inclusive $t\bar{t}$ production cross section in the lepton+jets channel in $pp$ collisions at $\sqrt{s}= 7$~Te\kern-0.1em V with the ATLAS detector using support vector machines}

\AtlasAbstract{
A measurement of the top quark pair-production cross section in the lepton+jets decay channel is presented.  It is based on 4.6~fb$^{-1}$ of $\sqrt{s} = 7$~Te\kern-0.1em V $pp$ collision data collected during 2011 by the ATLAS experiment at the CERN Large Hadron Collider.  A three-class, multidimensional event classifier based on support vector machines is used to differentiate $t\bar{t}$ events from backgrounds.  The $t\bar{t}$ production cross section is found to be $\sigma_{t\bar{t}}=168.5\pm0.7$(stat.)~$^{+6.2}_{-5.9}$(syst.)~$^{+3.4}_{-3.2}$(lumi.)~pb.  The result is consistent with the Standard Model prediction based on QCD calculations at next-to-next-to-leading order.}
 
\author{The ATLAS Collaboration}
 
\AtlasRefCode{TOPQ-2017-08}
 
\PreprintIdNumber{CERN-EP-2022-191}

\AtlasJournalRef{\PRD 108 (2023) 032014}
\AtlasJournal{Phys.\ Rev.\ D}
\AtlasDOI{10.1103/PhysRevD.108.032014}

\hypersetup{pdftitle={ATLAS document},pdfauthor={The ATLAS Collaboration}}
 
\begin{document}
 
\maketitle
 
\tableofcontents
 
\clearpage
\section{Introduction\label{sec:intro}}
 
In the Standard Model of particle physics~\cite{Glashow,Weinberg,Salam}, the top quark ($t$) and the bottom quark ($b$) belong to a doublet representation of the weak-isospin SU(2). The top quark is the most massive of the known elementary particles.  Because its mass is close to the electroweak symmetry breaking scale, it may play a fundamental role in the mechanism of breaking of the SU(2) symmetry of the electroweak interaction. Top quark production is also the dominant background in many analyses looking for physics beyond the Standard Model at high mass scales at the Large Hadron Collider (LHC), and a good understanding of top quark production is a necessary step in many \textquotedblleft new physics\textquotedblright\ searches.
 
The cross section is one of the simplest observables that can be measured in the \ttbar system.  It allows one to make important comparisons with theoretical predictions available at next-to-next-to-leading order in perturbative QCD, including the soft-gluon resummation at next-to-next-to-leading-log order (NNLO + NNLL);  see Ref.~\cite{Czakon:2013goa} and references therein.  For $pp$ collisions at a center-of-mass energy of $\sqrt{s}=7$~\TeV, the predicted \ttbar production cross section is $\sigma_{t\bar{t}}^{\text{NNLO+NNLL}}= 177^{+10}_{-11}$~pb.  This theoretical value was calculated with the \TOPpp[2.0] program~\cite{Czakon:2011xx}, including soft-gluon resummation, assuming a top quark mass value of 172.5~\GeV, and using the PDF4LHC~\cite{Botje:2011sn} procedures.
 
According to the Standard Model, top quarks from $pp$ collisions at the LHC are produced mostly via the strong interaction as \ttbar pairs, with each top quark decaying into a $W$ boson and a $b$-quark nearly $100\%$ of the time. The \ttbar events in which one of the $W$ bosons decays into a quark pair and the other into a charged lepton and a neutrino, are classified as \textquotedblleft lepton+jets\textquotedblright, as such events contain an electron or muon or $\tau$-lepton, a neutrino and typically four hadronic jets (two of which originate from the $b$-quark and $\bar{b}$-quark).
 
In this paper, a measurement of the top quark pair-production cross section at $\sqrt{s}=7$~\TeV\ using events with a single charged lepton (electron or muon) and jets in the final state is presented.  The previously published result from the ATLAS Collaboration for the lepton+jets channel~\cite{TOPQ-2011-02} uses 35~pb$^{-1}$ of data and obtains a precision of 12\%.  The most precise CMS \ttbar cross-section measurement in the same channel~\cite{CMS-TOP-11-003} has a precision of 7\%.  In the dilepton channel, the best ATLAS result~\cite{TOPQ-2013-04} achieves 3.5\% precision, while CMS~\cite{CMS-TOP-13-004} reaches 3.6\%.  These ATLAS and CMS dilepton results have been combined~\cite{TOPQ-2018-39}, resulting in an uncertainty of 2.6\% in $\sigma_{t\bar{t}}$.
 
The analysis presented in this paper is based on the full dataset collected with the ATLAS detector at the LHC in 2011, corresponding to an integrated luminosity of 4.6~fb$^{-1}$, and attains statistical and systematic uncertainties that are significantly lower than in previous ATLAS measurements in this final state. In an extension of the usual application of binary multivariate classifiers, this analysis uses a large number of variables to train three different support vector machines (SVMs).  The three SVMs are used to define a three-dimensional space in which a multi-class event discriminator is constructed to identify the \ttbar events through a simultaneous profile likelihood fit in four independent regions of this space.
 
\newcommand{\AtlasCoordFootnote}{
ATLAS uses a right-handed coordinate system with its origin at the nominal interaction point (IP) in the center of the detector and the $z$-axis along the beam pipe.  The $x$-axis points from the IP to the center of the LHC ring, and the $y$-axis points upwards.  Cylindrical coordinates $(r,\phi)$ are used in the transverse plane, $\phi$ being the azimuthal angle around the $z$-axis.  The pseudorapidity is defined in terms of the polar angle $\theta$ as $\eta = -\ln \tan(\theta/2)$, and the distance $\Delta R$ in the $\eta$--$\phi$ space is defined as $\Delta R \equiv \sqrt{(\Delta\eta)^{2} + (\Delta\phi)^{2}}$.}
 
\section{ATLAS detector\label{sec:atlas}}
 
The ATLAS detector is described in Ref.~\cite{PERF-2007-01}.  It is a multipurpose particle detector with forward-backward symmetry and a cylindrical geometry.\footnote{\AtlasCoordFootnote}  The inner tracking detectors are surrounded by a thin superconducting solenoid, electromagnetic and hadronic calorimeters, and a muon spectrometer with a magnetic field generated by three superconducting toroidal magnets of eight coils each.  The inner-detector system (ID), in combination with the 2~T magnetic field from the solenoid, provides precision momentum measurements for charged particles within the pseudorapidity range $\vert \eta \vert < 2.5$.  Moving radially outwards, it consists of a silicon pixel detector, a silicon microstrip detector, and a straw-tube tracker that also provides transition radiation measurements for electron identification.  The calorimeter system covers the pseudorapidity range $\vert \eta \vert < 4.9$.  A high-granularity liquid-argon (LAr) sampling calorimeter with lead absorber measures electromagnetic showers within $\vert \eta \vert < 3.2$.  In the region matched to the ID, $\vert \eta \vert < 2.5$, the innermost layer has fine segmentation in $\eta$ to improve the resolution of the shower position and direction measurements.  Hadronic showers are measured by an iron/plastic-scintillator tile calorimeter in the central region, $\vert \eta \vert < 1.7$, and by a LAr calorimeter in the end cap region, $1.5 <\vert \eta \vert < 3.2$.  In the forward region, measurements of both electromagnetic and hadronic showers are provided by a LAr calorimeter covering the pseudorapidity range $3.1 <\vert \eta \vert < 4.9$.  The muon spectrometer is instrumented with separate trigger and high-precision tracking chambers.  It provides muon identification for charged-particle tracks within $\vert \eta \vert < 2.7$.  The combination of all ATLAS detector systems provides charged-particle measurement along with lepton and photon measurement and identification in the pseudorapidity range $\vert \eta \vert < 2.5$.  Jets are reconstructed over the full range covered by the calorimeters, $\vert \eta \vert < 4.9$.
 
A three-level trigger system~\cite{PERF-2011-02} is used to select interesting events.  The first-level (L1) trigger is implemented in hardware and uses a subset of detector information to reduce the event rate to a design value of at most \SI{75}{\kHz}.  This is followed by two software-based trigger levels which together reduce the event rate to about \SI{200}{\Hz}.
 
An extensive software suite~\cite{ATL-SOFT-PUB-2021-001} is used in data simulation, in the reconstruction and analysis of real and simulated data, in detector operations, and in the trigger and data acquisition systems of the experiment.
 
\section{Object definitions\label{sec:object_def}}
 
\noindent \textbf{Electrons}
 
Electron candidates are selected using the offline identification with tight requirements~\cite{PERF-2010-04} within a fiducial region with transverse momentum $p_{\mathrm{T}} >25$~\GeV\ and $\vert \eta \vert< 2.47$, excluding the calorimeter transition region $1.37<\vert \eta \vert< 1.52$.  They are subjected to several other strict criteria including requirements on track quality, impact parameter, calorimeter shower shape, and track--cluster matching.  The electron candidates are also required to be isolated.  The transverse energy ($E_{\mathrm{T}}$) deposited in the calorimeter in a cone of size $\Delta R = 0.2$ around the electron is calculated.  Additionally, the scalar sum of the $p_{\mathrm{T}}$ of tracks in a cone of size $\Delta R = 0.3$ is determined.  Both of these quantities have selection requirements that depend on the $\eta$ and $E_{\mathrm{T}}$ of the electron candidate, and which ensure 90\% efficiency for electrons from $W$ boson or $Z$ boson decays~\cite{PERF-2013-03}.  Finally, electrons lying within $\Delta R = 0.4$ of a selected jet are discarded to reject leptons from heavy-flavor decays.
 
\noindent \textbf{Muons}
 
Muon candidates reconstructed from tracks in both the muon spectrometer and ID are selected with the MuID algorithm~\cite{PERF-2014-05}.  Only candidates satisfying $p_{\mathrm{T}} > 20$~\GeV\ and $\vert \eta \vert< 2.5$ are selected.  Muon candidates are required to have a sufficient number of hits in the ID.  The impact parameter with respect to the primary vertex in the longitudinal direction along the beam axis is required to satisfy $\vert z_0 \vert < 2$~mm.  The tight muon candidates used in this analysis are required to be isolated.  The sum of the calorimeter transverse energy within $\Delta R = 0.2$ of a muon is required to be below 4~\GeV, and the sum of the $p_{\mathrm{T}}$ of all the tracks within $\Delta R = 0.3$ (excluding the muon track) must be below 2.5~\GeV.  The efficiency of this combined isolation requirement varies between 95\% and 97\%, depending on the data-taking period.  In order to reduce the background from muons produced by heavy-flavor decays inside jets, muons are required to be separated by $\Delta R > 0.4$ from the nearest selected jet.
 
\noindent \textbf{Jets}
 
Jets are reconstructed from topological clusters~\cite{PERF-2014-07} formed from energy deposits in the calorimeters using the anti-$k_{t}$ algorithm~\cite{Cacciari:2008gp, Fastjet} with a radius parameter of 0.4.  Clusters are calibrated using the local cluster weighting (LCW), which differentiates between the energy deposits arising from electromagnetic and hadronic showers~\cite{PERF-2012-01}.  The jet reconstruction is done at the electromagnetic scale and then a scale factor is applied in order to obtain the jet energy at the hadronic scale.  The jet energy scale (JES) corrections account for the calorimeter response to the true jet energy by using \textquotedblleft truth jets\textquotedblright\ from simulation.  The truth jets are formed through the application of the anti-$k_{t}$ algorithm to stable particles, with the exception of final-state muons and neutrinos.  Jet calibration includes both the LCW and JES calibrations.  In addition, the jets are corrected for distortions due to multiple $pp$ collisions per bunch crossing (pileup) using a method which estimates the pileup activity on an event-by-event basis, as well as the sensitivity of a given jet to pileup.  With this method~\cite{PERF-2012-01}, a contribution to the jet transverse momentum equal to the product of the jet area in the $\eta$--$\phi$ plane and the average transverse energy density in a given event is subtracted.  The effects of additional collisions in either the same bunch crossing or those adjacent in time are taken into account using corrections which depend on the average number of interactions per bunch crossing and the number of primary vertices.  To avoid double counting of jets and electrons (which are reconstructed by independent algorithms), jets within $\Delta R=0.2$ of a reconstructed electron are removed.  For this analysis, only jets in the central region of the detector, $\vert \eta \vert< 2.5$, and with transverse momentum $p_{\mathrm{T}} >25$~\GeV\ are considered.

\noindent \textbf{Identification of {\textit{b}}-jets}
 
The identification of \textquotedblleft $b$-jets\textquotedblright\ (jets arising from the decay of $B$-hadrons) is performed with the MV1 algorithm~\cite{PERF-2012-04}, which combines the outputs of three different tagging algorithms into a multivariate discriminant.  Jets are defined to be \textquotedblleft $b$-tagged\textquotedblright\ if the MV1 discriminant value is larger than a threshold (operating point) corresponding to 70\% efficiency to identify $b$-quark jets in simulated \ttbar events.  Approximately 20\% of jets originating from charm quarks are identified as $b$-jets, while light-flavor jets are mistagged as $b$-jets at the 1\% level.
 
\noindent \textbf{Missing transverse momentum}
 
The missing transverse momentum is calculated~\cite{PERF-2011-07} as the complement of the vector sum, in the transverse plane, of calorimeter cell energies within $\vert \eta \vert< 4.9$, after all corrections are applied to the associated physics objects (including jets, electrons, and muons).  A correction for significant energy deposits not associated with high-$p_{\mathrm{T}}$ physics objects is also included.  The magnitude of the missing transverse momentum vector is denoted by \MET, while its direction in the transverse plane is either denoted by an azimuthal angle $\phi$ or inferred through its vector components \MET$_{x}$ and \MET$_{y}$.
 
\section{Event selection}
 
This analysis considers the single-lepton decay channel for the \ttbar pair.  The selected events are required to have exactly one lepton (either an electron or a muon), a large amount of \MET, and three or more hadronic jets.  The number of $b$-tagged jets in an event must be two or less.  Events must have at least one primary vertex, with five or more tracks with $p_{\textrm{T}} > 150$~\MeV.  If there is more than one primary vertex, the one with the largest $\sum^{\text{tracks}}{p_{\textrm{T}}^2}$ is chosen.  Events were collected using single-lepton triggers, and each lepton candidate must be matched to an appropriate lepton trigger.  In the muon channel, events are selected with a single-muon trigger with a $p_{\mathrm{T}}$ threshold of 18~\GeV.  For the electron channel, a single-electron trigger is required with $p_{\mathrm{T}} > 20$~\GeV.  This is increased to 22~\GeV\ during high instantaneous luminosity periods.  Three or more jets with $p_{\mathrm{T}}$ greater than 25~\GeV\ are required in each event.  A large amount of \MET\ is required to select events containing a neutrino.  For electron events the \MET\ must be greater than 30~\GeV, while for muon events the \MET\ is required to be greater than 20~\GeV.
 
In order to reduce the background due to multi-jet production containing misidentified or nonprompt leptons, an additional selection requirement is imposed.  Typically, in an event arising from this background, the missing transverse momentum vector points in the same direction in the transverse plane as the charged-lepton candidate.  Therefore, electron candidate events must pass a requirement that $m_{\mathrm{T}}(W) > 30$~\GeV, while muon candidate events must have $m_{\mathrm{T}}(W)+\MET>60$~\GeV.  Here, $m_{\mathrm{T}}(W)$ is defined as
 
\begin{equation*}
m_{\mathrm{T}}(W)=\sqrt{2 ~ p_{\mathrm{T}}^{\ell} ~ \met ~ (1-\cos\Delta\phi)}\,,
\label{eqn:m_t_of_W}
\end{equation*}
 
\noindent where $\Delta\phi$ is the difference in $\phi$ between the direction of the charged-lepton transverse momentum, $p_{\mathrm{T}}^{\ell}$, and the missing transverse momentum vector.
 
\section{Data samples\label{sec:samples}}
 
The data sample used in this analysis comes from the $\sqrt{s} = 7$~\TeV\ $pp$ collisions collected during LHC Run~1 in 2011, and was recorded during stable beam conditions with all relevant ATLAS subdetector systems operational.  It corresponds to an integrated luminosity of 4.6~fb$^{-1}$, with an uncertainty of 1.8\%~\cite{DAPR-2011-01}.

\section{Signal and background modeling\label{sec:mc_samples}}
 
Except for the background due to multi-jet production leading to misidentified leptons (which is estimated from the data), all signal and background samples are modeled using Monte Carlo (MC) simulated events in conjunction with factors to correct the simulations to data where required.
 
\subsection{Multi-jet background / fake leptons}
 
The background from multi-jet events, in which a jet is misidentified as a muon or electron, or where a nonprompt lepton within a jet passes the tight lepton selection requirements, is sizable because of the large multi-jet production cross section.  Events from these two sources of multi-jet background are referred to as fake-lepton events. This background is estimated using the so-called \textquotedblleft matrix method,\textquotedblright\ which is based on the measurement of the selection efficiencies of leptons using data event samples satisfying relaxed identification criteria~\cite{TOPQ-2010-01,ATLAS-CONF-2014-058}.  Loose electrons are electrons satisfying the baseline selection criteria where the requirements on particle identification using transition radiation measurements and on the energy-to-momentum ratio ($E$/$p$) are eased, and no requirement on the isolation is imposed. For loose muons, isolation is not required, but all other selection criteria are applied.  The matrix method is based on the measurement of the efficiencies for real and fake leptons in the loose lepton selection to pass the tight selection criteria.  The real-lepton efficiencies are measured in $Z \rightarrow \ell\ell$ data samples, while the fake-lepton efficiencies are determined in data control regions with selection requirements designed to enhance the multi-jet content (1 lepton, $\ge 1$ jet, and only a small amount of $\MET$).  These efficiencies depend on both the lepton kinematics and event characteristics.  To account for this, event weights are computed from the efficiencies parametrized as a function of a number of observables, which are then used to reweight the sample of data events with lepton candidates that satisfy the loose but not the tight selection criteria.  The sums of these weights provide estimates of the multi-jet background.
 
\subsection{Monte Carlo samples}
 
The samples used in this analysis were obtained from a simulation chain consisting of an event generator interfaced to a parton shower and hadronization model, the outputs of which were passed through a simulation of the ATLAS detector and trigger system~\cite{SOFT-2010-01}, and then reconstructed with the same algorithms as the data.  The ATLAS detector response was modeled with the ATLAS full simulation (FS) based on \GEANT~\cite{Agostinelli:2002hh}.  For \ttbar samples used to evaluate the signal modeling systematic uncertainties, the ATLAS fast simulation AtlasFast-II (AF)~\cite{SOFT-2010-01,ATLFAST} was used to model the response of the detector.
 
The \ttbar signal was simulated using the NLO \POWHEG-hvq (patch 4) matrix element generator~\cite{Frixione:2007nw} interfaced with \PYTHIA[6.425]~\cite{Sjostrand:2006za}, using parameter values set according to the \textsc{C} variant of the Perugia 2011 tune~\cite{Skands:2010ak} to model the underlying event and parton shower. The NLO \CT[10]~\cite{Lai:2010vv} parton distribution function (PDF) set was used  for the NLO matrix element part, and the LO PDF set CTEQ6L1~\cite{Pumplin:2002vw} was used with \PYTHIA.  The top quark mass was fixed at 172.5~\GeV.  This sample is referred to as \POWPY[6].  In order to evaluate the dependence on the choice of parton shower and fragmentation models, additional samples of \ttbar events were created.  The $\MCatNLO{+}\HERWIG$ sample was created with the NLO \MCatNLO~\cite{Frixione:2003ei} generator interfaced with \HERWIG~\cite{Corcella:2000bw} using the LO AUET2 tune~\cite{ATL-PHYS-PUB-2011-008}.  The \POWHER sample was created with \POWHEG interfaced to \HERWIG using the LO AUET2 tune.
 
The largest backgrounds to the \ttbar events in the selected sample are from $W$+jets and $Z$+jets production.  These were simulated with the LO event generator \ALPGEN[2.13]~\cite{Mangano:2002ea} with LO PDF set CTEQ6L1, and interfaced with \HERWIG[6.52].  \ALPGEN calculates matrix elements (ME) for final states with up to six partons.  The MLM~\cite{Alwall:2007fs} matching procedure was used to remove the overlaps between ME and parton shower products in samples with $N$ and $N+1$ final-state partons.  In addition to the inclusive parton-flavor processes, separate ME samples of $W$+$b\bar{b}$+jets, $W$+$c\bar{c}$+jets, $W$+$c$+jets, and $Z$+$b\bar{b}$+jets were generated.  The double counting of $b$- and $c$-quarks in $W$/$Z$+jets that occurs between the shower of the inclusive samples and the ME of heavy-flavor production was eliminated using an overlap-removal algorithm based on parton-to-jet $\Delta R$ matching~\cite{HFOR}.  The $W$+jets and $Z$+jets event samples are referred to as $\ALPGEN{+}\HERWIG$.
 
The single-top backgrounds were simulated at NLO using the \MCatNLO generator with the NLO PDF set CTEQ6.6~\cite{Pumplin:2002vw}, and interfaced with \HERWIG, except the $t$-channel samples, which were modeled with the LO \ACERMC[3.8] generator~\cite{AcerMC2013} interfaced with \PYTHIA.  Dibosons ($WW$, $WZ$, and $ZZ$) were generated with \HERWIG using the LO PDF set CTEQ6L1.  All samples generated with \HERWIG for the parton shower evolution and hadronization used \JIMMY[4.31]~\cite{localButterworth:1996zw} for the underlying-event model.
 
The effects of pileup were modeled by overlaying simulated minimum-bias events on the hard-scattering events.  The Monte Carlo events were then reweighted such that the distribution of the number of interactions per bunch crossing, $\langle \mu \rangle$, matched the shape and observed average of $\langle\mu\rangle= 9.1$ in the 2011 data.
 
\subsection{Signal and background classes}
 
The most challenging backgrounds (i.e., those which most resemble \ttbar~) are single-top and $W$/$Z$+$b\bar{b}$+jets.  Therefore, the \ttbar cross-section measurement is expected to be affected most by the modeling of these backgrounds.  To improve discrimination between the \ttbar signal and the different types of background, this analysis separates the background events into two classes and treats them independently.  The \textquotedblleft Heavy\textquotedblright\ background class includes the Monte Carlo samples for single-top, $W$+$b\bar{b}$+jets, and $Z$+$b\bar{b}$+jets.  All other types of background, including fake leptons, are assigned to the group designated as the \textquotedblleft Light\textquotedblright\ class.  Table~\ref{tab:2011_composition_info} summarizes the composition of the classes and lists the datasets which are used to model them.
 
\begin{table}[ht]
\caption{ \footnotesize  Class definitions and compositions are presented.  In the process column, \textquotedblleft lf\textquotedblright\ is defined as any partons that are not $b$-quarks.  The other columns show the source of the events and the fractional contribution to the given class.}
\label{tab:2011_composition_info}
\centering
\begin{tabular}{lllr} \toprule{Class} & {Process} & {Source} &  {Fraction} \\ \midrule
 
Signal&\ttbar&\POWPY[6]&100.0\%\\ \midrule
 
Light&$W$+lf+jets&$\ALPGEN{+}\HERWIG$&74.9\%\\
Light&$Z$+lf+jets&$\ALPGEN{+}\HERWIG$&7.6\%\\
Light&Dibosons&\HERWIG&1.4\%\\
Light&Fake $e$&Data&6.4\%\\
Light&Fake $\mu$&Data&9.7\%\\ \midrule
 
Heavy&$W$+$b\bar{b}$+jets&$\ALPGEN{+}\HERWIG$&43.3\%\\
Heavy&$Z$+$b\bar{b}$+jets&$\ALPGEN{+}\HERWIG$&6.3\%\\
Heavy&Single-top&\MCatNLO~/ \ACERMC&50.4\%\\ \bottomrule
 
\end{tabular}
\end{table}

The expected numbers of signal and background events in the selected sample are presented in Table~\ref{tab:MC_prediction}.  The uncertainties shown include theoretical uncertainties in the production cross sections of the processes~\cite{Czakon:2013goa, Anastasiou:2003ds, Campbell:1999ah, singleTopXsec, Aliev:2010zk, Kant:2014oha}.  The $W$/$Z$+jets and diboson uncertainties include a contribution derived from an event yield comparison with \SHERPA~\cite{Gleisberg:2008ta} Monte Carlo samples.  The uncertainty in the number of events with fake leptons is estimated to be 20\% for muons and 50\% for electrons~\cite{TOPQ-2012-05, TOPQ-2012-21}.  The observed number of events in data is in good agreement with the prediction.

\begin{table}[ht]
\caption{\footnotesize The observed and expected numbers of events in the selected sample is shown.  The first two columns list the contributions by physics process, and the two rightmost columns present events by class.  The Heavy class includes the $W$/$Z$+$b\bar{b}$+jets and single-top processes, while the Light class includes all other backgrounds.  Predicted values are rounded with respect to their individual uncertainties.}
\label{tab:MC_prediction}
\centering
\sisetup{group-minimum-digits=4,separate-uncertainty}
\begin{tabular}{l S[table-format=6.0(5)] l S[table-format=6.0(5)]}
\toprule{Process} & {Events} & {Class} & {Events} \\
\midrule{\ttbar} & 86400 \pm 5700 & {Signal} & 86400 \pm 5700 \\
\midrule{$W$+jets} & 184000 \pm 44000 && \\
{$Z$+jets} & 19000 \pm 4300 && \\
{Dibosons} & 3200 \pm 1600 && \\
{Single-top} & 11040 \pm 670 && \\
{Fake leptons} &  37500 \pm 8700 && \\
\midrule
&& {Light} & 233000 \pm 44000  \\
&& {Heavy} & 21900 \pm 1600  \\
\midrule
\midrule{Total} & 341000 \pm 45000 && \\
{Observed} & 344520 && \\
\bottomrule
\end{tabular}
\end{table}

\section{Analysis method\label{sec:AnalysisMethod}}
 
The SVM is a binary learning algorithm~\cite{SVM}.  For any two classes of events, the signed distance from a hyperplane that separates the events is the SVM discriminant.  For the analysis presented in this paper, a system of three support vector machines is used to create a three-dimensional multi-class event classifier to distinguish signal events from two classes of background (i.e., Light and Heavy).  For events from any dataset, the distances from the three hyperplanes, trained to distinguish between Signal vs.\ Light (SvL), Signal vs.\ Heavy (SvH), and Light vs.\ Heavy (LvH), are treated as the coordinates of points in a 3D \textit{decision space}.  The resulting templates of the prediction model are used in a binned likelihood fit to the analogous 3D distribution of the data events.
 
The SVM was chosen as the binary classifier because it is linear, it has firm mathematical foundations, and it offers a simple geometrical interpretation.  Because the SVM method provides the solution to a straightforward convex optimization problem, the minimum it finds is a global one.  The stopping point at the training stage is well defined, which therefore makes the method robust against overtraining.  The method also works well in problems involving a large number of observables.
 
\subsection{The SVM discriminant\label{sec:SVMs}}
 
Each event is described by $N$ observables (i.e., features), and can be represented as a point, $\vv{z}$, in an $N$-dimensional feature space.  A linear binary classifier finds a hyperplane of dimension $N-1$ that separates the two classes.
Once the separating hyperplane is found, its reconstruction only requires the vectors that lie closest to the plane. These are the \textit{support vectors} from whence the method derives its name.
If the two classes to be discriminated are not linearly separable in the original feature space, this $N$-dimensional space can be mapped, $\vv{z} \rightarrow \vv{\varphi}(\vv{z})$, into a higher-dimensional space in which the problem is linearly separable.  Detailed knowledge of this mapping is not required when it is known how to calculate the inner product of the mapped vectors~\cite{Mercer,Aronszajn}.  The distribution of classes in the mapped space can be probed directly by analyzing the multidimensional space which takes as its mathematical basis the SVM solutions for different class pairings.
 
The soft-margin SVMs~\cite{soft} used in this analysis are constructed using a variant of the sequential minimal optimization algorithm~\cite{platt} that includes the improvements suggested in Ref.~\cite{keerthi}.  
 
In the case of a three-class problem like the one considered in this analysis, three different SVM classifiers are trained.  Each SVM has the form
 
\begin{equation*}
f(\vv{z})=\langle w|z\rangle - b=\sum_{i}^{\text{SVs}}\lambda_{i}y_{i} K(\vv{v_{i}},\vv{z}) - b\,,
\label{eqn:learned_func}
\end{equation*}
 
\noindent which is the generalized equation of a plane.  The $j^{\text{th}}$ SVM (with $j$ in \{1, 2, 3\}) has a normal vector given by $|w_j\rangle$, and a constant offset $b_j$ from the origin.  The vectors $\vv{v}$ are the \textit{support vectors} from training (all other training vectors find their $\lambda=0$), the $y$'s are their \textquotedblleft truth\textquotedblright\ values ($\pm 1$), and the $\lambda$'s are parameters found in the training process along with $b_j$.  Hence, $|w_j\rangle$ is a linear combination of training vectors mapped by $\vv{\varphi}$ to an alternative vector space.  The bra-ket notation here serves as a reminder that these vectors belong to this mapped space.  The inner product of two vectors in the mapped space given their non-mapped vectors $\vv{x_{1}}$ and $\vv{x_{2}}$ is determined via the kernel function $K(\vv{x_{1}}, \vv{x_{2}})$.  The SVMs in this analysis use the Gaussian kernel:
 
\begin{equation*}
K(\vv{x_{1}}, \vv{x_{2}})=\vv{\varphi}(\vv{x_{1}}) \cdot \vv{\varphi}(\vv{x_{2}}) = \exp(-|\vv{x_{1}} - \vv{x_{2}}|^2/2 \sigma^2)\,.
\label{eqn:gaussian_kernel}
\end{equation*}
 
The width $\sigma$ is an input parameter of the training process, along with an additional positive constant $C$ which limits the range of the $\lambda$'s and is necessary for soft-margin SVMs.
 
In order to construct an orthonormal basis from the three trained SVMs, the Gram--Schmidt procedure~\cite{Gram-Schmidt} is used with their $|w_j\rangle$ vectors:
 
\begin{equation*}
|{w^\prime_{1}}\rangle=|w_{1}\rangle
\quad\text{,}\quad
|{w^\prime_{2}}\rangle=|w_{2}\rangle-\frac{\langle w_{1}|w_{2}\rangle}{\langle w_{1}|w_{1}\rangle}|w_{1}\rangle
\quad\text{,}\quad
|{w^\prime_{3}}\rangle=|w_{3}\rangle-\frac{\langle w_{1}|w_{3}\rangle}{\langle w_{1}|w_{1}\rangle}|w_{1}\rangle-\frac{\langle w^{\prime}_{2}|w_{3}\rangle}{\langle w^{\prime}_{2}|w^{\prime}_{2}\rangle}|w^{\prime}_{2}\rangle\,.
\label{eqn:gs_w3prime}
\end{equation*}
 
Using this basis, 3-tuples (X,Y,\,Z) for a \textit{decision space} are created:
 
\begin{equation*}
\textrm{X}(\vv{z})=\frac{\langle z|w^{\prime}_{1}\rangle}{\sqrt{\langle w^{\prime}_{1}|w^{\prime}_{1}\rangle}}
\quad\text{,}\quad
\textrm{Y}(\vv{z})=\frac{\langle z|w^{\prime}_{2}\rangle}{\sqrt{\langle w^{\prime}_{2}|w^{\prime}_{2}\rangle}}
\quad\text{,}\quad
\textrm{Z}(\vv{z})=\frac{\langle z|w^{\prime}_{3}\rangle}{\sqrt{\langle w^{\prime}_{3}|w^{\prime}_{3}\rangle}}\,.
\label{eqn:orthoZ}
\end{equation*}
 
In this way, an input vector $\vv{z}$ describing an event has coordinates in the XYZ space given by calculating ( X$(\vv{z})$, Y$(\vv{z})$, Z$(\vv{z})$ ).  It is these new coordinates in the \textit{decision space} which are used to describe all events, and this is the space in which the 3D templates of the likelihood function are created.
 
\subsection{Physics observables}
 
In this analysis, $21$ physics observables are used to distinguish \ttbar events from background events (see Table~\ref{tab:svms_features}). Twenty are kinematic variables, and one comprises the $b$-tagging information of the event.  These include the electron or muon momentum, the number of jets in the event, the magnitude and direction of the missing transverse momentum vector, sums of the jet momenta components, the first five Fox--Wolfram moments (FWM), $H_{\mathrm{T}}$, the two largest eigenvalues of the normalized momentum tensor, and the mass of the lepton+jets system ($m_{\ell\mathrm{j}}$). The $H_{\mathrm{T}}$ is the scalar sum of \MET, electron $p_{\mathrm{T}}$ or muon $p_{\mathrm{T}}$, and the $p_{\mathrm{T}}$ of all jets passing the selection requirements.
 
Fox--Wolfram moments~\cite{FWM} were originally introduced for $e^+e^-$ colliders.  The FWMs correspond to a decomposition of the event's phase space into Fourier modes on the surface of a sphere.  They were modified for use at the Tevatron and the LHC to characterize the complex shapes of final states at hadron colliders~\cite{FWinHEP2,FWinHEP3}.  They form a set of event shape variables, and the $l^{\text{th}}$ FWM ($H_{l}$) is defined in the following way:
 
\begin{equation*}
H_{l}=\frac{4\pi}{2 l + 1}\sum_{m=-l}^{l}{\left|\sum_{i}^{\text{jets}} \frac{E_{\mathrm{T}}(i)}{E_{\mathrm{T}}(\text{total})} Y_l^m(\theta_{i},\phi_{i})\right|^{2}}\,.
\label{eqn:fox_wolfram}
\end{equation*}
 
\noindent The $Y_l^m$'s are the spherical harmonics, $i$ runs over all selected jets in the event, and $E_{\mathrm{T}}{(\text{total})}$ represents the sum of the transverse energy from selected jets.  The angles $\theta_{i}$ and $\phi_{i}$ indicate the direction of the $i^{\text{th}}$ jet.  This analysis makes use of $H_{1}$ through $H_{5}$.
 
The normalized momentum tensor uses the \MET\ and the momenta of the lepton and up to five jets, and has the following form:
 
\begin{equation*}
P_{ij}= \frac{{\met}_{i}~{\met}_{j} }{|\met|^2} + \sum_{}^{\text{lep}+5\text{jets}}\frac{p_{i} p_{j}}{|p|^2}\,.
\label{eqn:p_tensor}
\end{equation*}
 
\noindent Here $i$ and $j$ run over the $x$, $y$, and $z$ components of momentum (for \MET, only the $x$ and $y$ components are nonzero).  The two largest eigenvalues of this \textquotedblleft $p$-tensor\textquotedblright\ are used as SVM inputs.
 
Because the lepton+jets decays are rotationally invariant in $\phi$, some variables are calculated with respect to the lepton direction in the plane transverse to the beam.  Hence, for the momenta of jets, $p_{\parallel}$ and $p_{\bot}$ denote the components which are parallel and perpendicular to the direction of the lepton in the transverse plane.  Similarly, the $\phi(\MET)$ variable is then the angle between the transverse momentum of the lepton and the missing transverse momentum vector, and $p_{\parallel}$ for the lepton corresponds to its entire transverse momentum.
 
The SVMs treat each variable as one of the coordinates of a point in a $21$-dimensional space.  The algorithm requires that each variable should fall roughly in the same numeric range so that all features contribute a similar weight when evaluating the distance from the separating hyperplane.  The variables which have values outside the range of ($-1$, $+1$) are transformed such that they approximately meet this requirement.  All input variables and the values that were used to scale them are listed in Table~\ref{tab:svms_features}.
 
\begin{table}[ht]
\caption{\footnotesize List of the 21 variables used as input to the SVMs.  The variables were divided by the given values to make them all of similar magnitude.}
\begin{center}
\begin{tabular}{clr} \toprule{Number} & {Feature} & {Divided by}  \cr \midrule
\phantom{1}1&\MET~[\GeV]&250\\
\phantom{1}2&$\phi(\MET)$ [radians]&$2\pi$\\
\phantom{1}3&Lepton $E$ [\GeV]&400\\
\phantom{1}4&Lepton $p_\parallel$ [\GeV]&400\\
\phantom{1}5&Lepton $p_{\mathrm{z}}$ [\GeV]&400\\
\phantom{1}6&Mass(lepton+jets) [\GeV]&750\\
\phantom{1}7&Fox--Wolfram moment 1&1\\
\phantom{1}8&Fox--Wolfram moment 2&1\\
\phantom{1}9&Fox--Wolfram moment 3&1\\
10&Fox--Wolfram moment 4&1\\
11&Fox--Wolfram moment 5&1\\
12&Sum all jets $E_{\mathrm{T}}$ [\GeV]&500\\
13&Sum all jets $E$ [\GeV]&750\\
14&Sum all jets $p_\parallel$ [\GeV]&750\\
15&Sum all jets $p_\bot$ [\GeV]&750\\
16&Sum all jets $p_{\mathrm{z}}$ [\GeV]&750\\
17&$H_{\mathrm{T}}$  [\GeV]&500\\
18&$p$-tensor eigenvalue 1&1\\
19&$p$-tensor eigenvalue 2&1\\
20&Number of jets&10\\
21&Number of $b$-tags&10\\
\bottomrule
 
\end{tabular}
\end{center}
\label{tab:svms_features}
\end{table}
 
\subsection{SVM training\label{sec:training}}
 
The SVMs are trained to separate three classes of events: the Light and Heavy backgrounds and the \ttbar signal.  In order to train the SVMs, the Monte Carlo simulation samples and a data sample representing the multi-jet background are split into two subsamples.  For training purposes, events from each class are randomly selected from those passing the selection requirements.  The remaining events are used to test how well the trained SVMs perform.  Also, it is only these remaining events that are utilized in the subsequent analysis.
 
The training process aims to find the set of \textit{support vectors} that forms the optimal decision plane in the mapped space induced by the kernel function.  As described in Section~\ref{sec:SVMs}, there are two free parameters that need to be specified when training.  These are the $\sigma$ parameter of the Gaussian kernel and $C$, the positive constant which constrains the $\lambda$'s in the solution.  A search grid over the values of these parameters was implemented, and the performance of a given training was then evaluated based upon the area of the resulting receiver operating characteristic (ROC) curve created with the events not used in the training.  As a result of this study, the values of 1.2 for $\sigma$ and 2.0 for $C$ were settled upon.  The \ttbar Signal class and the two background classes, Light and Heavy, each used 8,000 events for training, which is a small fraction of the total available events.  Increasing the number of training events was not found to improve performance.  It was also verified that the trained SVMs were not overtrained (i.e., that their discriminant distributions generalize well from the training set to the full class dataset).

\subsection{Class templates}
Different physics processes can be distinguished by their distinctly different distributions in the XYZ \textit{decision space}.  These distributions are obtained by applying the Gram--Schmidt procedure to each event's SVM output values.  Histogramming the different physics processes in the resulting 3D \textit{decision space} creates probability distribution functions (i.e., templates) that can then be used in the likelihood fit.
 
\begin{figure}[htbp]
\centering
\begin{subfigure}[b]{0.49\textwidth}
\includegraphics[width=\linewidth]{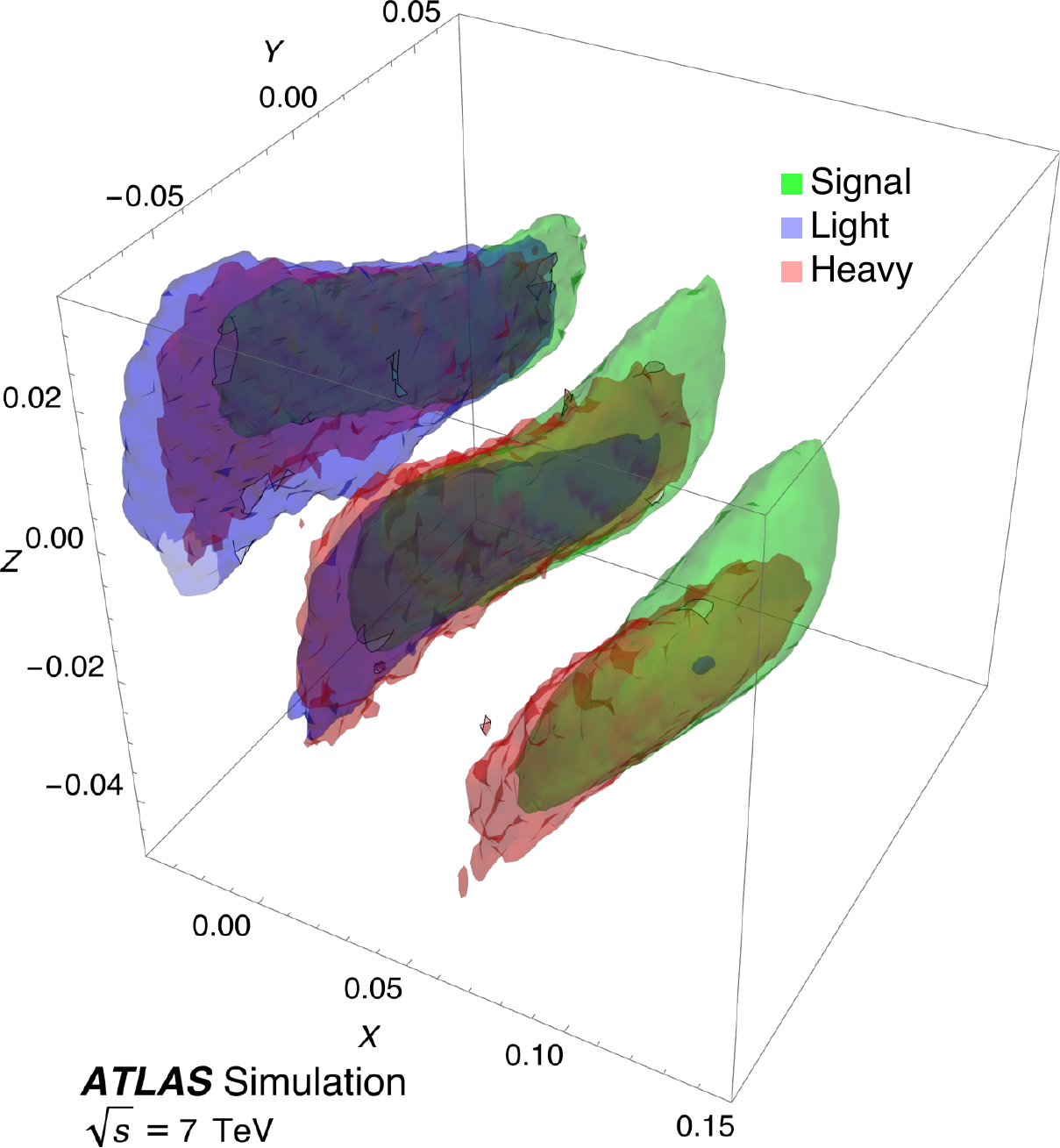}
\caption{}
\label{fig:template_contour_left}
\end{subfigure}
\begin{subfigure}[b]{0.49\textwidth}
\includegraphics[width=\linewidth]{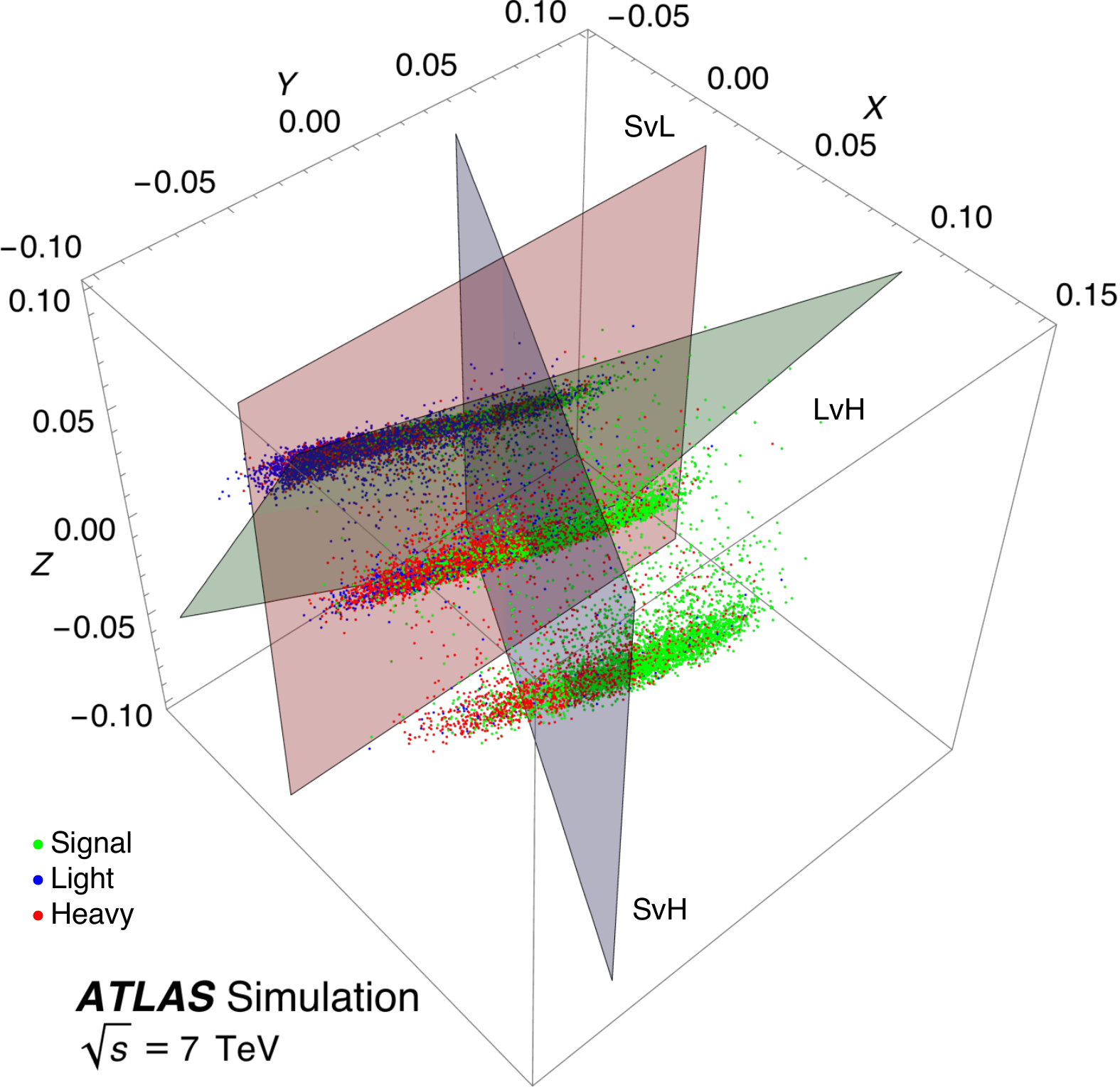}
\caption{}
\label{fig:template_contour_right}
\end{subfigure}
\caption{\label{fig:template_contour} \footnotesize Distribution of the three classes of events shown in the 3D orthonormalized \textit{decision space}.  \subref{fig:template_contour_left} Contours from each class are shown together.  \subref{fig:template_contour_right} A view of the SVM \textit{decision space} is shown with a random sampling of 10,000 events from each class.  The three planes corresponding to each of the three trained SVMs are also depicted.
}
\end{figure}

Figure~\ref{fig:template_contour_left} shows a contour at a fixed value of the template function from each of the classes.  This highlights the different regions in 3D \textit{decision space} where the different event types congregate.  As an alternative way of illustrating these distributions, Figure~\ref{fig:template_contour_right} shows a sampling of 10,000 events from each class.  Also shown are the three trained decision planes, which serve to demonstrate the nonorthogonal nature of the basis defined by the SVMs.  The SvL plane at $\text{X} = 0$ is seen to separate Signal from Light as it extends downwards in Z.  Similarly, the SvH and LvH planes separate their training classes.  The multiple band structure in the templates arises because of training with the number of $b$-tags, which is a strong discriminant and is discrete.  These template bands represent groupings of events with 0, 1, or 2 $b$-tags.

In order to minimize the potential effect of small fluctuations in the modeling, a small set of wider bins was constructed.  The full 3D XYZ \textit{decision space} was organized into four quadrants by dividing the space at $\text{Y}=0$ and $\text{Z}=-0.01$.  Each quadrant was then further divided into bins along the X axis.  The quadrants are designated YZ, Yz, yZ, and yz, where the capitalization of the letters indicates where the quadrant is located (e.g., Yz means $\text{Y} > 0.00$ and $\text{Z} \le -0.01$).  The division points for X were chosen to keep a minimum of approximately 1,000 events in each bin while preserving the shapes of the distributions.  It is these four binned distributions from the quadrants that are used for the final profile likelihood fit.
 
\subsection{Cross-section measurement}
 
A binned profile likelihood function is used in a fit to determine the \ttbar cross section from the data.  In the likelihood fit, four templates are used: \ttbar, $W$/$Z$, single-top, and fake lepton.  In evaluating the systematic uncertainties, particularly with respect to the modeling of \ttbar, it was observed that a large uncertainty occurs because of the similarity between the final states of \ttbar events and single-top background events.  To alleviate this effect the single-top backgrounds, which arise from electroweak processes rather than the strong interactions responsible for the production of \ttbar pairs, are combined into a single template normalized to their predicted cross sections.  The $W$/$Z$+$b\bar{b}$+jets, the light-flavor $W$/$Z$+jets, and the diboson backgrounds are combined into a $W$/$Z$ template.  The normalizations of the \ttbar, $W$/$Z$, and fake-lepton templates are free parameters of the fit.
 
The grouping of physics processes used when constructing the templates for the fit can differ from the class definitions used for training.  At the training stage, the events are arranged in order to create SVMs that can distinguish between the \ttbar signal, the Light backgrounds, and the Heavy backgrounds.  After training, each physics process can be reassigned to templates.  The chosen allocation of the physics processes to four templates (\ttbar, $W$/$Z$, single-top, and fake lepton) results in smaller expected uncertainties in the \ttbar cross section.
 
The likelihood function uses the templates (projections onto the X axis from each of the quadrants) that have been built in the XYZ \textit{decision space}.  Each template has an associated strength parameter $\theta$ in the likelihood.  The maximum value of the likelihood is obtained in determining the central values of the $\theta$ parameters.  The systematic uncertainties of the fit results are also included in the likelihood as nuisance parameters (NPs, $\alpha$'s, or collectively $\vv{\alpha}$) with Gaussian constraints.  Each template is a function of the nuisance parameters in the likelihood, which is then able to capture the effects due to each source of systematic uncertainty.
 
The likelihood of an unknown sample for an $n_\text{T}$ template problem is defined as
 
\begin{equation*}
L = \prod_{i}^{\text{bins}} P( \mathscr{E}_{i}, o_{i} )\times \prod_{j}^{\text{NPs}} G(\alpha_j,\sigma_j) \,.
\label{eqn:likelihood1}
\end{equation*}
 
\noindent Here $G(\alpha_j,\sigma_j)$ is the Gaussian constraint for the $j^{\text{th}}$ NP, $\alpha_j$, with the corresponding uncertainty $\sigma_j$; and $P( \mathscr{E}_{i}, o_{i} )$ is the Poisson probability mass function for the $i^{\text{th}}$ bin given the observed number of events, $o_{i}$, and the expected number of events, $\mathscr{E}_{i}$ :

\begin{equation*}
\mathscr{E}_{i} =\mathscr{E}(\vv{\theta},\vv{\alpha})_{i} = \sum_{j}^{n_\text{T}}{ \theta_{j} N(\vv{\alpha})_{j} T_{j}(i,\vv{\alpha}) }\,.
\label{eqn:likelihood2}
\end{equation*}
 
\noindent The templates are constructed such that $T_{j}(i,\vv{\alpha})$ gives the fractional number of events in template $T_{j}$'s $i^{\text{th}}$ bin.  Consequently, the sum over all bins of a given template is equal to $1$.  The $N(\vv{\alpha})_{j}$ are defined to be the total number of events expected from the $j^{\text{th}}$ template assuming an integrated luminosity $\mathscr{L}$.  To calculate this value, the following sum over all modeled processes belonging to the $j^{\text{th}}$  template is computed:
 
\begin{equation*}
N(\vv{\alpha})_{j} =  \sum_{k}^{\text{processes}}{ \sigma_{k} \epsilon(\vv{\alpha})_{k} \mathscr{L} }\,.
\label{eqn:likelihood3}
\end{equation*}
 
\noindent The $\sigma_{k}$ and $\epsilon_{k}$ are the cross section and acceptance for the $k^{\text{th}}$ physics process.  They are derived from MC simulation.  For the multi-jet background, $N_{j}$ is taken from the fake-lepton estimate.
 
A maximum-likelihood fit is performed to extract the values of the $\theta$ and $\alpha$ parameters.  The $1 \sigma$ uncertainty for a given parameter is taken to be the change in the value of that parameter which causes $\ln L$ to decrease by $0.5$ away from $\ln L_{0}$, when $\ln L$ is maximized with respect to all other free parameters and where $\ln L_{0}$ is the global maximum.  All $\theta$ and $\alpha$ parameters have both their $+1 \sigma$ and $-1 \sigma$ uncertainties determined in this way.
 
The $\theta$ for the \ttbar class, multiplied by the assumed cross section, gives the measured value of the \ttbar cross section.  Similarly, the uncertainty in $\theta_{t\bar{t}}$ from the fit, multiplied by the assumed cross section, gives the uncertainty in the $\sigma_{t\bar{t}}$ measurement.

\subsection{Systematic uncertainties\label{sec:Systematic}}
 
All systematic uncertainties were evaluated using the profile likelihood fit.  The systematic effects are incorporated into the templates, and each template is associated with appropriate nuisance parameters ($\alpha$'s) in the likelihood.  A nuisance parameter that takes a value of $0$ in the fit keeps the nominal template, while a value of $+1$ or $-1$ changes the template to look like the $+1\sigma$ / $-1\sigma$ effect.  Templates at intermediate values of the $\alpha$'s are linearly interpolated.  A Gaussian constraint is also applied to each $\alpha$ in order to propagate its controlled uncertainty when the data have no preference for that systematic effect.  The profile likelihood fit then provides a simultaneous measurement of the $\theta$ and $\alpha$ parameters.  In this way, the systematic effects are converted into a statistical framework that properly takes into account correlations and which can potentially lower the uncertainties in the measurement.
 
The individual effects of various sources of systematic uncertainty are displayed in Table~\ref{tab:2011_M22_TRExFitter_FINAL_syst_full_breakdown}.  They are obtained by leaving groupings of nuisance parameters out of the fit, and calculating the square of each effect as the difference of the squares of the total error and the residual error.
 
\noindent \textbf{Object modeling}
 
Systematic uncertainties in the lepton selection arise from uncertainties in lepton identification, reconstruction, and triggering.  These are evaluated by applying tag-and-probe methods to $Z\rightarrow\ell\ell$ events~\cite{PERF-2013-03}.  Uncertainties due to the energy scale and resolution are also considered for electrons and muons.  These effects are evaluated by assigning each of them a separate nuisance parameter in the likelihood so as to allow the error source to be shifted both upwards and downwards by its uncertainty.  The resulting systematic effects are summarized in Table~\ref{tab:2011_M22_TRExFitter_FINAL_syst_full_breakdown} as \textit{Leptons}.

For jets, the main source of uncertainty is the jet energy scale (JES).  The JES and its uncertainty are evaluated using a combination of test-beam data, LHC collision data, and simulation~\cite{PERF-2012-01}.  As a result of the \emph{in situ} analyses for the calibration of the full 2011 dataset, the correlations between various JES uncertainty components are encoded in 21 subcomponents.  These include statistical and method uncertainties, detector uncertainties, modeling and theory uncertainties, mixed detector and modeling uncertainties, and pileup.  The JES uncertainty is evaluated by assigning a separate NP to each of these 21 JES subcomponents.  The jet energy resolution is separated by process ($t\bar{t}$, single-top, and $W$/$Z$+jets) and is assigned three corresponding NPs.  These extra degrees of freedom allow differences in the kinematics and prevalence of $b$-quark, light-quark, and gluon jets in these processes to be better represented in the profile likelihood.  The resulting uncertainties in $\sigma_{t\bar{t}}$ from these sources are indicated in Table~\ref{tab:2011_M22_TRExFitter_FINAL_syst_full_breakdown} as \textit{Jets}.

The jet-flavor-dependent efficiencies of the $b$-tagging algorithm are calibrated using dijet events, and dilepton \ttbar events from data.  Differences in the $b$-tagging efficiency as well as $c$-jet and light-jet mistag rates between data and simulation are parametrized using correction factors, which are functions of $p_{\mathrm{T}}$ and $\eta$~\cite{PERF-2012-04}.  The $b$-tag systematic uncertainties were evaluated by constructing nine NPs that correspond to unique bins in jet $p_{\mathrm{T}}$, as the uncertainties at low and high jet $p_{\mathrm{T}}$ should be largely uncorrelated.  Single NPs were used for the \textit{$c$-tag} and \textit{Mistag} systematic uncertainties.  These systematic effects appear in Table~\ref{tab:2011_M22_TRExFitter_FINAL_syst_full_breakdown} as three uncertainties labeled \textit{$b$-tag}, \textit{$c$-tag}, and \textit{Mistag}.

During the variation of nuisance parameters related to jets and leptons, the \MET\ is recalculated in accordance with the changes caused by those systematic effects.  In this way, the jet and lepton uncertainties are propagated to the \MET.  However, the \MET\ uncertainty due to calorimeter cells not assigned to any other physics object is evaluated individually.  Also, an additional 6.6\% uncertainty due to pileup is applied to \MET.  Both of these are given separate NPs in the profile likelihood, and they are listed in Table~\ref{tab:2011_M22_TRExFitter_FINAL_syst_full_breakdown} under \textit{Missing transverse momentum}.
 
\noindent \textbf{Modeling of $\boldsymbol{\ttbar}$ events}

Systematic uncertainties due to the choice of \ttbar MC generator are evaluated by taking the full difference between \POWHER (AF) and $\MCatNLO{+}\HERWIG$ (AF).  The systematic uncertainty due to the choice of parton shower model is taken as the full difference between \POWHER (AF) and \POWPY[6] (AF).  These are listed as \textit{Generator} and \textit{Shower/hadronization} in Table~\ref{tab:2011_M22_TRExFitter_FINAL_syst_full_breakdown}, respectively.

The systematic error due to uncertainties in the modeling of initial- and final-state radiation (ISR/FSR) is evaluated using \ALPGEN interfaced to \PYTHIA[6].  Monte Carlo samples were created in which the parameter that controls the amount of ISR/FSR in \ALPGEN was either halved or doubled.  Half of the spread between the \ALPGEN samples with raised and lowered ISR/FSR parameter values is taken as the systematic error.

The uncertainty due to renormalization and factorization scales is evaluated with two modified samples, generated with \MCatNLO interfaced with \HERWIG, in which parameters controlling the renormalization and factorization scales, introduced to cure the ultraviolet and infrared divergences in ME calculations, are simultaneously either halved or doubled.  The full difference between the two samples is taken as the \textit{Renormalization/factorization} error.

Each of the major $t\bar{t}$ modeling systematic uncertainties (\textit{Generator}, \textit{Shower/hadronization}, \textit{ISR/FSR}, and \textit{Renormalization/factorization}) is given a shape NP in each quadrant.  The uncertainty in the normalization of events is assigned two NPs.  One of these is used to track the migration of events between quadrants, and it mirrors the change in the relative normalization of the quadrants as observed when comparing the nominal and systematically shifted samples.  The second NP is taken as an overall normalization error which corresponds to the normalization difference seen for the full event selection (where all four quadrants are combined).  Therefore, each of the $t\bar{t}$ modeling uncertainties mentioned above has six NPs (four for shape, and two for normalization).

The underlying-event modeling error is evaluated by comparing two different \ttbar MC event samples produced with varied parameters in \POWPY[6].  One was generated with the Perugia 2011 central tune, and the other with Perugia 2011 \textsc{mpiHI}~\cite{Skands:2010ak}.  Both of these samples use the P2011 CTEQ5L \PYTHIA tune, and not P2011C CTEQ6L1, which applies to the nominal \ttbar MC sample.  Their full difference is used as the measurement uncertainty for the underlying event, using a single NP in the profile likelihood.
 
All particles in the final state from the LHC $pp$ collisions must be color singlets.  Different schemes for the color reconnection (CR) of the beam remnant and other outgoing hard collision objects are examined.  The \ttbar cross-section uncertainty due to this effect is estimated by comparing two different \ttbar MC samples produced with \POWPY[6].  A reference sample was obtained using the Perugia 2011 central tune.  The other sample was generated with Perugia 2011 \textsc{noCR}~\cite{Skands:2010ak}, and has modified color reconnection parameters.  These samples use the P2011 CTEQ5L \PYTHIA tune.  The full difference between these two samples is taken as the CR uncertainty, using a single NP.

To estimate the uncertainty due to the choice of parton distribution function, the CT10 PDF set parametrization is examined using its 26 upwards and downwards systematic variations.  Each of the 26 CT10 eigenvector components is assigned a separate NP in the profile likelihood.

\noindent \textbf{Background modeling}
 
To estimate the error due to the shape of the $W$/$Z$+jets backgrounds and to assess the effect of any mismodeling, background samples are reweighted to match data for each of the following variables, taken one at a time, in a signal-depleted control region:  lepton $E$, $\phi(\MET)$, $m_{\ell\mathrm{j}}$, $\sum_{\text{jets}}{p_\parallel}$, and $\sum_{\text{jets}}{p_\bot}$.  This control region was defined as events matching the nominal selection, but containing exactly three jets, none of which are $b$-tagged.  The reweighting functions were applied only to $W$/$Z$+jets samples (leaving $t\bar{t}$, single-top, and fake lepton untouched).  Five NPs in the profile likelihood implement these functions such that the NPs turn the reweighting effects on and off, each according to the differences seen for these five variables in the data.  In Table~\ref{tab:2011_M22_TRExFitter_FINAL_syst_full_breakdown}, these effects appear under the heading \textit{$W$/$Z$ reweighting}.  For the single-top shape, \ACERMC samples with raised and lowered ISR/FSR parameter values are compared, and a single NP is assigned to this systematic uncertainty.
 
The effects due to the uncertainty of the single-top, $W$+jets, and $Z$+jets cross sections are investigated by varying these cross sections within their theoretical errors. For the $W$+jets background, a 4\% uncertainty is applied to the inclusive $W$ boson cross section, with an additional 24\% uncertainty added in quadrature at each ascending jet multiplicity~\cite{Alwall:2007fs, TOPQ-2011-01}.  This method is also applied to the $Z$+jets cross sections.  To evaluate the systematic uncertainty in the \ttbar cross section due to the theoretical uncertainties in the single-top cross section, the single-top cross section is varied in accordance with the theoretical results, taken from Refs.~\cite{singleTopXsec, Aliev:2010zk, Kant:2014oha}.  The relative normalization within the $W$/$Z$+jets MC sample is varied by raising and lowering the corresponding nominal relative yields of each jet multiplicity by their respective errors.  Similarly, the relative normalization between fake-electron and fake-muon events is varied by raising and lowering their nominal predictions by their errors.  The resulting effects are evaluated using appropriate NPs added in the profile likelihood fit.  This uncertainty is quoted as \textit{$W$/$Z$ and fakes relative normalization} in Table~\ref{tab:2011_M22_TRExFitter_FINAL_syst_full_breakdown}.  Also, the uncertainty due to variations of the $W$+jets heavy-flavor fraction is included via three NPs in the profile likelihood.  These NPs place an additional $25$\% uncertainty on each of the assumed $W$+$b\bar{b}$+jets, $W$+$c\bar{c}$+jets, and $W$+$c$+jets cross sections.  Table~\ref{tab:2011_M22_TRExFitter_FINAL_syst_full_breakdown} summarizes these in the row labeled \textit{Heavy-flavor fraction}.
 
\noindent \textbf{Template statistics / luminosity}
 
For the profile likelihood fit, an additional fit parameter is introduced for each bin.  These parameters are used to represent the Poisson fluctuation of the predicted number of events in each bin as estimated from the size of the Monte Carlo samples.  The error propagated to $\theta_{t\bar{t}}$ from these additional parameters is then an appropriate representation of the MC statistical error.
 
The integrated luminosity measurement has an uncertainty of 1.8\%~\cite{DAPR-2011-01}, and therefore each physics process is assigned an uncertainty of this magnitude.  This systematic error is controlled by a single nuisance parameter in the likelihood.
 
The total measurement uncertainty, including individual groups of contributions, is listed in Table~\ref{tab:2011_M22_TRExFitter_FINAL_syst_full_breakdown}.  The largest uncertainties are due to the lepton selection and luminosity, followed by the uncertainties due to JES, $b$-tagging,  ISR/FSR, and other \ttbar modeling.
 
\noindent \textbf{Beam energy}
 
The LHC beam energy during the 2012 $\sqrt{s} = 8$~\TeV\ $pp$ run was measured to be within $0.1$\% of the nominal value of 4~\TeV\ per beam, using the revolution frequency difference of protons and lead ions during $p+Pb$ runs in early 2013 combined with the magnetic model errors~\cite{PhysRevAccelBeams.20.081003}.  A similar uncertainty in the beam energy is applicable to the 2011 LHC run.  The approach used in Ref.~\cite{TOPQ-2015-02} was therefore applied to the measurement using the $\sqrt{s} =  7$~\TeV\ dataset.  The uncertainty in the \ttbar theoretical cross section due to this energy difference was calculated to be 0.27\%, using the \TOPpp[2.0] program~\cite{Czakon:2011xx} and assuming that the relative change of the \ttbar cross section for a $0.1$\% change in $\sqrt{s}$ is as predicted by the NNLO + NNLL calculation.  It is negligible compared to other sources of systematic uncertainty.

\begin{table}[ht]
\caption{ \footnotesize Summary table of the measurement uncertainties.  Because the profile likelihood fit accounts for correlations, the total error is not simply the components added in quadrature.  Individual effects were obtained by leaving groupings of NPs out of the fit, and calculating the square of each effect as the difference of the squares of the total error and the residual error.}
\label{tab:2011_M22_TRExFitter_FINAL_syst_full_breakdown}
\centering
\begin{tabular}{lcccc} \toprule
\textsc{Source} & {$-1\sigma$ [pb]} & {$+1\sigma$ [pb]} & {$-1\sigma$ [\%]} & {$+1\sigma$ [\%]} \cr \midrule

\textsc{Object Modeling}\\
\hline
Leptons&$-3.1$&$+3.3$&$-1.8$&$+2.0$\\
Jets&$-2.9$&$+3.0$&$-1.7$&$+1.8$\\
$b$-tag&$-1.9$&$+2.0$&$-1.1$&$+1.2$\\
$c$-tag&$-0.4$&$+0.4$&$-0.3$&$+0.3$\\
Mistag&$-0.3$&$+0.3$&$-0.2$&$+0.2$\\
Missing transverse momentum&$-0.2$&$+0.2$&$-0.1$&$+0.1$\\
\hline
\textsc{$t\bar{t}$ Modeling}\\
\hline
Generator&$-1.6$&$+1.8$&$-1.0$&$+1.1$\\
Shower/hadronization&$-2.4$&$+2.6$&$-1.4$&$+1.5$\\
Renormalization/factorization&$-1.4$&$+1.4$&$-0.8$&$+0.9$\\
ISR/FSR&$-2.4$&$+2.5$&$-1.4$&$+1.5$\\
Underlying event&$-0.7$&$+0.8$&$-0.4$&$+0.5$\\
Color reconnection&$-0.5$&$+0.5$&$-0.3$&$+0.3$\\
PDF&$-1.8$&$+1.9$&$-1.1$&$+1.1$\\
\hline
\textsc{Background Modeling}\\
\hline
$W$/$Z$ reweighting&$-1.0$&$+1.0$&$-0.6$&$+0.6$\\
$W$/$Z$/fakes relative normalization&$-1.2$&$+1.2$&$-0.7$&$+0.7$\\
Heavy-flavor fraction&$-1.1$&$+1.2$&$-0.7$&$+0.7$\\
Single-top&$-1.0$&$+1.0$&$-0.6$&$+0.6$ \\
\hline
\textsc{Other}\\
\hline
Data statistics&$-0.7$&$+0.7$&$-0.4$&$+0.4$\\
Template statistics&$-1.0$&$+1.0$&$-0.6$&$+0.6$\\
Luminosity&$-3.2$&$+3.4$&$-1.9$&$+2.0$\\ \midrule
\textsc{Total}&$-6.8$&$+7.1$&$-4.0$&$+4.2$\\ \bottomrule
 
\end{tabular}
\end{table}
 
\clearpage
 
\section{Results}
 
The top quark pair-production cross section for $pp$ collisions at a center-of-mass energy of $\sqrt{s} = 7$~\TeV\ in the lepton+jets channel is found to be
\begin{center}
$\sigma_{t\bar{t}}=168.5\pm0.7$(stat.)~$^{+6.2}_{-5.9}$(syst.)~$^{+3.4}_{-3.2}$(lumi.)~pb.
\end{center}
This result includes all systematic uncertainties as evaluated with the profile likelihood fit, with the statistical and luminosity errors listed separately.
 
\begin{figure}[htbp]
\centering
\includegraphics[width=7.5cm,clip]{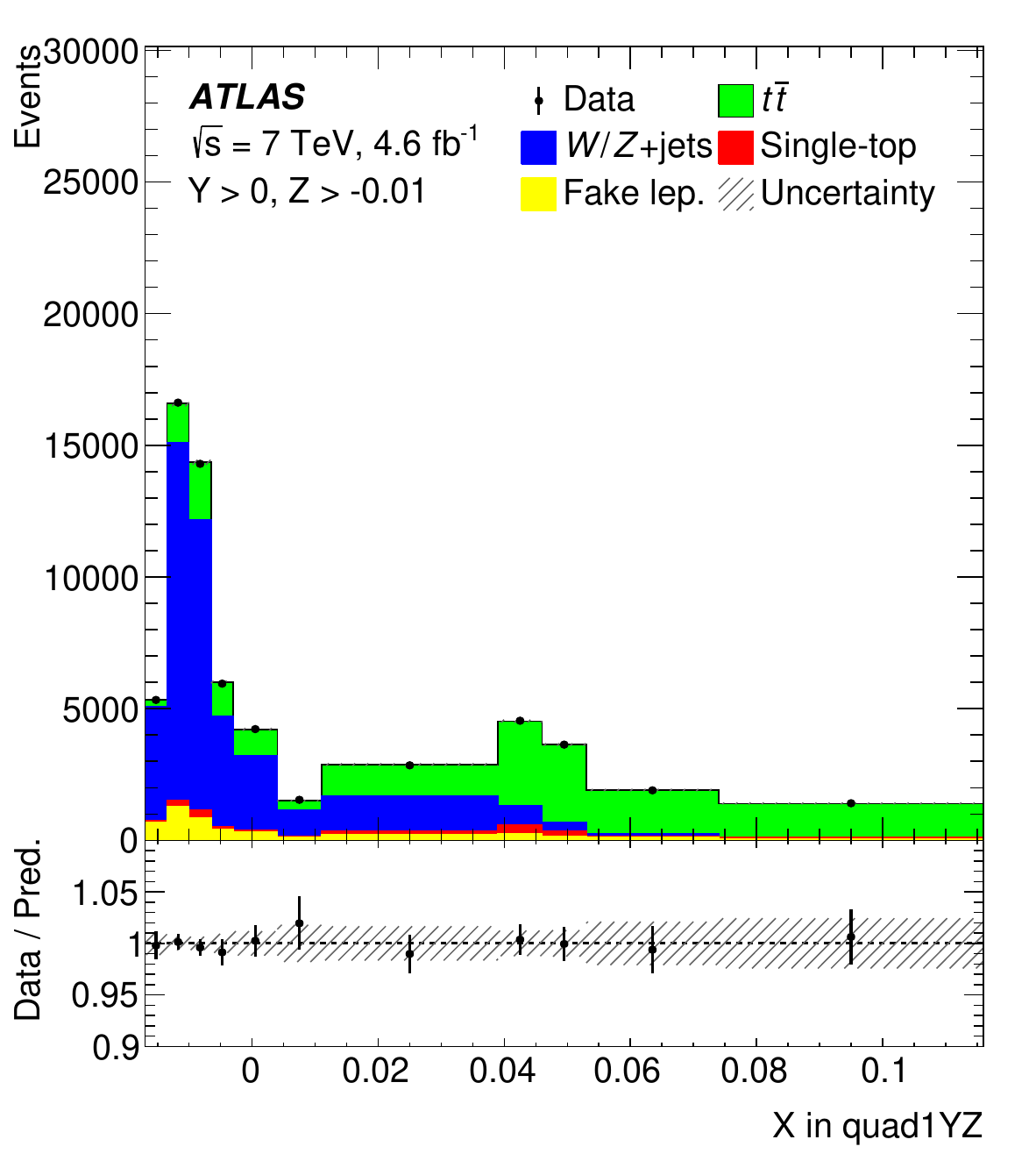}
\includegraphics[width=7.5cm,clip]{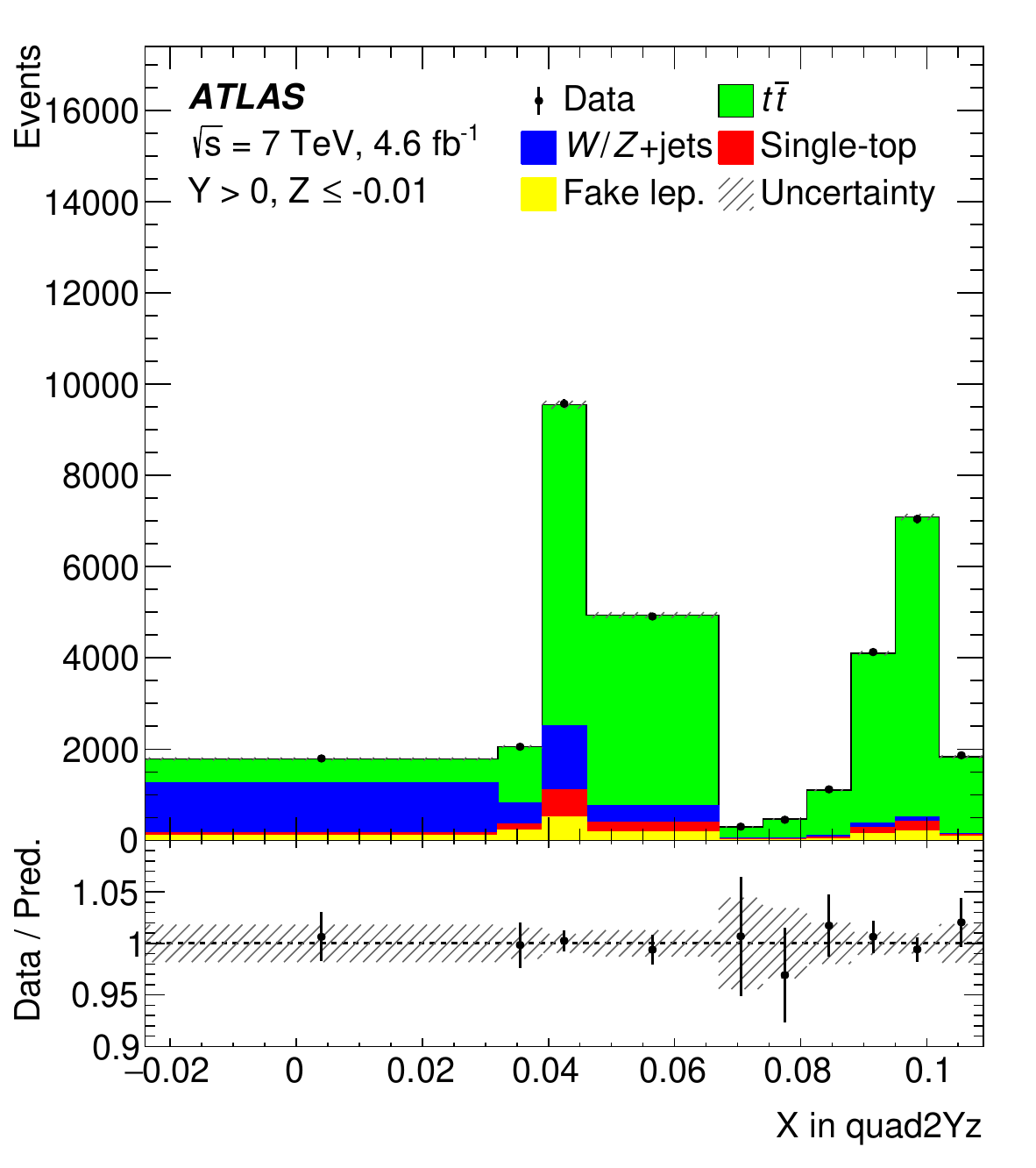}
\includegraphics[width=7.5cm,clip]{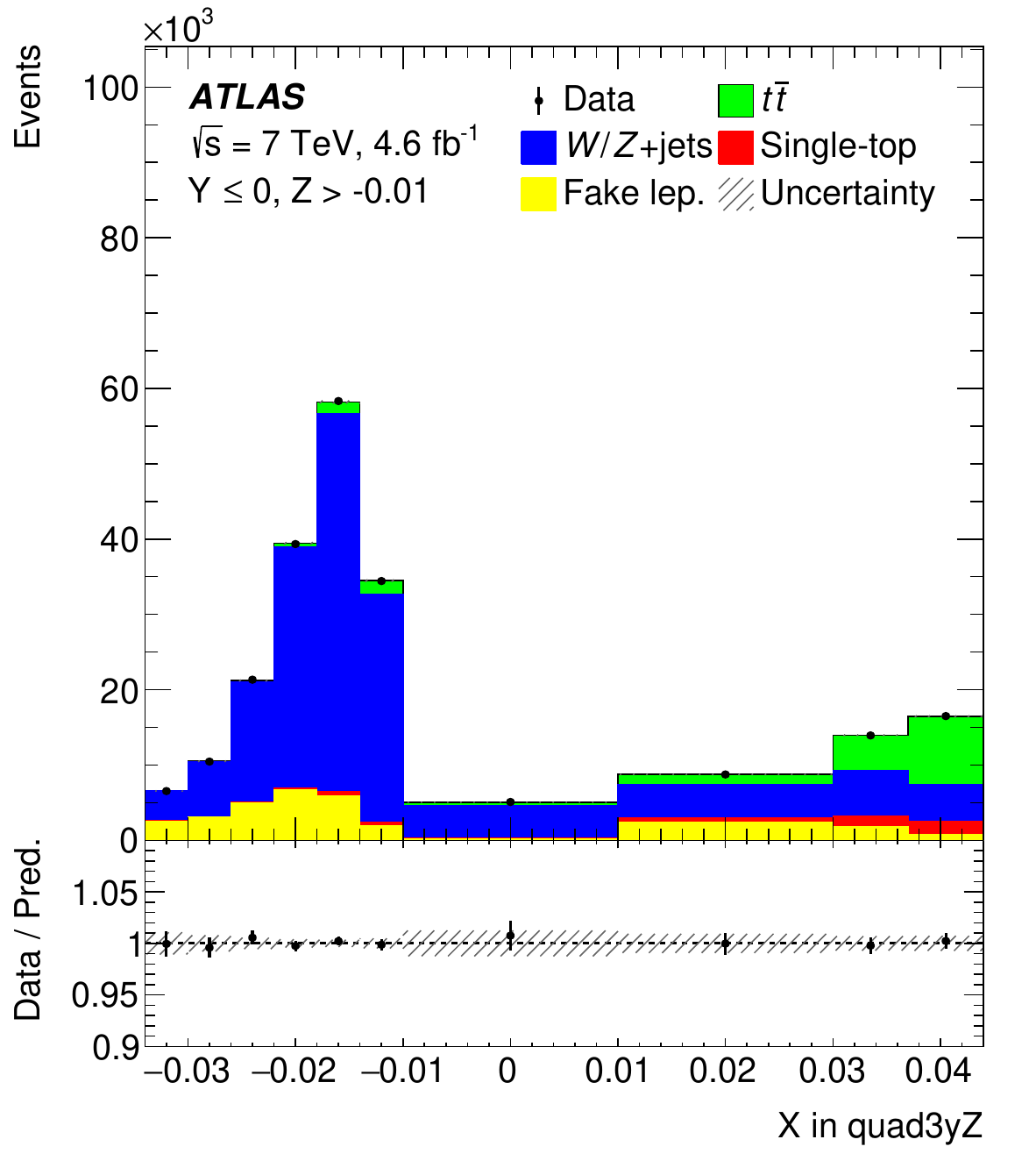}
\includegraphics[width=7.5cm,clip]{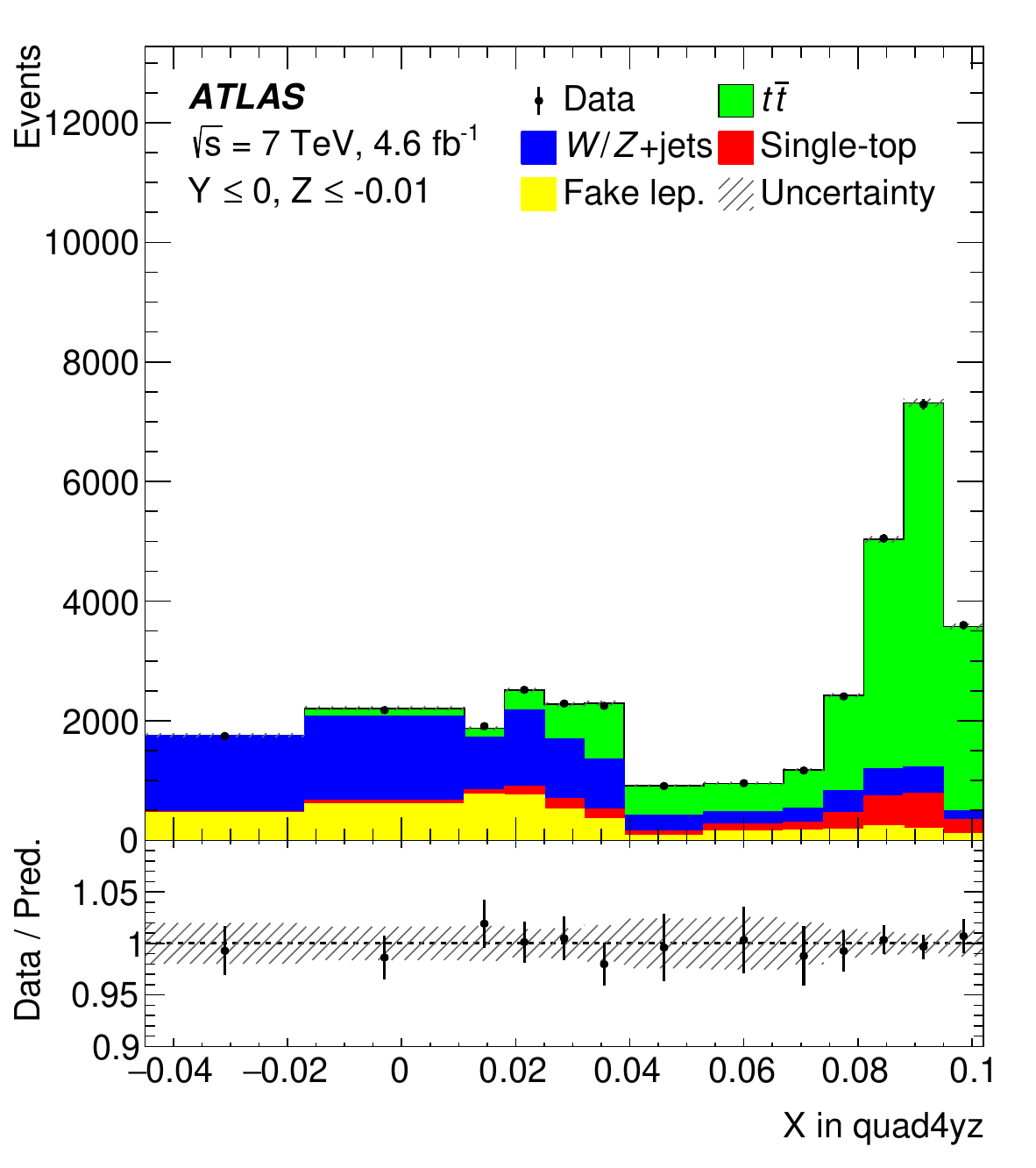}
\caption{\label{fig:dataMC2011fitQuadsTREx_final} \footnotesize The fitted yields of signal and background processes compared with data, shown in four YZ quadrants divided along the X axis, as used in the fit. They are labeled quad1YZ, quad2Yz, quad3yZ, and quad4yz (the boundary letters are appended for easy reference).  The lower panel shows the ratio of data to fit prediction. The shaded regions correspond to the statistical and systematic uncertainties.  The first and last bins also contain any events found outside the range of the horizontal axis.}
\end{figure}

Figure~\ref{fig:dataMC2011fitQuadsTREx_final} shows a comparison between the observed and fitted numbers of events in each of the quadrants.  A correlated $\chi^2$ test was used to check that there is good agreement between the data and the fit results within the combined statistical and systematic error bands.

\begin{figure}[htbp]
\centering
\begin{subfigure}[b]{0.35\textwidth}
\includegraphics[width=\linewidth]{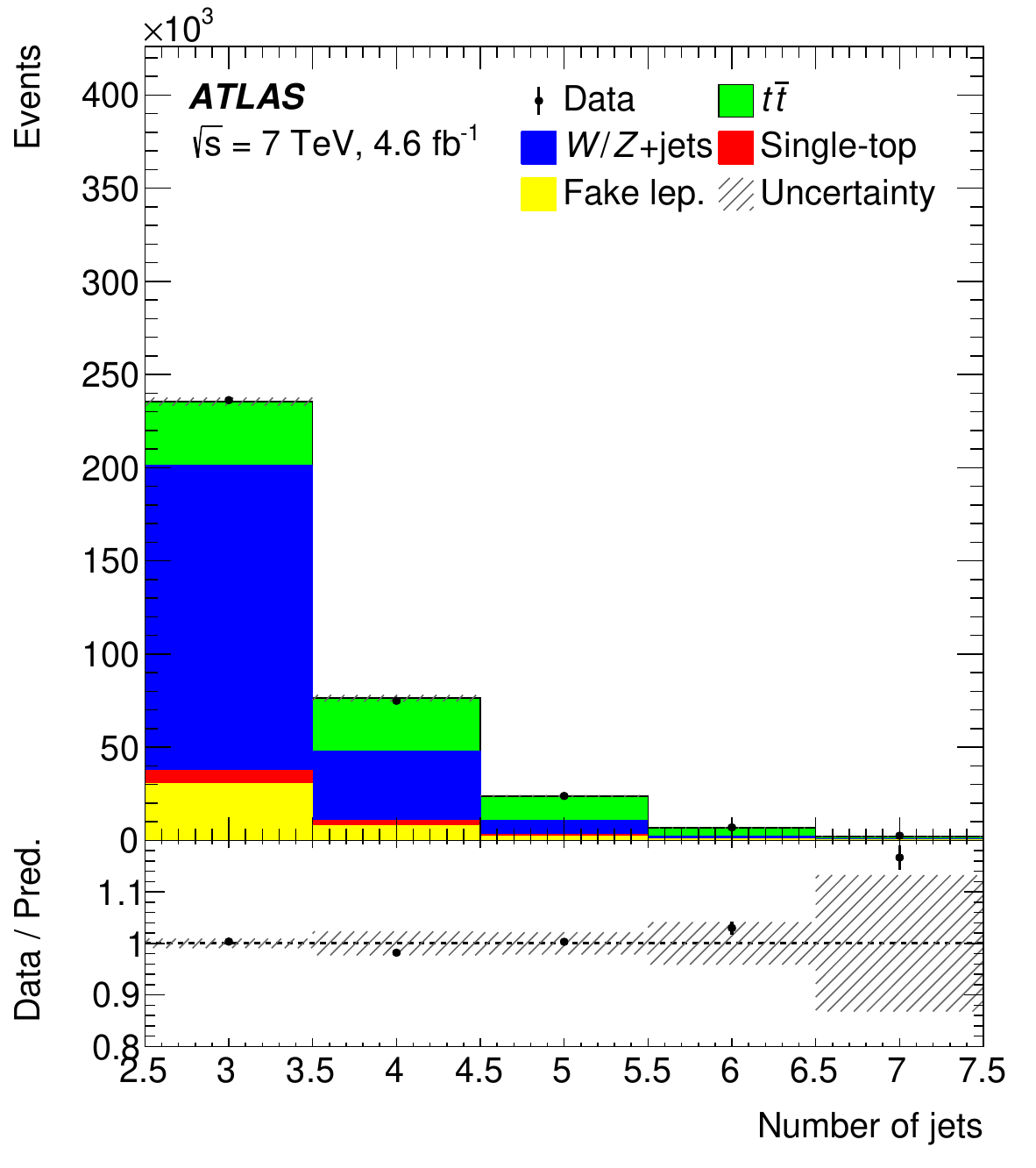}
\caption{}
\label{fig:dataMC2011fitProcessSelected_numJets}
\end{subfigure}
\begin{subfigure}[b]{0.35\textwidth}
\includegraphics[width=\linewidth]{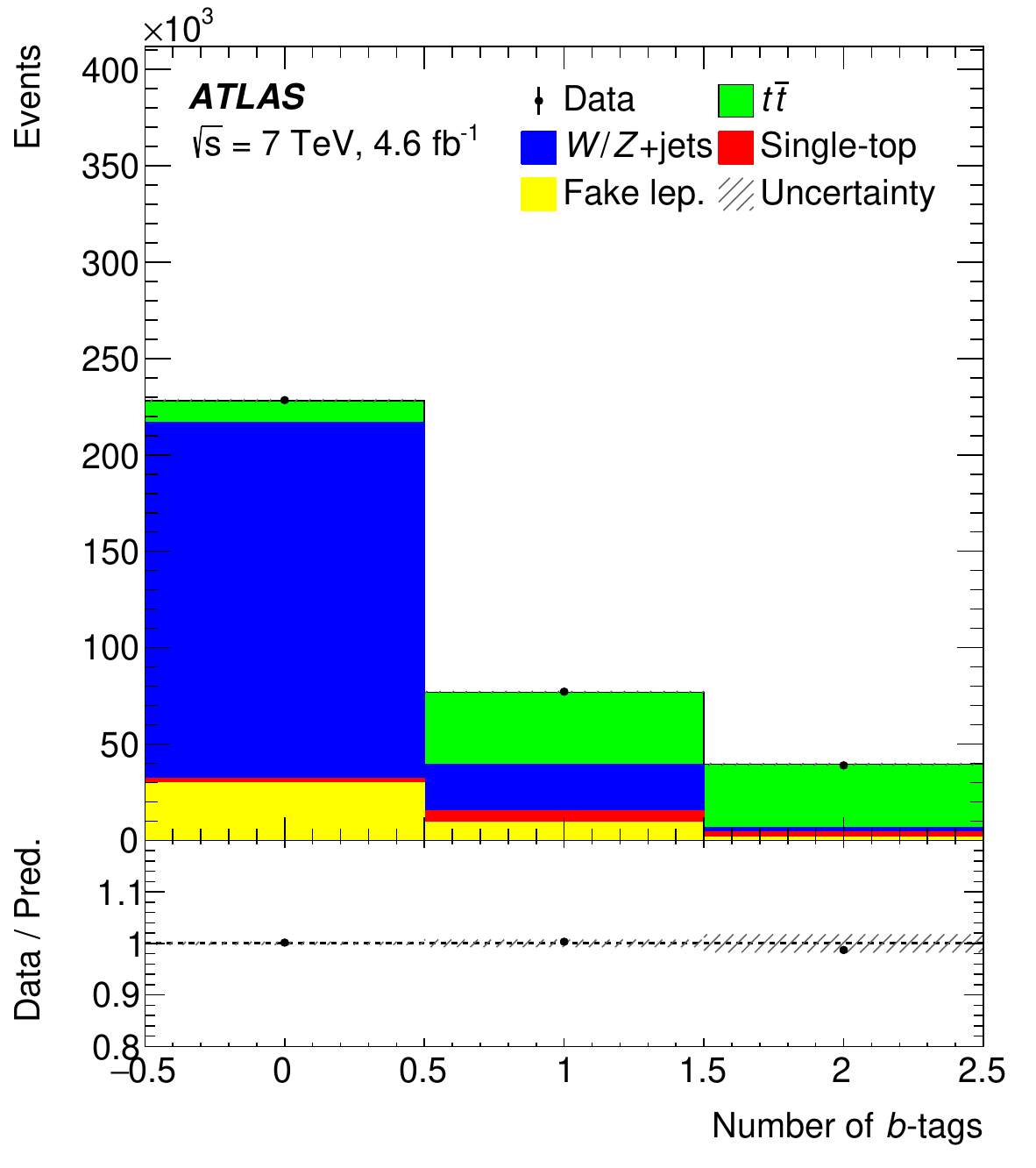}
\caption{}
\label{fig:dataMC2011fitProcessSelected_MV1}
\end{subfigure}
\begin{subfigure}[b]{0.35\textwidth}
\includegraphics[width=\linewidth]{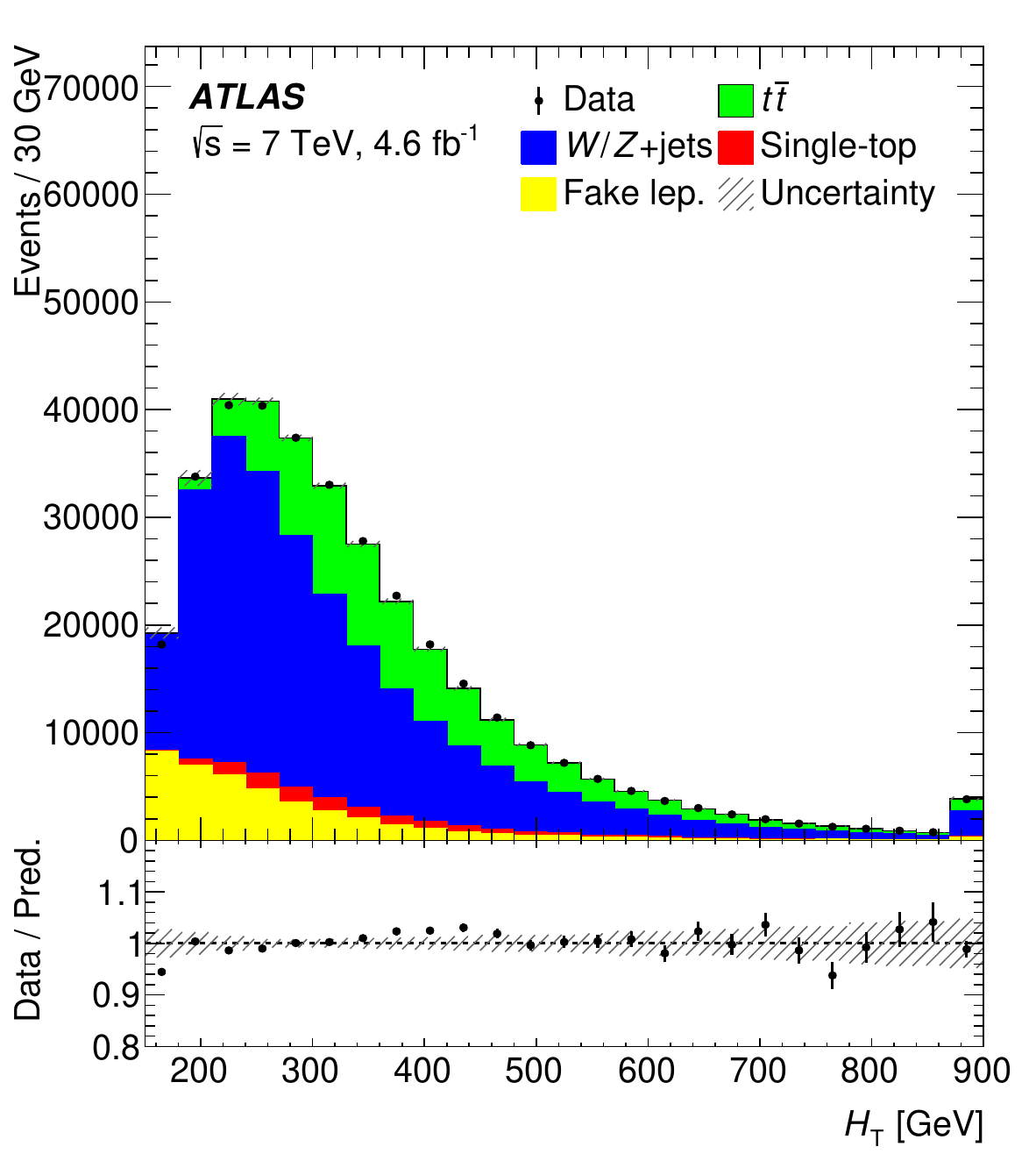}
\caption{}
\label{fig:dataMC2011fitProcessSelected_HT}
\end{subfigure}
\begin{subfigure}[b]{0.35\textwidth}
\includegraphics[width=\linewidth]{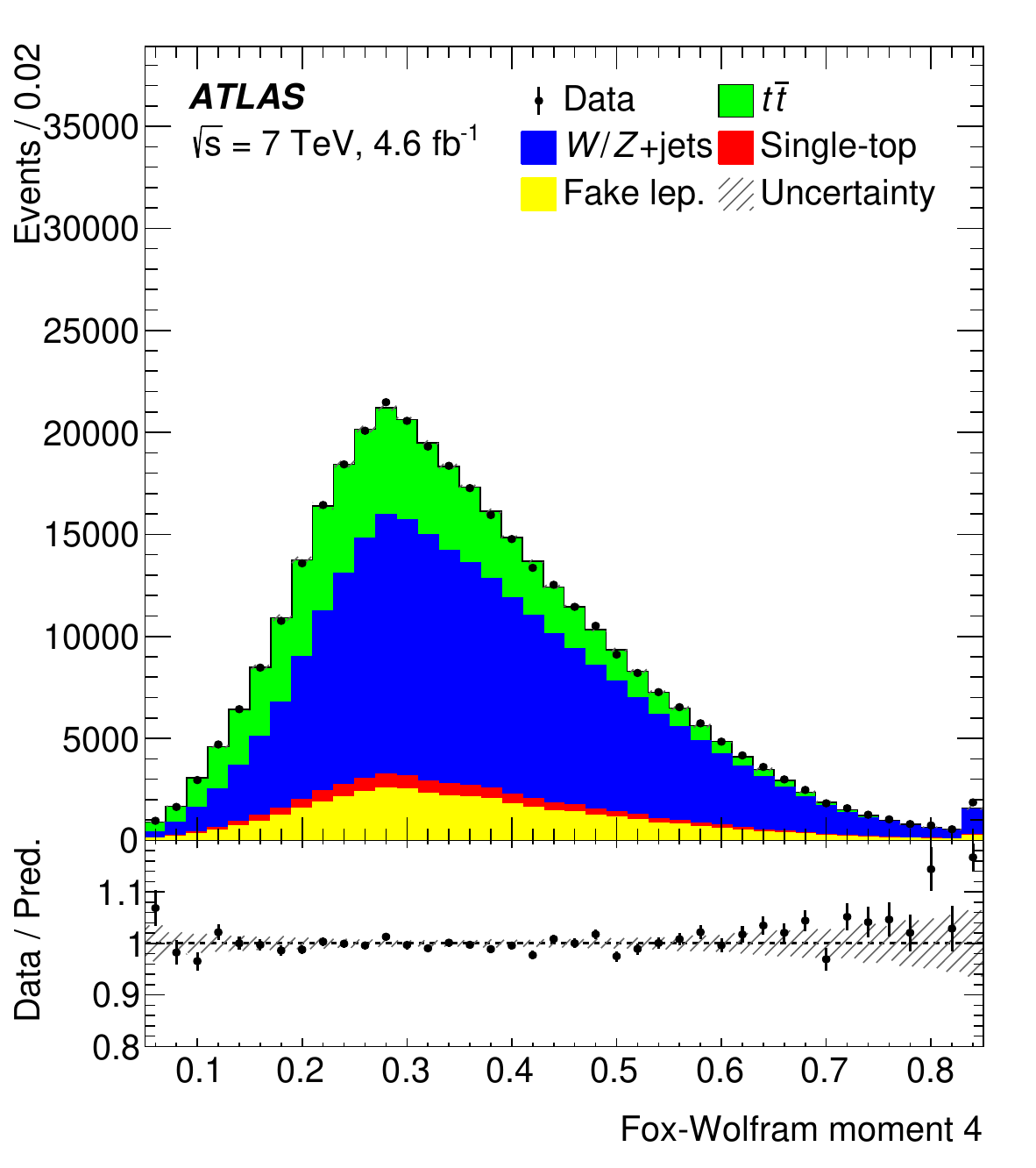}
\caption{}
\label{fig:dataMC2011fitProcessSelected_FW4}
\end{subfigure}
\begin{subfigure}[b]{0.35\textwidth}
\includegraphics[width=\linewidth]{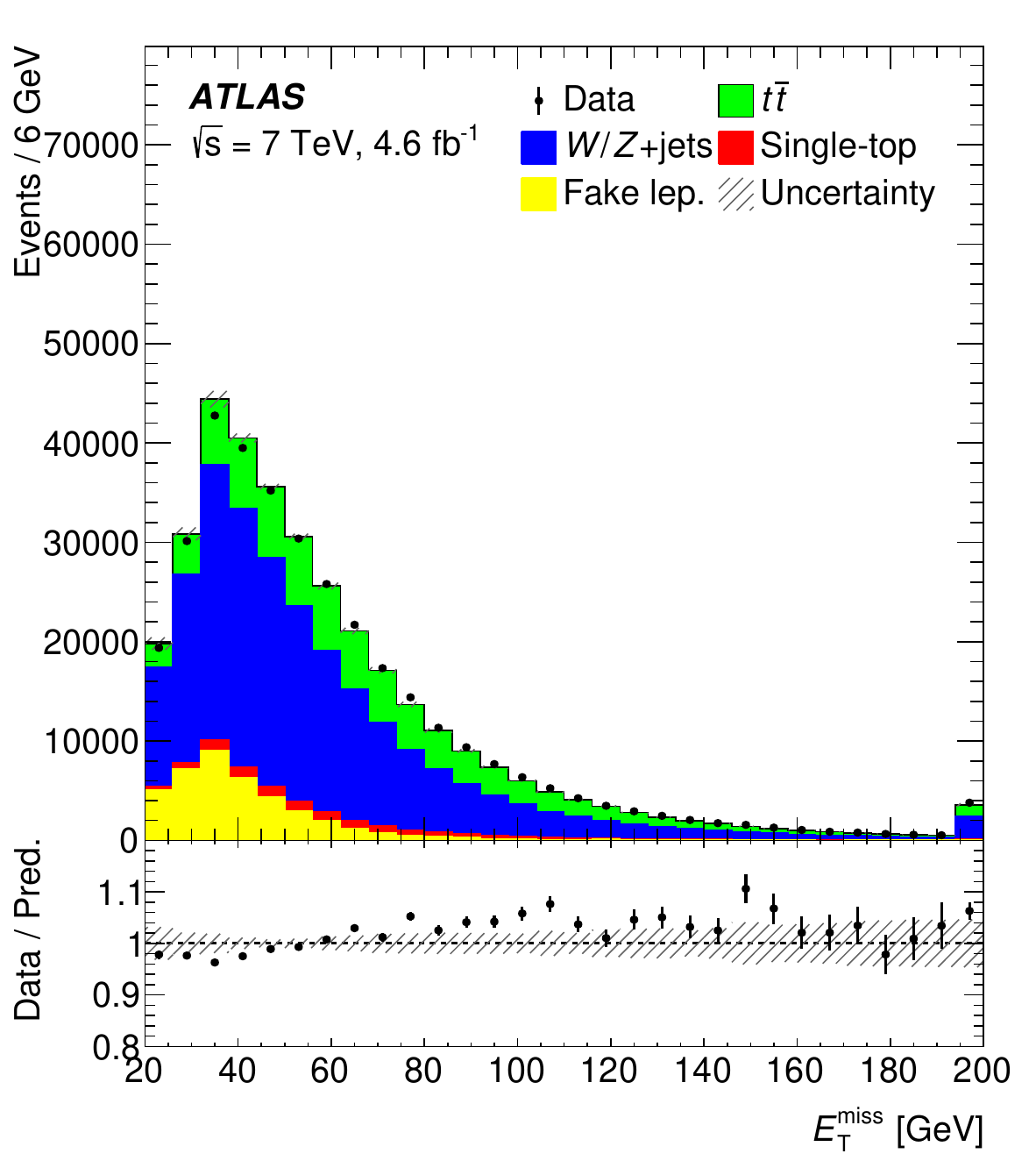}
\caption{}
\label{fig:dataMC2011fitProcessSelected_MET}
\end{subfigure}
\begin{subfigure}[b]{0.35\textwidth}
\includegraphics[width=\linewidth]{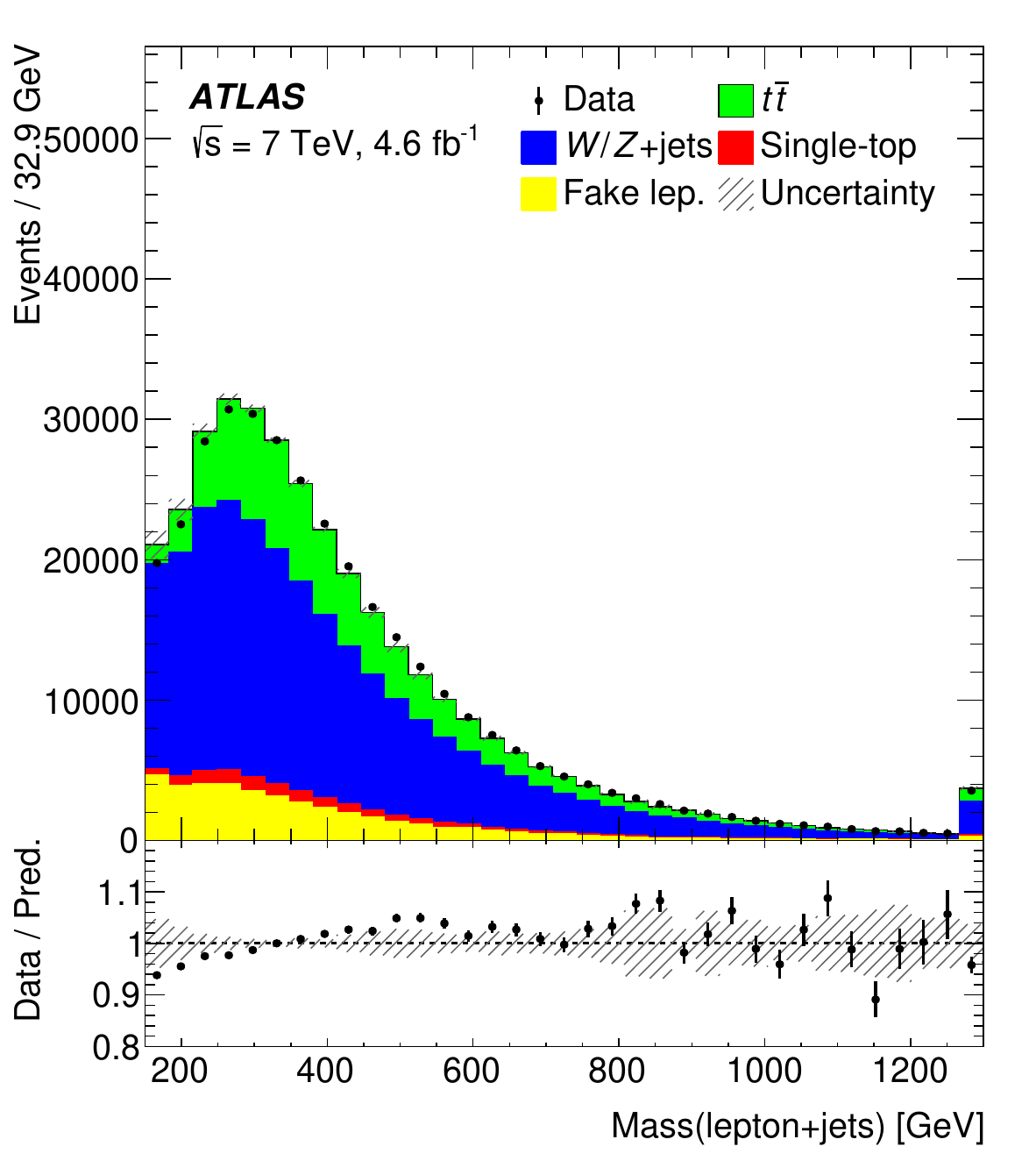}
\caption{}
\label{fig:dataMC2011fitProcessSelected_LJ-Mass}
\end{subfigure}
\caption{\label{fig:dataMC2011fitProcessSelected} \footnotesize The data distributions of six selected input variables are shown with their post-fit predictions in the selected sample. The predicted signal fraction is 24.8\%.  Shown are \subref{fig:dataMC2011fitProcessSelected_numJets} the number of jets, \subref{fig:dataMC2011fitProcessSelected_MV1} the number of $b$-tagged jets, \subref{fig:dataMC2011fitProcessSelected_HT} $H_{\mathrm{T}}$, \subref{fig:dataMC2011fitProcessSelected_FW4} the $4^{\text{th}}$ Fox--Wolfram moment, \subref{fig:dataMC2011fitProcessSelected_MET} $\MET$, and \subref{fig:dataMC2011fitProcessSelected_LJ-Mass} the mass of the lepton and jets.  Data are shown with the overlaid dots.  The predicted events are shown for each of the templates	used in modeling the data.  The statistical and systematic error bands are given by the shaded regions.  The first and last bins contain events found outside the range of the horizontal axis.}
\end{figure}

\begin{figure}[htbp]
\centering
\begin{subfigure}[b]{0.35\textwidth}
\includegraphics[width=\linewidth]{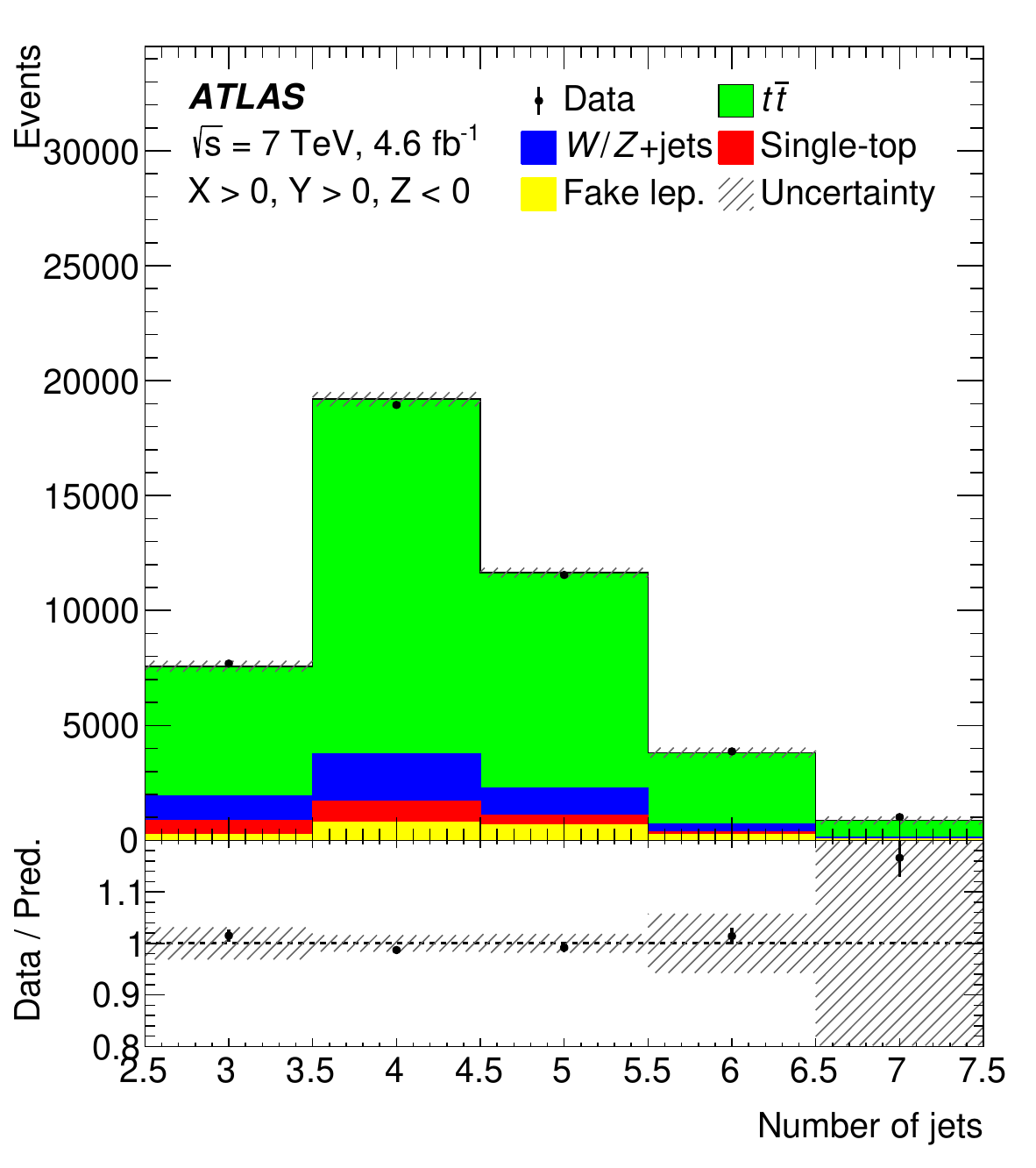}
\caption{}
\label{fig:dataMC2011fitProcessSelected_Signal_numJets}
\end{subfigure}
\begin{subfigure}[b]{0.35\textwidth}
\includegraphics[width=\linewidth]{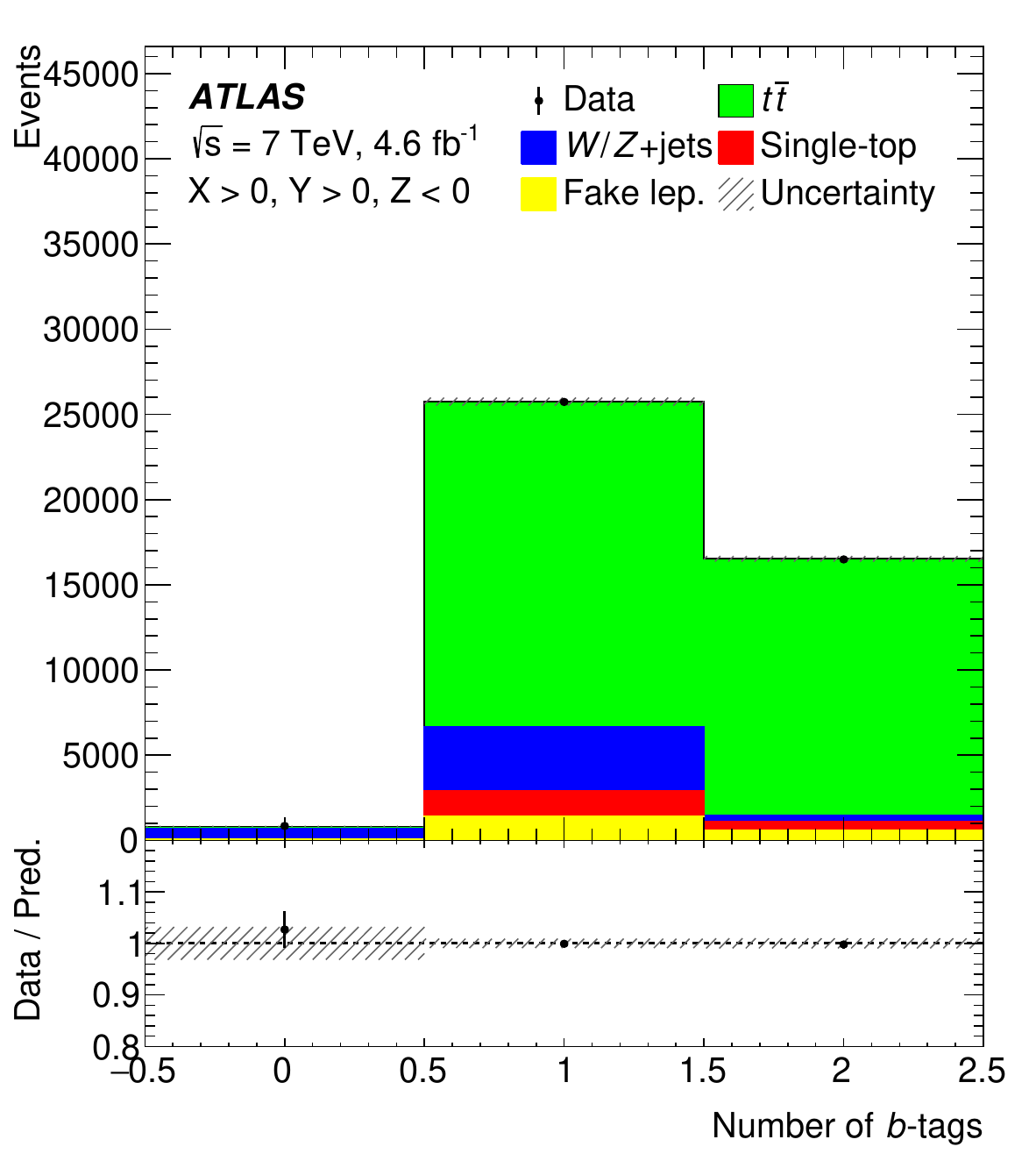}
\caption{}
\label{fig:dataMC2011fitProcessSelected_Signal_MV1}
\end{subfigure}
\begin{subfigure}[b]{0.35\textwidth}
\includegraphics[width=\linewidth]{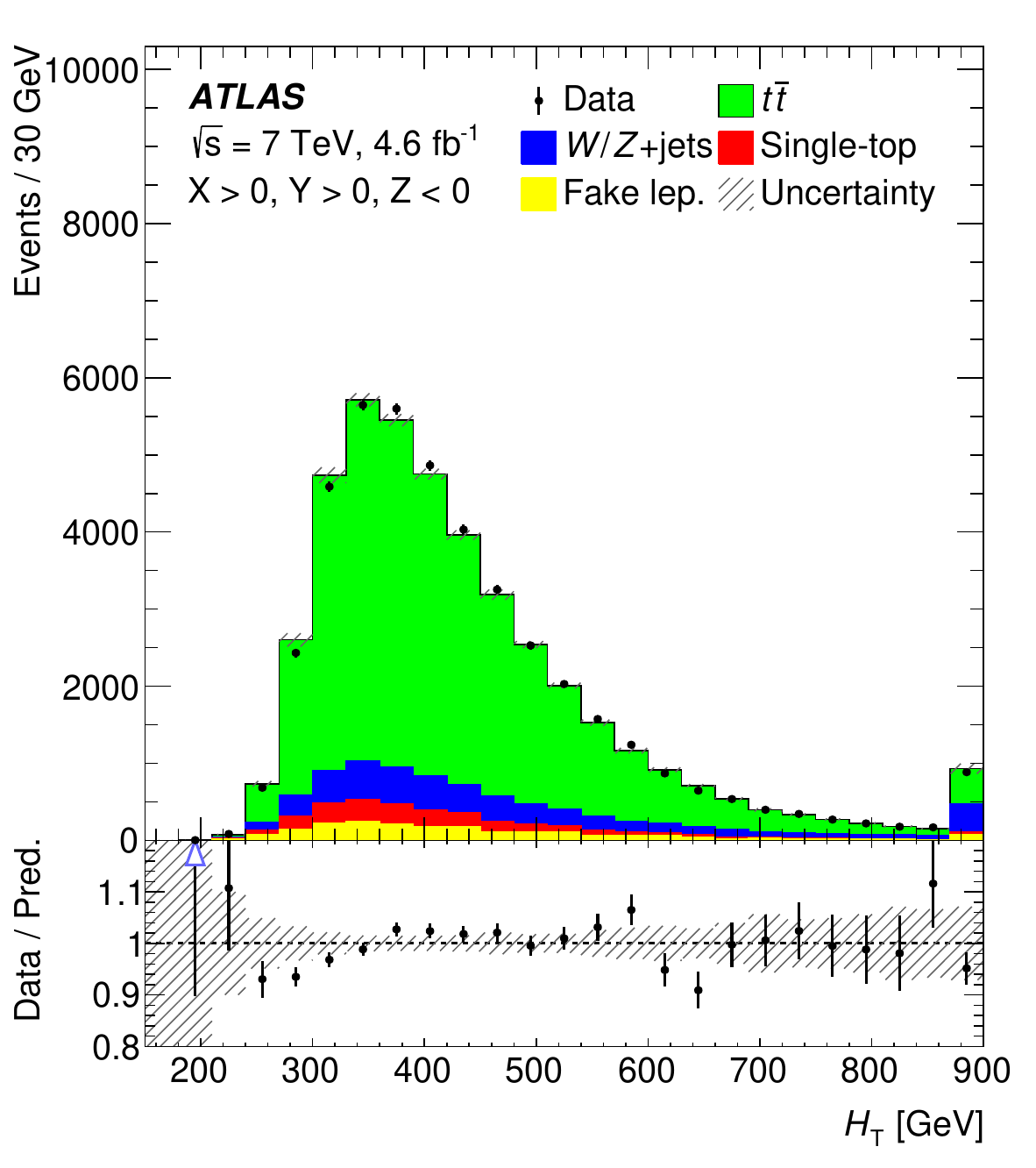}
\caption{}
\label{fig:dataMC2011fitProcessSelected_Signal_HT}
\end{subfigure}
\begin{subfigure}[b]{0.35\textwidth}
\includegraphics[width=\linewidth]{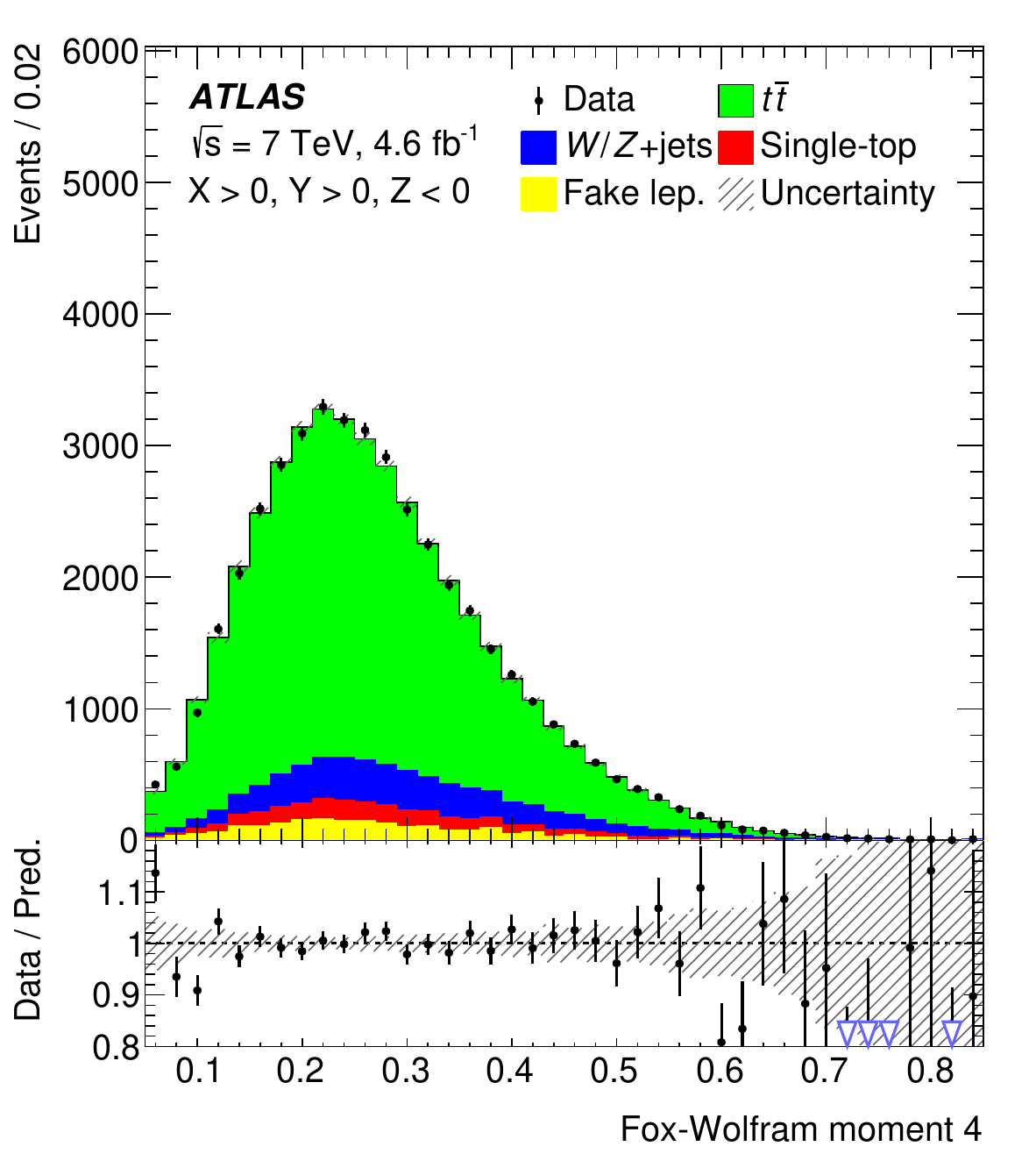}
\caption{}
\label{fig:dataMC2011fitProcessSelected_Signal_FW4}
\end{subfigure}
\begin{subfigure}[b]{0.35\textwidth}
\includegraphics[width=\linewidth]{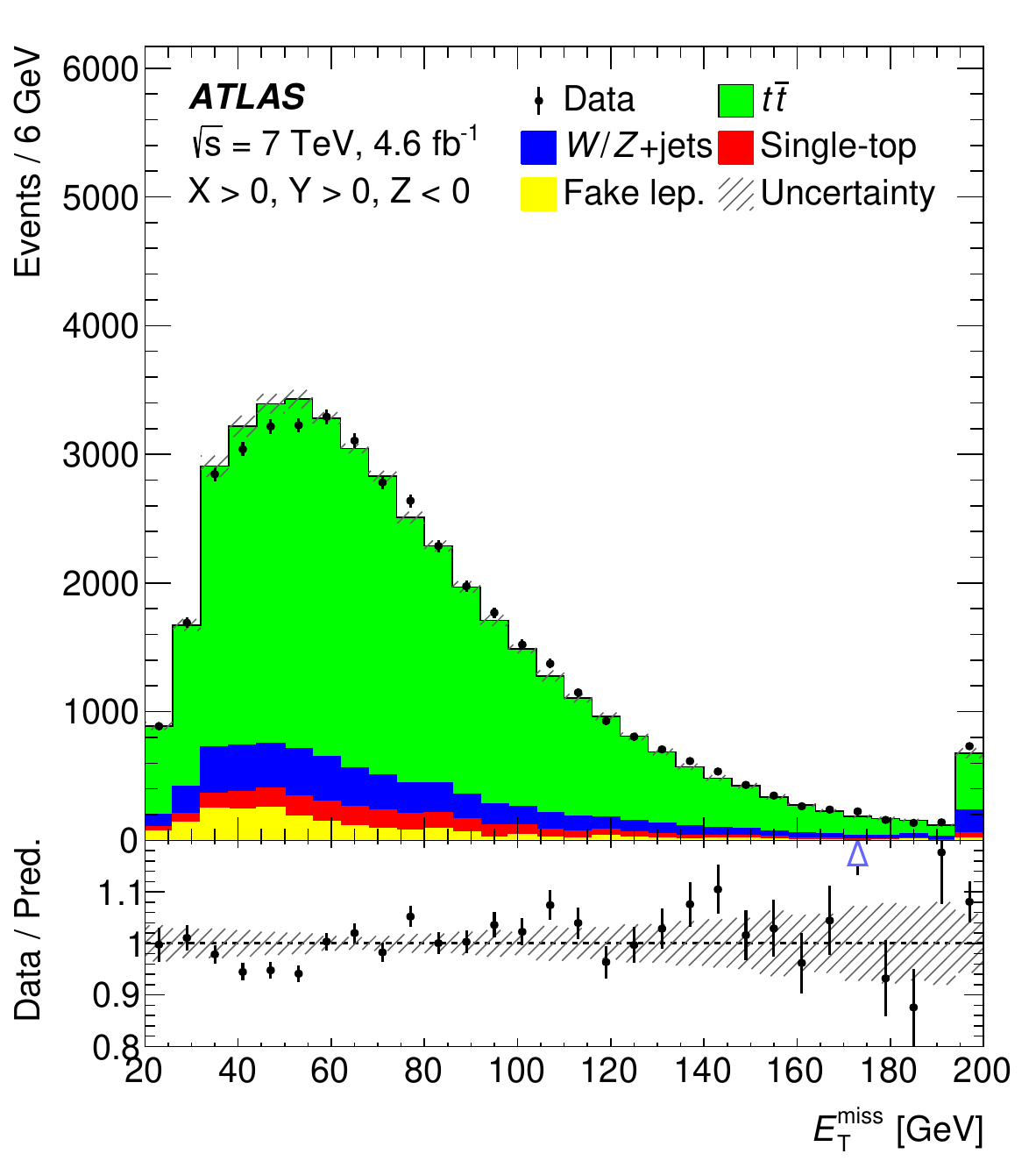}
\caption{}
\label{fig:dataMC2011fitProcessSelected_Signal_MET}
\end{subfigure}
\begin{subfigure}[b]{0.35\textwidth}
\includegraphics[width=\linewidth]{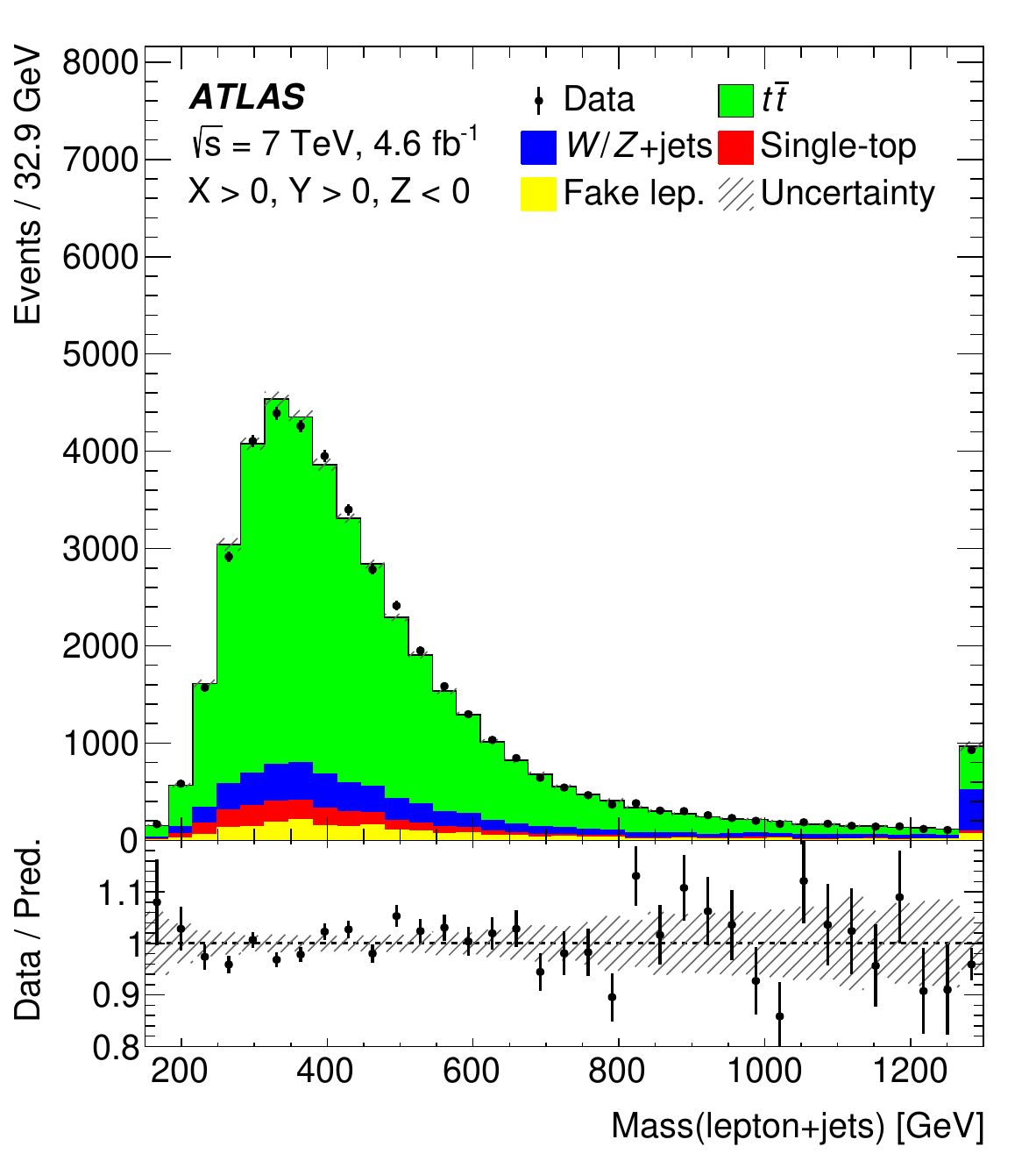}
\caption{}
\label{fig:dataMC2011fitProcessSelected_Signal_LJ-Mass}
\end{subfigure}
\caption{\label{fig:dataMC2011fitProcessSelected_Signal} \footnotesize The data distributions of six selected input variables are shown with their post-fit predictions in the selected sample for the signal-rich region with $\text{X} > 0$, $\text{Y} > 0$, and $\text{Z} < 0$.  The predicted signal fraction in this region is 79.3\%.  Shown are \subref{fig:dataMC2011fitProcessSelected_Signal_numJets} the number of jets, \subref{fig:dataMC2011fitProcessSelected_Signal_MV1} the number of $b$-tagged jets, \subref{fig:dataMC2011fitProcessSelected_Signal_HT} $H_{\mathrm{T}}$, \subref{fig:dataMC2011fitProcessSelected_Signal_FW4} the $4^{\text{th}}$ Fox--Wolfram moment, \subref{fig:dataMC2011fitProcessSelected_Signal_MET} $\MET$, and \subref{fig:dataMC2011fitProcessSelected_Signal_LJ-Mass} the mass of the lepton and jets.  Data are shown with the overlaid dots.  The predicted events are shown for each of the templates used in modeling the data.  The statistical and systematic error bands are given by the shaded regions.  The first and last bins contain events found outside the range of the horizontal axis.}
\end{figure}

\begin{figure}[htbp]
\centering
\begin{subfigure}[b]{0.35\textwidth}
\includegraphics[width=\linewidth]{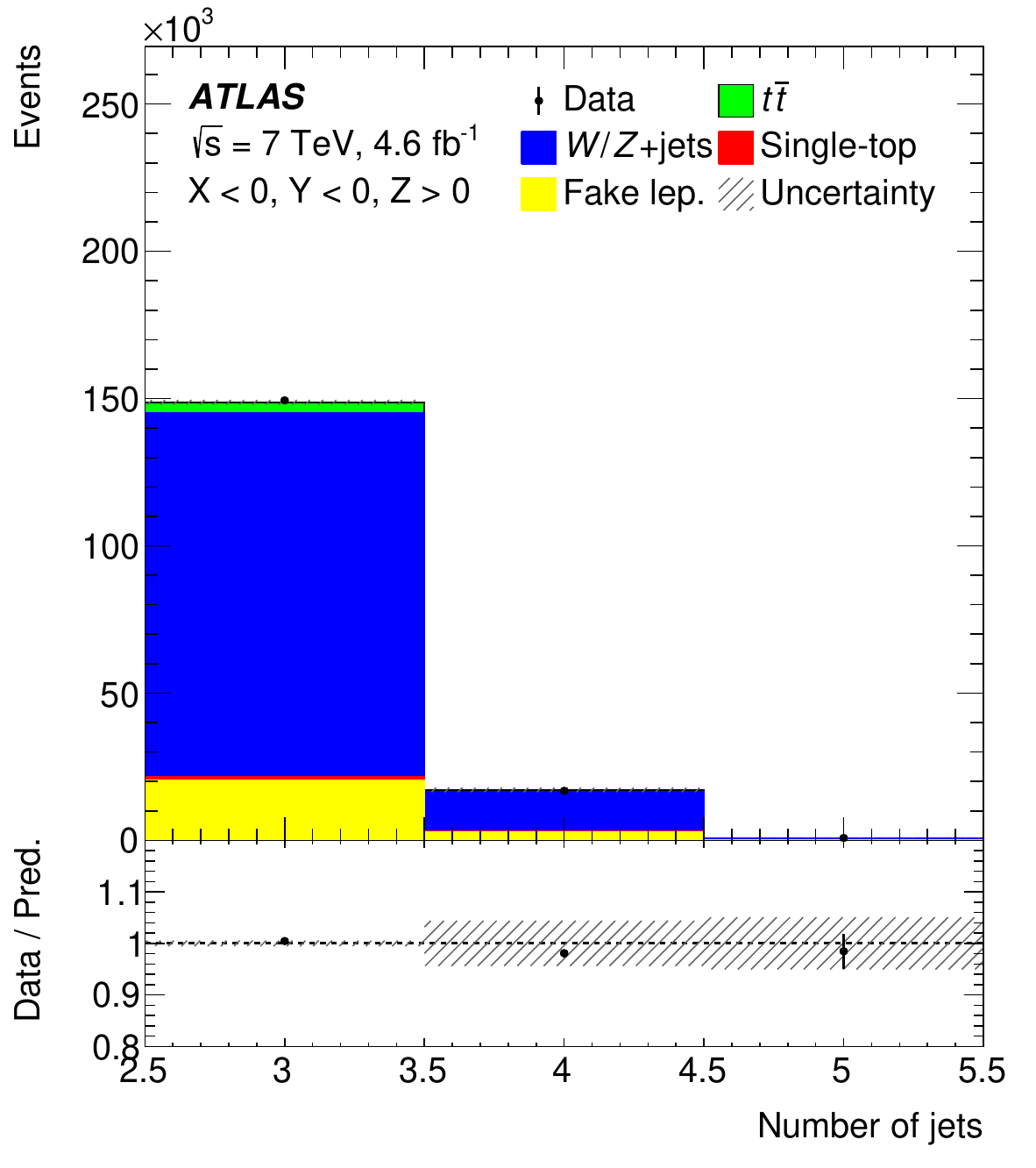}
\caption{}
\label{fig:dataMC2011fitProcessSelected_Background_numJets}
\end{subfigure}
\begin{subfigure}[b]{0.35\textwidth}
\includegraphics[width=\linewidth]{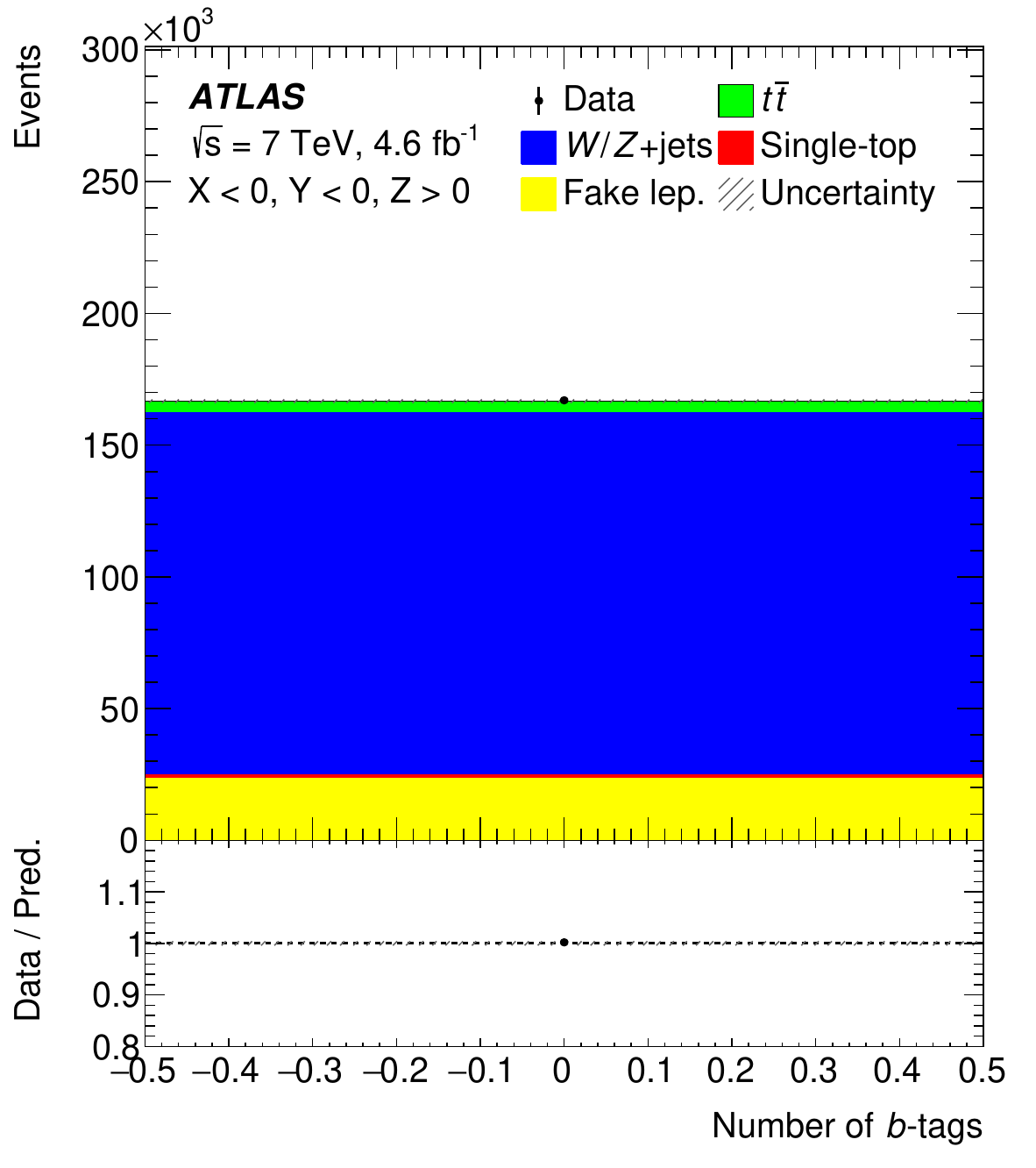}
\caption{}
\label{fig:dataMC2011fitProcessSelected_Background_MV1}
\end{subfigure}
\begin{subfigure}[b]{0.35\textwidth}
\includegraphics[width=\linewidth]{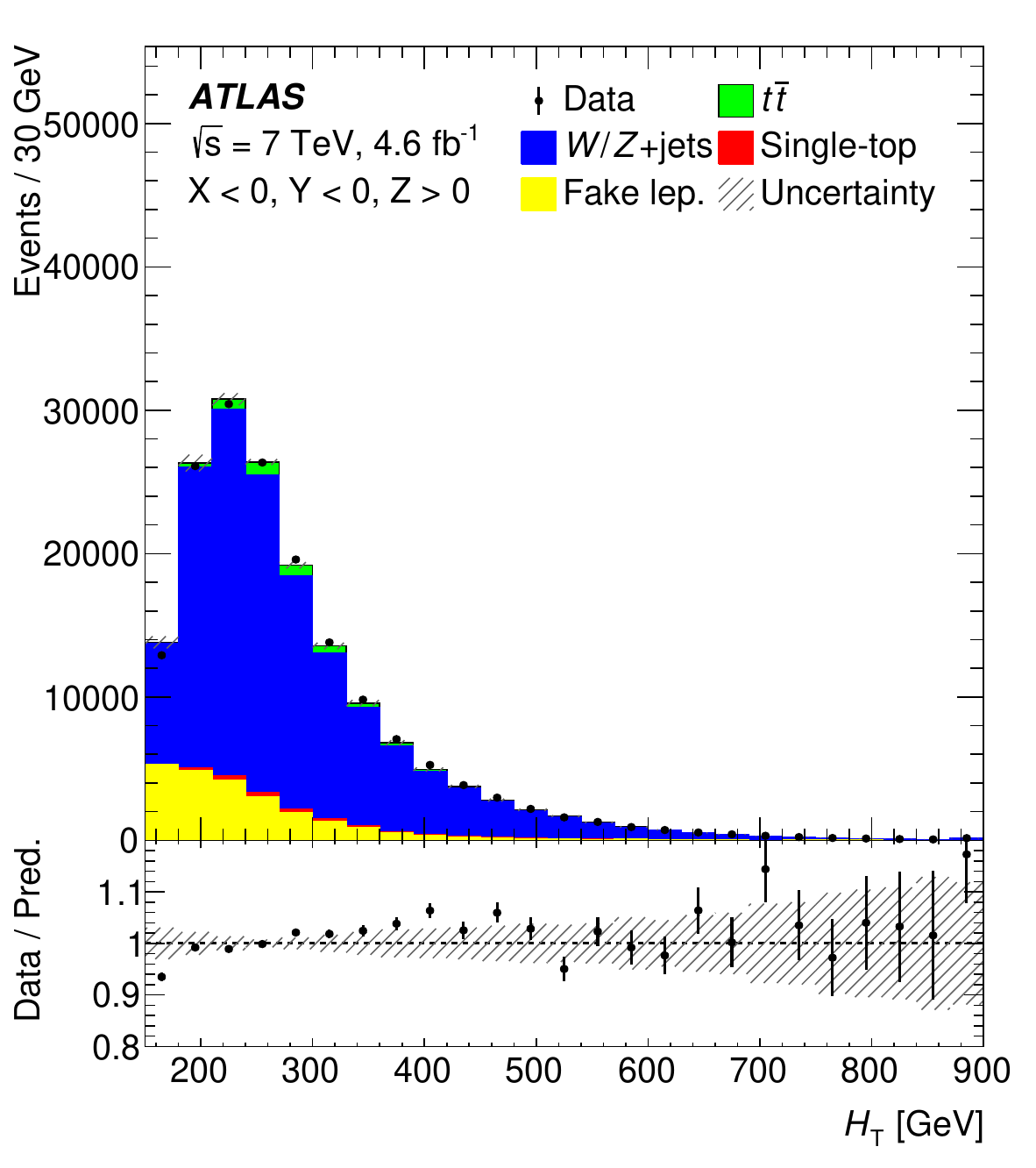}
\caption{}
\label{fig:dataMC2011fitProcessSelected_Background_HT}
\end{subfigure}
\begin{subfigure}[b]{0.35\textwidth}
\includegraphics[width=\linewidth]{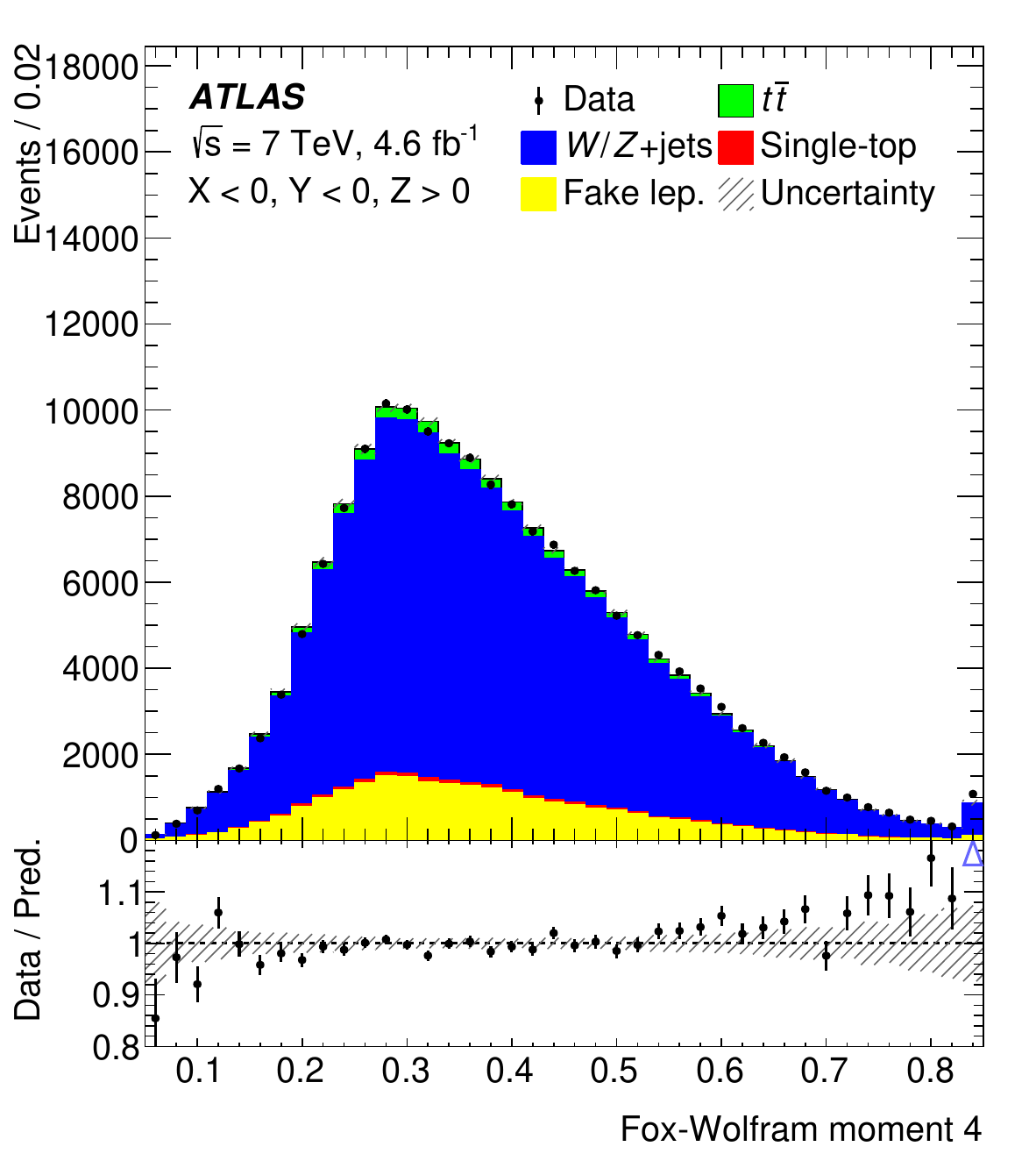}
\caption{}
\label{fig:dataMC2011fitProcessSelected_Background_FW4}
\end{subfigure}
\begin{subfigure}[b]{0.35\textwidth}
\includegraphics[width=\linewidth]{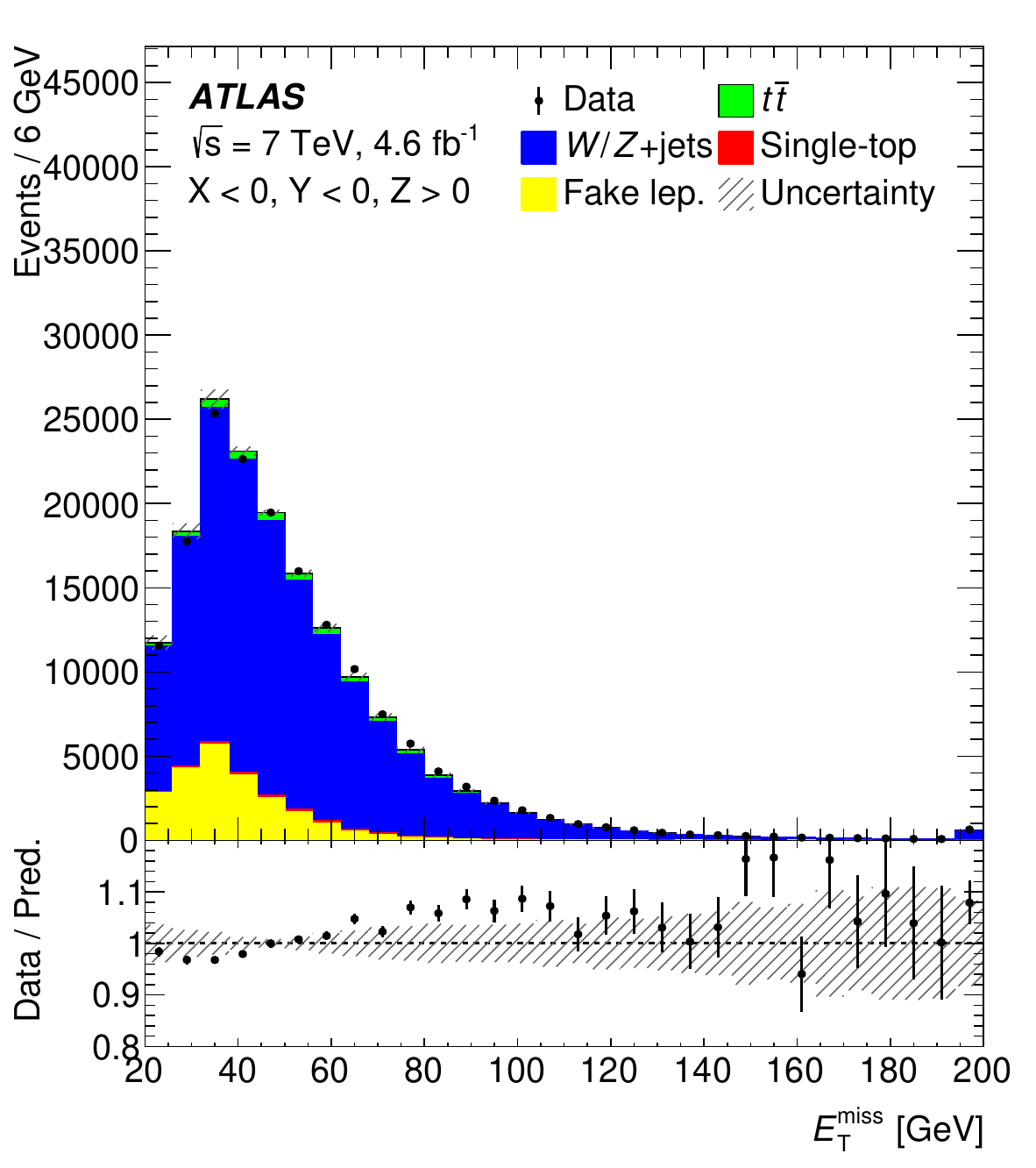}
\caption{}
\label{fig:dataMC2011fitProcessSelected_Background_MET}
\end{subfigure}
\begin{subfigure}[b]{0.35\textwidth}
\includegraphics[width=\linewidth]{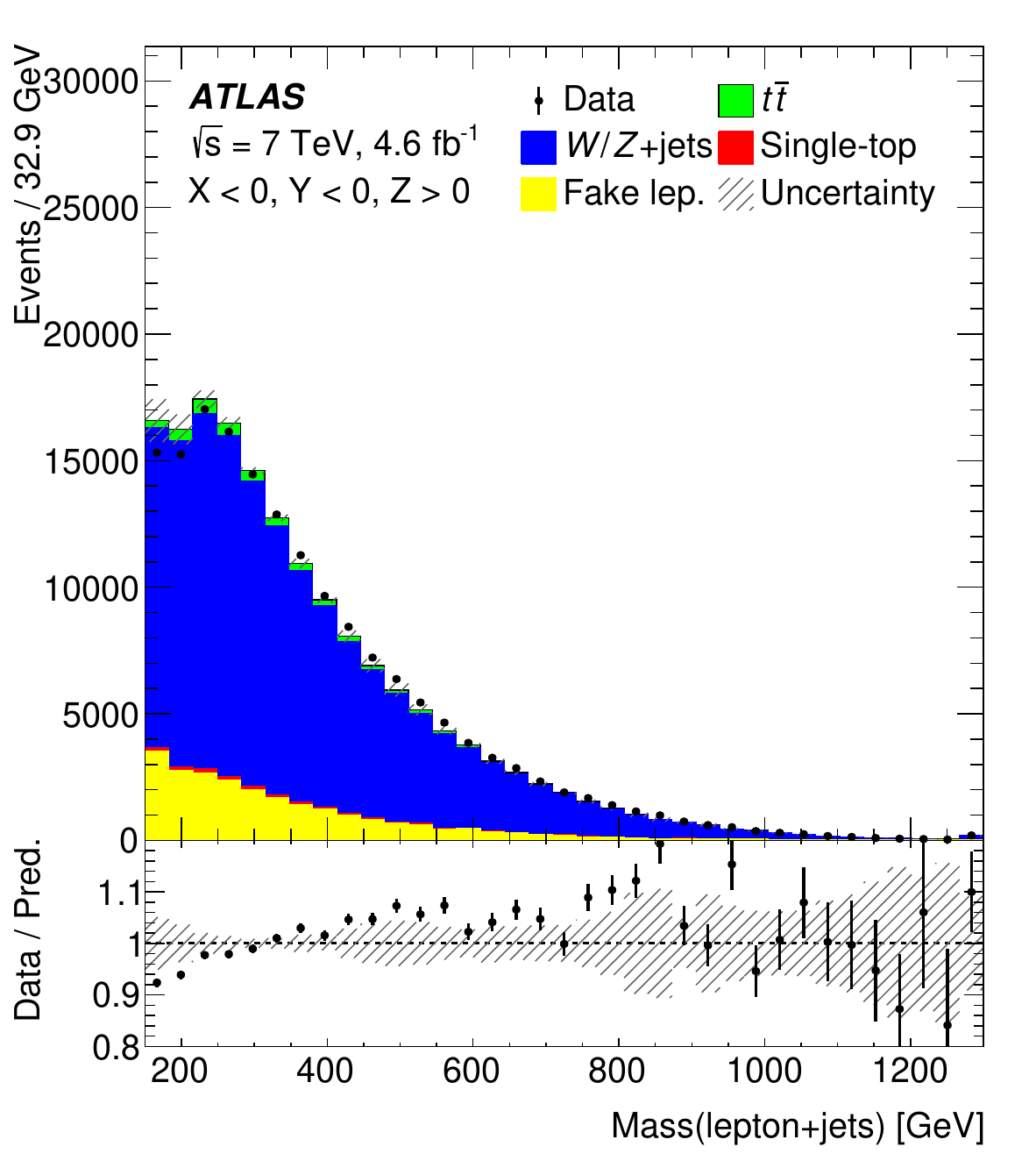}
\caption{}
\label{fig:dataMC2011fitProcessSelected_Background_LJ-Mass}
\end{subfigure}
\caption{\label{fig:dataMC2011fitProcessSelected_Background} \footnotesize The data distributions of six selected input variables are shown with their post-fit predictions in the selected sample for the background-rich region with $\text{X} < 0$, $\text{Y} < 0$, and $\text{Z} > 0$.  The predicted signal fraction in this region is 3.1\%.  Shown are \subref{fig:dataMC2011fitProcessSelected_Background_numJets} the number of jets, \subref{fig:dataMC2011fitProcessSelected_Background_MV1} the number of $b$-tagged jets, \subref{fig:dataMC2011fitProcessSelected_Background_HT} $H_{\mathrm{T}}$, \subref{fig:dataMC2011fitProcessSelected_Background_FW4} the $4^{\text{th}}$ Fox--Wolfram moment, \subref{fig:dataMC2011fitProcessSelected_Background_MET} $\MET$, and \subref{fig:dataMC2011fitProcessSelected_Background_LJ-Mass} the mass of the lepton and jets.  Data are shown with the overlaid dots.  The predicted events are shown for each of the templates used in modeling the data.  The statistical and systematic error bands are given by the shaded regions.  The first and last bins contain events found outside the range of the horizontal axis.}
\end{figure}

Comparison plots between data and the fit prediction are shown for a few selected input variables in Figure~\ref{fig:dataMC2011fitProcessSelected} for all events.  Analogous comparisons in signal-rich and background-rich regions of the XYZ space are shown in Figures~\ref{fig:dataMC2011fitProcessSelected_Signal} and~\ref{fig:dataMC2011fitProcessSelected_Background}, respectively.  The signal-rich region is defined by $\text{X} > 0$, $\text{Y} > 0$, and $\text{Z} < 0$, while the background-rich region lies in the opposite octant of XYZ space, and has $\text{X} < 0$, $\text{Y} < 0$, and $\text{Z} > 0$.  The X dimension corresponds to the Signal vs.\ Light decision hyperplane, while the Y and Z dimensions are linear combinations of the other SVM hyperplanes and are the directions orthogonal to X.  Based on a correlated $\chi^2$ test, the data and the fit agree well within the combined statistical and systematic error bands for all 21 variables.
 
The measured \ttbar cross section is in good agreement with the theoretical predictions based on the NNLO + NNLL calculations of $\sigma_{t\bar{t}}^{\text{NNLO+NNLL}}= 177^{+5}_{-6}\:\mathrm{(scale)}\,\pm 9\:\mathrm{(PDF+}\alphas )$~pb~$ = 177~^{+10}_{-11}$~pb for $pp$ collisions at a center-of-mass energy of $\sqrt{s} = 7$~\TeV\ and a top quark mass of 172.5~\GeV~\cite{Czakon:2013goa}.
 
The ATLAS measurement of the $t\bar{t}$ cross section at 7~\TeV\ in the dilepton channel~\cite{TOPQ-2013-04} is $\sigma_{t\bar{t}}=182.9\pm 3.1$(stat.)~$\pm$~4.2(syst.)~$\pm$~3.6(lumi.)~pb.  Depending upon the assumptions made for the systematic uncertainty correlations between these two measurements, the significance of their discrepancy was found to be in the $1.9\sigma$ to $2.1\sigma$ range.

\clearpage
\subsection{Top quark mass dependence}
 
The result of the profile likelihood fit depends on the assumed mass of the top quark through differences in \ttbar acceptance owing to lepton kinematics, and also from minor variations in the shape of the discriminant.  The analysis in this paper assumes a top quark mass of $m_{\text{ref}}=172.5$~\GeV.  The 2014 average of Tevatron and LHC Run~1 measurements of the top quark mass~\cite{TopMass} gives a value of $m_{t} = 173.34 \pm 0.27$(stat.)~$\pm~0.71$(syst.)~\GeV.  The current ATLAS average from 2019~\cite{TOPQ-2017-03} yields $m_{t} = 172.69 \pm 0.25$(stat.)~$\pm~0.41$(syst.)~\GeV.
 
The dependence of the \ttbar cross section on the mass of the top quark was determined through alternative profile likelihood fits that assume different top quark masses.  Monte Carlo samples for both \ttbar and single-top with top quark mass values of 165.0, 170.0, 172.5, 175.0, 177.5, and 180.0~\GeV\ were employed to measure the \ttbar cross section assuming each of these masses.  These measurements were then fitted to obtain the \ttbar cross section's top quark mass dependence.  When constrained to go through this measurement's central value at 172.5~\GeV, the best-fit second-order polynomial for the \ttbar cross section as a function of $\Delta m_{t}=m_{t} - m_{\text{ref}}$ is $\sigma_{\ttbar}(\Delta m_{t})=0.016 \cdot \Delta m_{t}^2-0.75 \cdot \Delta m_{t}+168.5$~pb, with $\Delta m_{t}$ in \GeV.
 
\section{Summary}
 
A measurement of the top quark pair-production cross section in the lepton+jets channel was performed with the ATLAS experiment at the LHC, using a multivariate technique based on support vector machines.  The measurement was obtained with a three-class, multidimensional event classifier.  It is based on $4.6 \text{~fb}^{-1}$ of data collected during 2011 in $pp$ collisions at $\sqrt{s}= 7$~\TeV.  The \ttbar cross section is found to be $\sigma_{t\bar{t}}=168.5\pm0.7$(stat.)~$^{+6.2}_{-5.9}$(syst.)~$^{+3.4}_{-3.2}$(lumi.)~pb, which has a relative uncertainty of $^{+4.2}_{-4.0}$\%. This measurement is consistent with the Standard Model prediction based on QCD calculations at next-to-next-to-leading order.

\section*{Acknowledgments}


We thank CERN for the very successful operation of the LHC, as well as the
support staff from our institutions without whom ATLAS could not be
operated efficiently.
 
We acknowledge the support of
ANPCyT, Argentina;
YerPhI, Armenia;
ARC, Australia;
BMWFW and FWF, Austria;
ANAS, Azerbaijan;
CNPq and FAPESP, Brazil;
NSERC, NRC and CFI, Canada;
CERN;
ANID, Chile;
CAS, MOST and NSFC, China;
Minciencias, Colombia;
MEYS CR, Czech Republic;
DNRF and DNSRC, Denmark;
IN2P3-CNRS and CEA-DRF/IRFU, France;
SRNSFG, Georgia;
BMBF, HGF and MPG, Germany;
GSRI, Greece;
RGC and Hong Kong SAR, China;
ISF and Benoziyo Center, Israel;
INFN, Italy;
MEXT and JSPS, Japan;
CNRST, Morocco;
NWO, Netherlands;
RCN, Norway;
MEiN, Poland;
FCT, Portugal;
MNE/IFA, Romania;
MESTD, Serbia;
MSSR, Slovakia;
ARRS and MIZ\v{S}, Slovenia;
DSI/NRF, South Africa;
MICINN, Spain;
SRC and Wallenberg Foundation, Sweden;
SERI, SNSF and Cantons of Bern and Geneva, Switzerland;
MOST, Taiwan;
TENMAK, T\"urkiye;
STFC, United Kingdom;
DOE and NSF, United States of America.
In addition, individual groups and members have received support from
BCKDF, CANARIE, Compute Canada and CRC, Canada;
PRIMUS 21/SCI/017 and UNCE SCI/013, Czech Republic;
COST, ERC, ERDF, Horizon 2020 and Marie Sk{\l}odowska-Curie Actions, European Union;
Investissements d'Avenir Labex, Investissements d'Avenir Idex and ANR, France;
DFG and AvH Foundation, Germany;
Herakleitos, Thales and Aristeia programmes co-financed by EU-ESF and the Greek NSRF, Greece;
BSF-NSF and MINERVA, Israel;
Norwegian Financial Mechanism 2014-2021, Norway;
NCN and NAWA, Poland;
La Caixa Banking Foundation, CERCA Programme Generalitat de Catalunya and PROMETEO and GenT Programmes Generalitat Valenciana, Spain;
G\"{o}ran Gustafssons Stiftelse, Sweden;
The Royal Society and Leverhulme Trust, United Kingdom.
 
The crucial computing support from all WLCG partners is acknowledged gratefully, in particular from CERN, the ATLAS Tier-1 facilities at TRIUMF (Canada), NDGF (Denmark, Norway, Sweden), CC-IN2P3 (France), KIT/GridKA (Germany), INFN-CNAF (Italy), NL-T1 (Netherlands), PIC (Spain), ASGC (Taiwan), RAL (UK) and BNL (USA), the Tier-2 facilities worldwide and large non-WLCG resource providers. Major contributors of computing resources are listed in Ref.~\cite{ATL-SOFT-PUB-2021-003}.


\FloatBarrier

\appendix
\numberwithin{equation}{section}
\addcontentsline{toc}{part}{Appendix:  Fit visualization}
\part*{Appendix:  Fit visualization}
 
The fit to the data can be difficult to visualize because of the 3D nature of the \textit{decision space}.  A series of 2D and 1D projections of the 3D space have been created in order to better illustrate its characteristics.
 
\begin{figure}[htbp]
\centering
\begin{tabular}{c}
\includegraphics[width=13.0cm,clip]{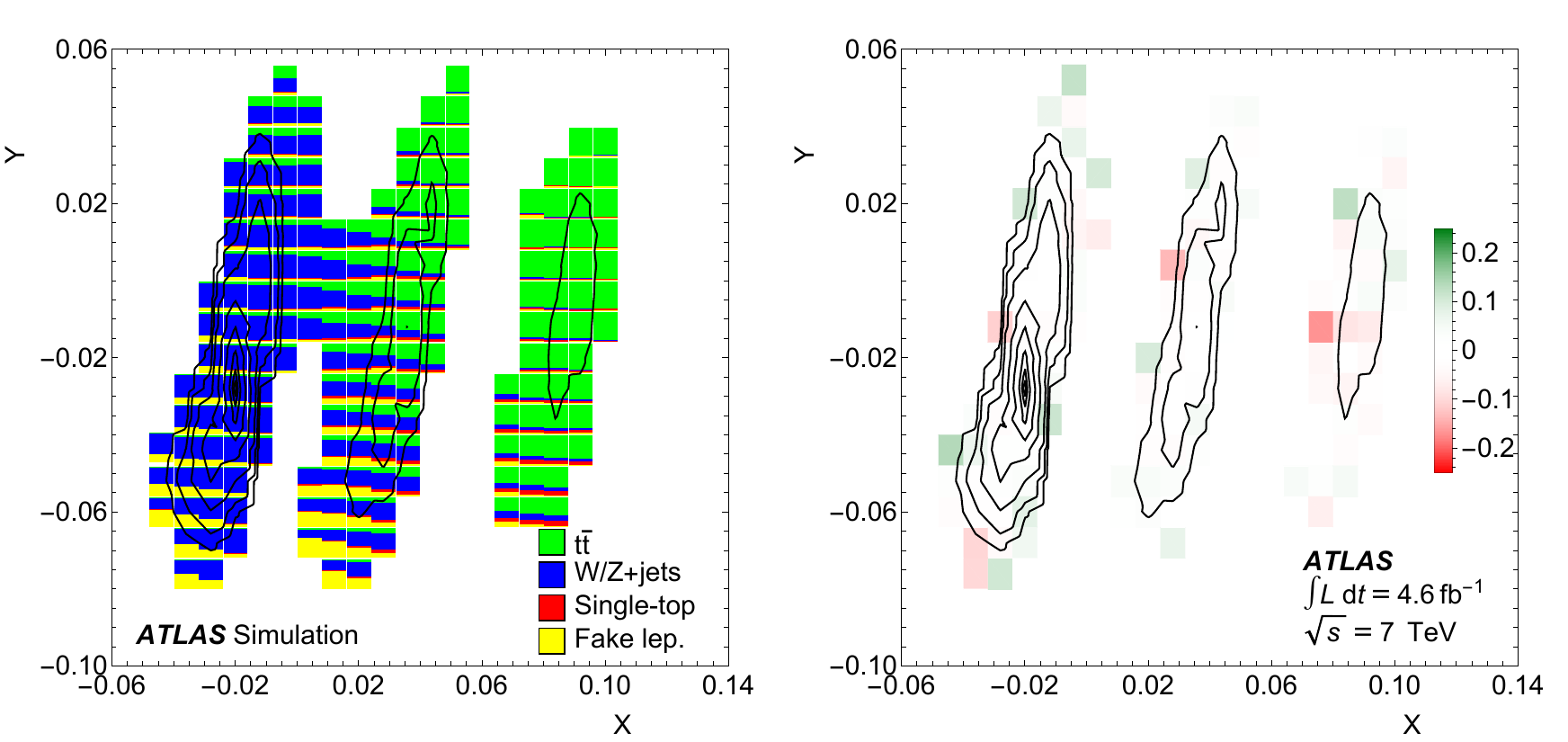} \\
\includegraphics[width=13.0cm,clip]{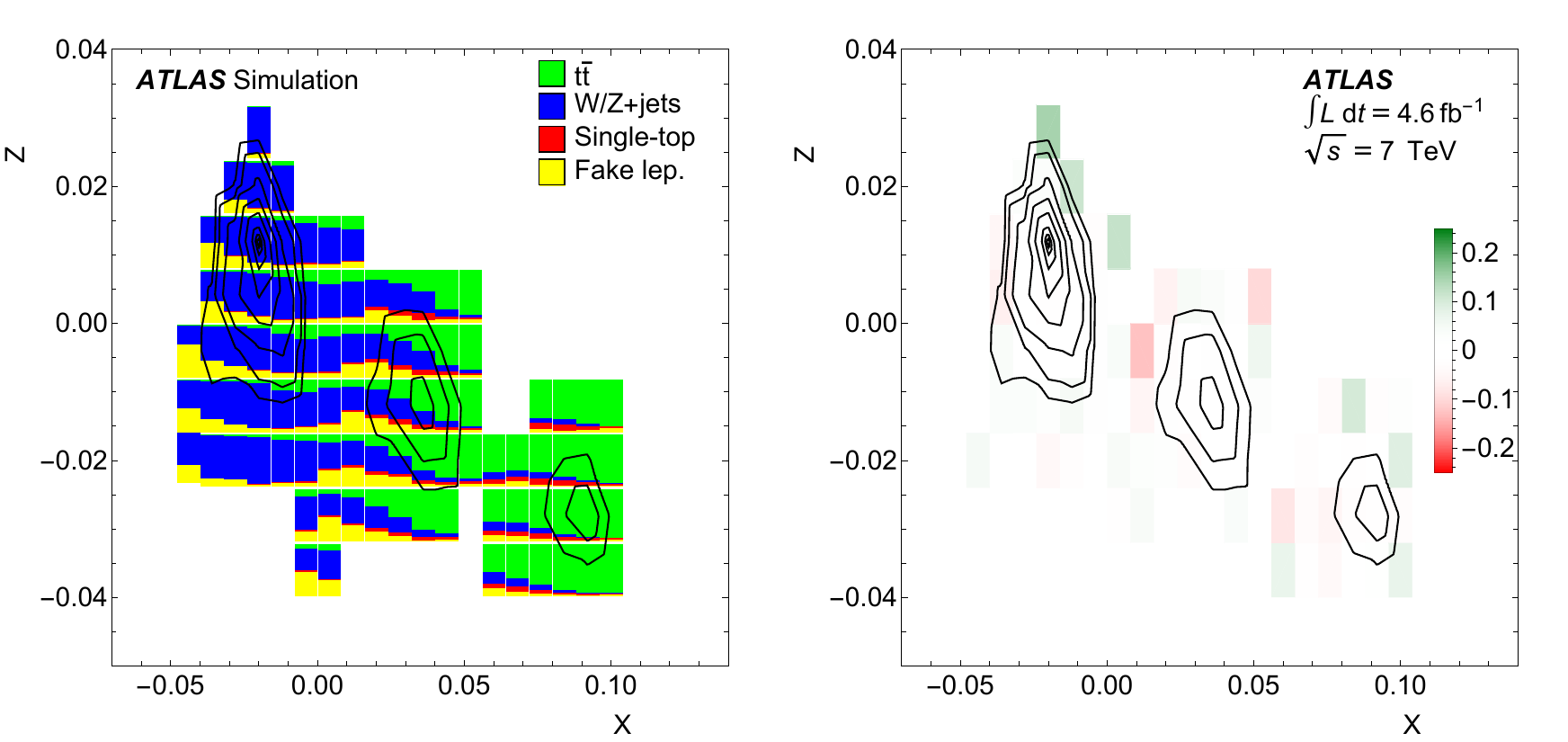} \\
\includegraphics[width=13.0cm,clip]{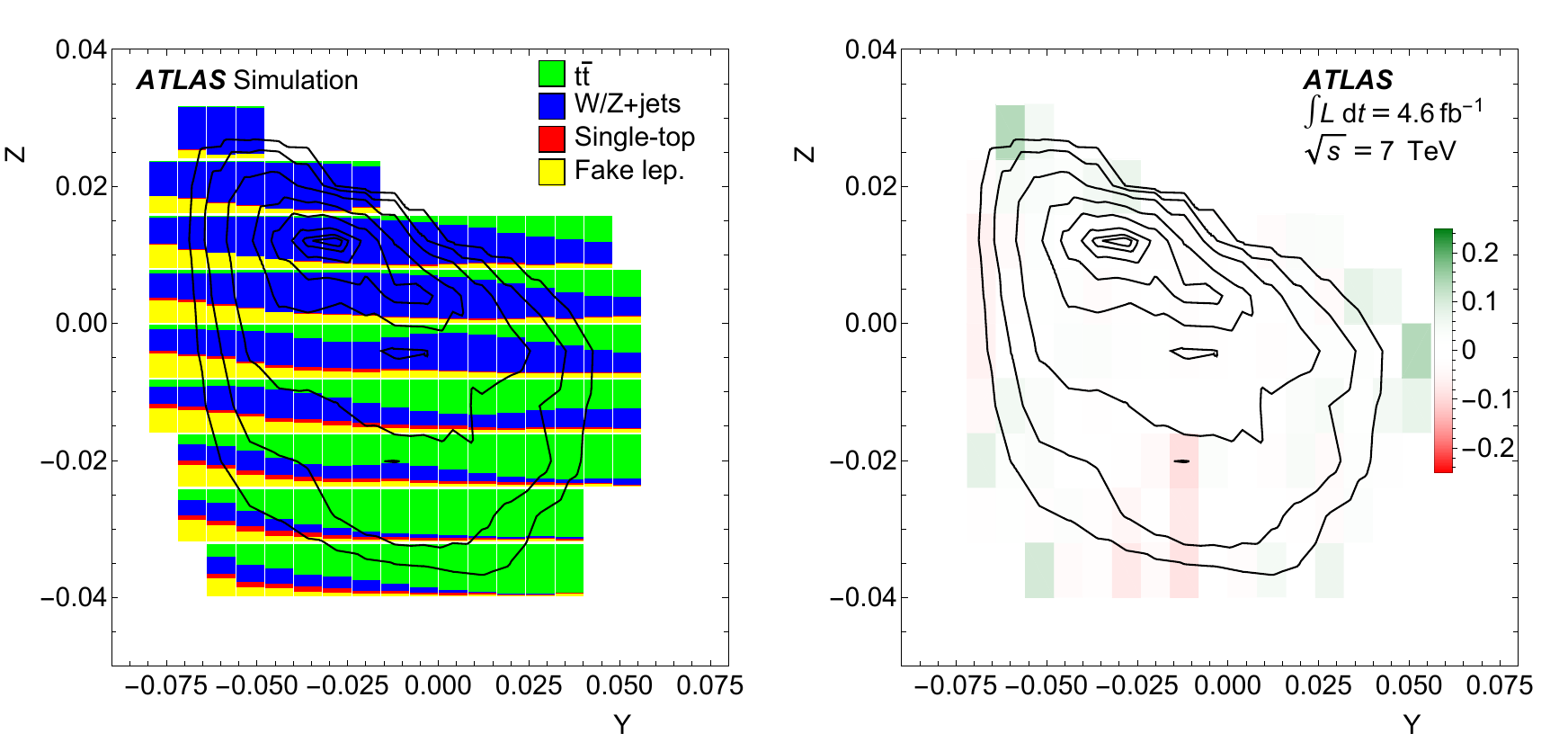} \\
\end{tabular}
\caption{\label{fig:2011_M22_2D_fits_TREx_final} \footnotesize (left) These 2D projections show the composition of each bin according to the template fit results.  Bins are shaded in the same ratio as the fit prediction.  The contours drawn on top of the bins represent the overall concentration of events.  Contours are provided at 5\%, 10\%, 20\%, 40\%, 60\%, 80\%, 90\%, and 95\% of the maximal 2D bin height.  (right) The 2D bins are shaded in accordance with the fractional difference between the observed data and the number of events predicted by the fit.  The contours shown are the same as in the left column and provide a reference.}
\end{figure}

Projections are provided in the XY, XZ, and YZ planes to give a qualitative comparison of the fit results with the data in XYZ space, and also to help visualize the definitions of the signal-rich and background-rich regions employed in Figures~\ref{fig:dataMC2011fitProcessSelected_Signal} and~\ref{fig:dataMC2011fitProcessSelected_Background}.  The 2D projections of the 3D \textit{decision space} use the standard fit results in XYZ space, but project the constant binning obtained with cubes of edge length 0.008 in that space.  On the right side of Figure~\ref{fig:2011_M22_2D_fits_TREx_final} the 2D projection plots show the fractional relative difference between the number of events predicted by the fit and the observed data.  The 2D plots on the left side of Figure~\ref{fig:2011_M22_2D_fits_TREx_final} show the class composition of each bin when projected onto the XY, XZ, and YZ planes.  The rectangle representing each 2D bin is colored from top to bottom such that the colors that fill it are in the same ratio as the predictions for each class in that bin.  Contour lines, which each represent the concentration of events in the plane, are drawn on top of the colored bins.  Contours exist for each of the following fractional values of the maximal 2D bin height: (0.05, 0.10, 0.20, 0.40, 0.60, 0.80, 0.90, 0.95).
 
\begin{figure}[htbp]
\centering
\begin{subfigure}[b]{0.49\textwidth}
\includegraphics[width=\linewidth]{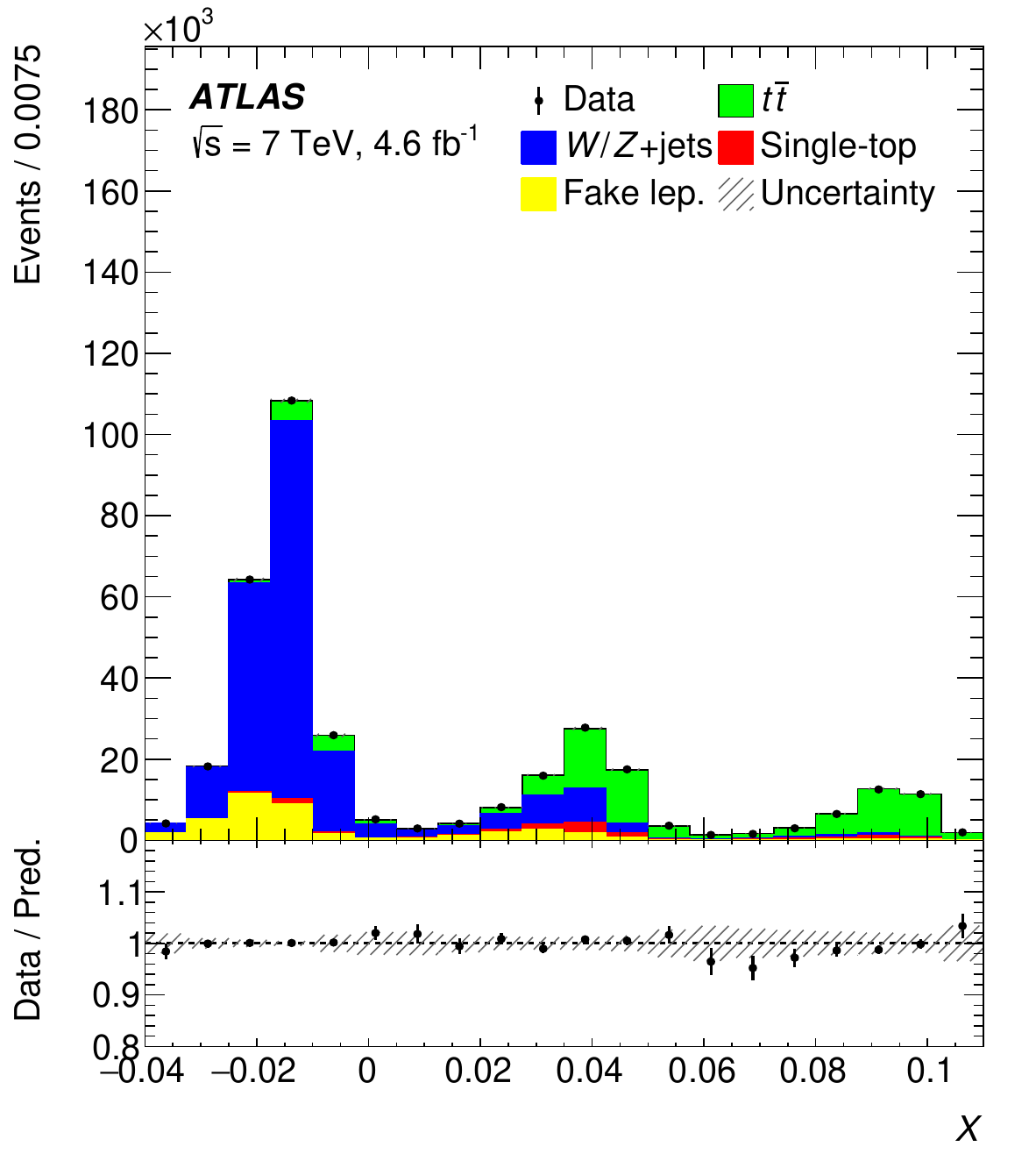}
\caption{}
\label{fig:2011_M22_1D_projections_X}
\end{subfigure}
\begin{subfigure}[b]{0.49\textwidth}
\includegraphics[width=\linewidth]{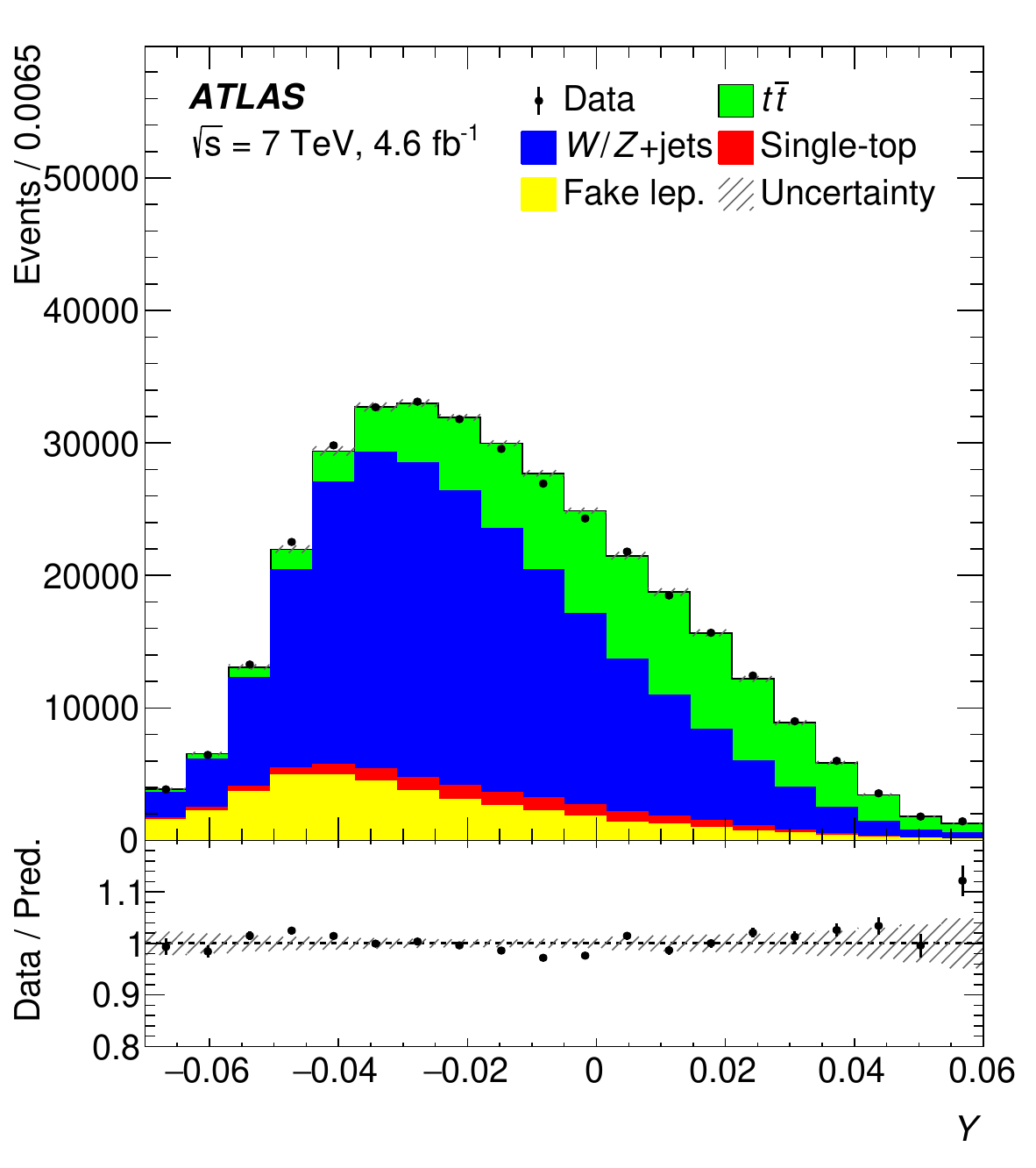}
\caption{}
\label{fig:2011_M22_1D_projections_Y}
\end{subfigure}
\begin{subfigure}[b]{0.49\textwidth}
\includegraphics[width=\linewidth]{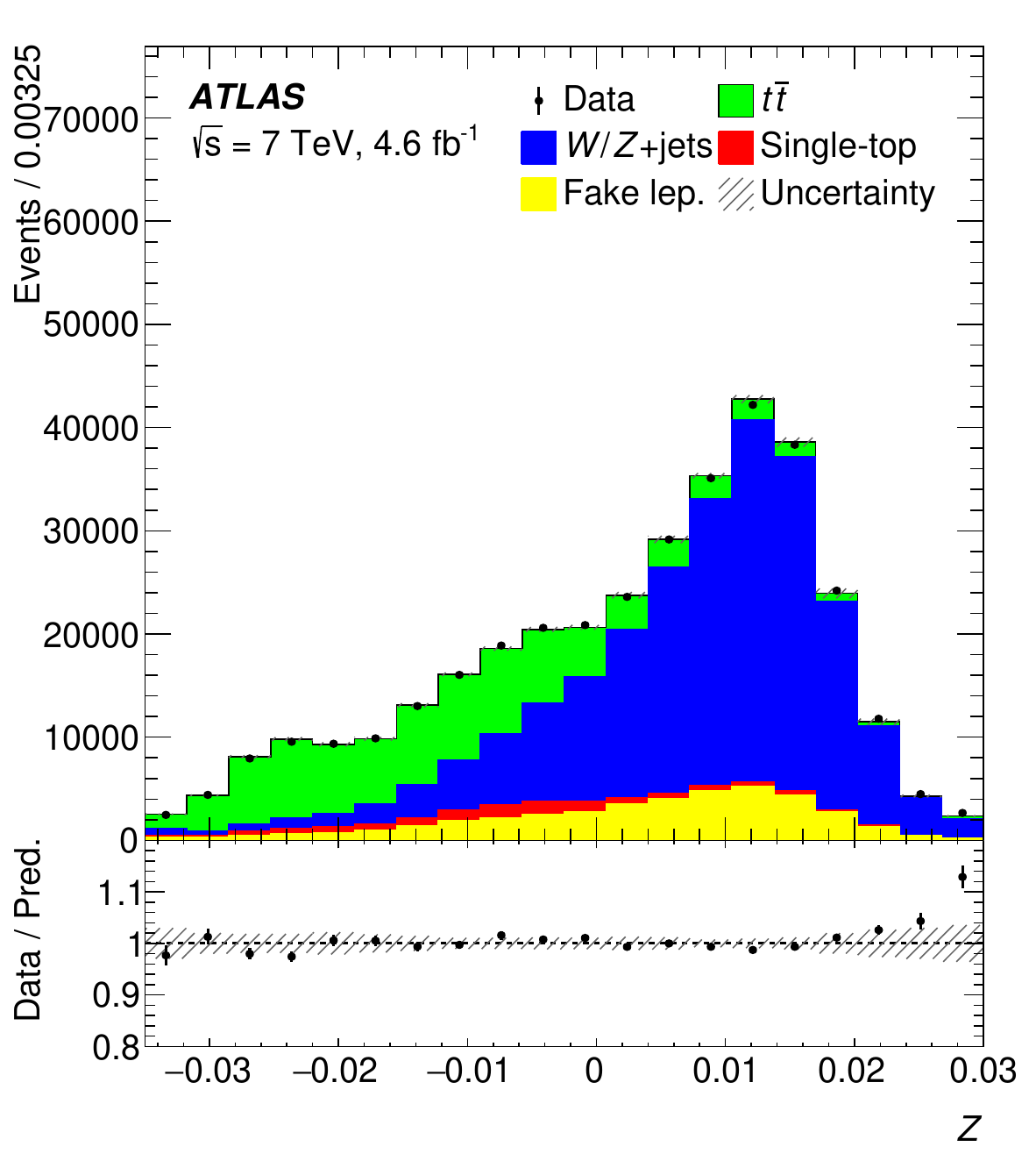}
\caption{}
\label{fig:2011_M22_1D_projections_Z}
\end{subfigure}
\caption{\label{fig:2011_M22_1D_projections} \footnotesize The 1D projections of the data and expected events after the likelihood fit.  X, Y, and Z form an orthogonal basis in the \textit{decision space} defined by the three trained SVMs.  X corresponds to the Signal vs.\ Light plane.  Y and Z are the directions orthogonal to X found with the remaining Signal vs.\ Heavy and Light vs.\ Heavy SVMs.  The projections use the fit parameters obtained with the profile likelihood fit to four quadrants, but project a constant binning obtained with cubes of edge length 0.008 in XYZ space.  In each of these plots, the prediction is broken down by the templates used in the fit.  The first and last bins contain events found outside the range of the horizontal axis.}
\end{figure}

The 1D projections of the 3D \textit{decision space} onto the X, Y, and Z axes are shown in Figure~\ref{fig:2011_M22_1D_projections} for the results of the fit to the data.  For visualization purposes, the full 3D \textit{decision space} is projected onto each of these axes.  These projections have the systematic error bands superimposed, and also include the ratio of data to fit prediction.

\clearpage
 
\printbibliography

@article{ATLFAST,
	Author = {Wolfgang Lukas},
	Journal = {J. Phys.: Conf. Ser.},
	Number = {2},
	Pages = {022031},
	Title = {Fast Simulation for ATLAS: Atlfast-II and ISF},
	doi = {10.1088/1742-6596/396/2/022031},
	Volume = {396},
	Year = {2012}}

@article{HFOR,
	Archiveprefix = {arXiv},
	Author = {M. L. Mangano and M. Moretti and F. Piccinini and M. Treccani},
	Eprint = {hep-ph/0611129},
	Journal = {JHEP},
	Pages = {013},
	Primaryclass = {hep-ph},
	Title = {Matching matrix elements and shower evolution for top-quark production in hadronic collisions},
	Volume = {01},
	Year = {2007},
	doi = "10.1088/1126-6708/2007/01/013"}

@book{Gram-Schmidt,
	Author = {Reed, Michael and Simon, Barry},
	Publisher = {New York: Academic Press},
	Title = {Methods of Modern Mathematical Physics. I: Functional Analysis},
	Year = {1972},
	doi = {10.1016/B978-0-12-585001-8.X5001-6}}

@booklet{TopMass,
	Archiveprefix = {arXiv},
	Author = {{The ATLAS, CDF, CMS and D0 Collaborations}},
	Eprint = {1403.4427},
	Howpublished = {ATLAS-CONF-2014-008},
	Primaryclass = {hep-ph},
	Title = {First combination of Tevatron and LHC measurements of the top-quark mass},
	Year = {2014}}

@article{AcerMC2013,
	Author = {B.P. Kersevan and E. Richter-Was},
	Journal = {Comput. Phys. Commun.},
	Number = {3},
	Pages = {919-985},
	Title = {The Monte Carlo event generator AcerMC versions 2.0 to 3.8 with interfaces to PYTHIA 6.4, HERWIG 6.5 and ARIADNE 4.1},
	Volume = {184},
	Year = {2013},
	doi = {10.1016/j.cpc.2012.10.032}}

@booklet{Botje:2011sn,
	Archiveprefix = {arXiv},
	Author = {Botje, Michiel and Butterworth, Jon and Cooper-Sarkar, Amanda and de Roeck, Albert and Feltesse, Joel and Forte, Stefano and Glazov, Alexander and Huston, Joey and McNulty, Ronan and Sjostrand, Torbjorn and Thorne, Robert},
	Eprint = {1101.0538},
	Primaryclass = {hep-ph},
	Slaccitation = {%%CITATION = ARXIV:1101.0538;%%},
	Title = {{The PDF4LHC Working Group Interim Recommendations}},
	Year = {2011}}

@conference{Salam,
	Address = {Stockholm},
	Author = {A. Salam},
	Booktitle = {Eighth Nobel Symposium},
	Editor = {N. Svartholm},
	Pages = {367},
	Publisher = {Almquist and Wiksell},
	Title = {Elementary Particle Physics: Relativistic Groups and Analyticity},
	Year = {1968}}

@article{Weinberg,
	Author = {S. Weinberg},
	Journal = {Phys. Rev. Lett.},
	Pages = {1264-1266},
	Title = {A model of leptons},
	Volume = {19},
	Year = {1967},
	doi = {10.1103/PhysRevLett.19.1264}}

@article{Glashow,
	Author = {S.L. Glashow},
	Journal = {Nucl. Phys.},
	Pages = {579-588},
	Title = {Partial-symmetries of weak interactions},
	Volume = {22},
	Year = {1961},
	doi = {10.1016/0029-5582(61)90469-2}}

@booklet{singleTopXsec,
	Archiveprefix = {arXiv},
	Author = {Kidonakis, Nikolaos},
	Eprint = {1205.3453},
	Howpublished = {DESY-PROC-2012-02/251},
	Primaryclass = {hep-ph},
	Title = {Differential and total cross section for top pair and single top production},
	doi = {10.3204/DESY-PROC-2012-02/251}}

@booklet{platt,
	Author = {J.C. Platt},
	Howpublished = {MSR-TR-98-14},
	Title = {Sequential Minimal Optimization: A Fast Algorithm for Training Support Vector Machines},
	Url = {https://citeseerx.ist.psu.edu/viewdoc/summary?doi=10.1.1.43.4376},
	Year = {1998}}

@article{keerthi,
	Author = {S.S. Keerthi},
	Journal = {Neural Comp.},
	Number = {3},
	Pages = {637},
	Title = {Improvements to Platt's SMO Algorithm for SVM Classifier Design},
	Volume = {13},
	Year = {2001},
	doi = {10.1162/089976601300014493}}

@article{FWinHEP3,
	Archiveprefix = {arXiv},
	Author = {Spiller, L. A.},
	Eprint = {1508.03144v1},
	Journal = {JHEP},
	Pages = {027},
	Primaryclass = {hep-ph},
	Title = {Modification of Fox-Wolfram Moments for Hadron Colliders},
	Volume = {03},
	Year = {2016},
	doi = {10.1007/JHEP03(2016)027}}

@article{FWinHEP2,
	Archiveprefix = {arXiv},
	Author = {C. Bernaciak and M. S. A. Buschmann and A. Butter and T. Plehn},
	Eprint = {1212.4436v1},
	Journal = {Phys. Rev. D},
	Pages = {073014},
	Primaryclass = {hep-ph},
	Title = {Fox-Wolfram Moments in Higgs Physics},
	Volume = {87},
	Year = {2013},
	doi = {10.1103/PhysRevD.87.073014}}

@article{FWM,
	Author = {G. C. Fox and S. Wolfram},
	Journal = {Nucl. Phys. B},
	Volume = {149},
	Pages = {413},
	Title = {Event Shapes in $e^+e^-$ Annihilation},
	Year = {1979},
	doi = {10.1016/0550-3213(79)90003-8}}

@article{soft,
	Author = {C. Cortes and V. Vapnik},
	Journal = {Machine Learning},
	Number = {3},
	Pages = {273},
	Title = {Support-vector networks},
	Volume = {20},
	Year = {1995},
	doi = {10.1007/BF00994018}}

@inproceedings{SVM,
	Author = {B. E. Boser and I. M. Guyon and V. N. Vapnik},
	booktitle = {Proceedings of the Fifth Annual Workshop on Computational Learning Theory (COLT'92)},
	Pages = {144-152},
	Title = {A Training Algorithm for Optimal Margin Classifiers},
	Year = {1992},
	location = {Pittsburgh},
	doi = {10.1145/130385.130401},
	url = {https://doi.org/10.1145/130385.130401}}

@article{Alwall:2007fs,
	Archiveprefix = {arXiv},
	Author = {J. Alwall and others},
	Eprint = {0706.2569},
	Journal = {Eur. Phys. J.},
	Pages = {473-500},
	Primaryclass = {hep-ph},
	Reportnumber = {SLAC-PUB-12604, CERN-PH-TH-2007-066, LU-TP-07-13, KA-TP-06-2007, DCPT-07-62, IPPP-07-31},
	Slaccitation = {%%CITATION = ARXIV:0706.2569;%%},
	Title = {{Comparative study of various algorithms for the merging of parton showers and matrix elements in hadronic collisions}},
	Volume = {C53},
	Year = {2008},
	doi = "10.1140/epjc/s10052-007-0490-5"}

@article{Mercer,
	Author = {J. Mercer},
	Journal = {Philos. Trans. Roy. Soc. London Ser. A},
	Pages = {415--446},
	Title = {Functions of Positive and Negative Type, and their Connection with Theory of Integral Equations},
	Volume = {209},
	Year = {1909},
	doi = {10.1098/rsta.1909.0016}}

@article{Aronszajn,
	Author = {N. Aronszajn},
	Journal = {Trans. Amer. Math. Soc.},
	Pages = {337-404},
	Title = {Theory of Reproducing Kernels},
	Volume = {68},
	Year = {1950},
	doi = {10.1090/S0002-9947-1950-0051437-7}}

@Article{localButterworth:1996zw,
      author         = "Butterworth, J. M. and Forshaw, J. R. and Seymour, M. H.",
      title          = "{Multiparton interactions in photoproduction at HERA}",
      journal        = "Z. Phys. C",
      volume         = "72",
      pages          = "637",
      doi            = "10.1007/s002880050286",
      year           = "1996",
      eprint         = "hep-ph/9601371",
      archivePrefix  = "arXiv",
      reportNumber   = "CERN-TH-96-82, M-C-TH-96-05, UCL-HEP-96-02",
      SLACcitation   = "%%CITATION = HEP-PH/9601371;%%",
}

@Article{PhysRevAccelBeams.20.081003,
  author = {Todesco, E. and Wenninger, J.},
  title = {Large Hadron Collider momentum calibration and accuracy},
  journal = {Phys. Rev. Accel. Beams},
  volume = {20},
  pages = {081003},
  year = {2017},
  doi = {10.1103/PhysRevAccelBeams.20.081003},
}

@Article{Cacciari:2008gp,
     author    = "Cacciari, Matteo and Salam, Gavin P. and Soyez, Gregory",
     title     = "{The anti-\(k_{t}\) jet clustering algorithm}",
     journal   = "JHEP",
     volume    = "04",
     year      = "2008",
     pages     = "063",
     eprint    = "0802.1189",
     archivePrefix = "arXiv",
     primaryClass  =  "hep-ph",
     doi       = "10.1088/1126-6708/2008/04/063",
     SLACcitation  = "%%CITATION = 0802.1189;%%"
}

@Article{Fastjet,
      author         = "Cacciari, Matteo and Salam, Gavin P. and Soyez, Gregory",
      title          = "{FastJet user manual}",
      journal        = "Eur. Phys. J. C",
      volume         = "72",
      year           = "2012",
      pages          = "1896",
      doi            = "10.1140/epjc/s10052-012-1896-2",
      eprint         = "1111.6097",
      archivePrefix  = "arXiv",
      primaryClass   = "hep-ph",
      reportNumber   = "CERN-PH-TH-2011-297",
      SLACcitation   = "%%CITATION = ARXIV:1111.6097;%%"
}

@Article{Pumplin:2002vw,
     author    = "Pumplin, J. and others",
     title     = "{New Generation of Parton Distributions with Uncertainties from Global QCD Analysis}",
     journal   = "JHEP",
     volume    = "07",
     year      = "2002",
     pages     = "012",
     doi       = "10.1088/1126-6708/2002/07/012",
     eprint    = "hep-ph/0201195",
     archivePrefix = "arXiv",
     SLACcitation  = "%%CITATION = HEP-PH/0201195;%%"
}

@article{Lai:2010vv,
      author         = "Lai, H.-L. and others",
      title          = "{New parton distributions for collider physics}",
      journal        = "Phys. Rev. D",
      volume         = "82",
      pages          = "074024",
      doi            = "10.1103/PhysRevD.82.074024",
      year           = "2010",
      eprint         = "1007.2241",
      archivePrefix  = "arXiv",
      primaryClass   = "hep-ph",
      reportNumber   = "MSUHEP-100707, SMU-HEP-10-10",
      SLACcitation   = "%%CITATION = ARXIV:1007.2241;%%",
}

@Article{Mangano:2002ea,
      author         = "Mangano, Michelangelo L. and others",
      title          = "{ALPGEN, a generator for hard multiparton processes in
                        hadronic collisions}",
      journal        = "JHEP",
      volume         = "07",
      pages          = "001",
      doi            = "10.1088/1126-6708/2003/07/001",
      year           = "2003",
      eprint         = "hep-ph/0206293",
      archivePrefix  = "arXiv",
      reportNumber   = "CERN-TH-2002-129, FTN-T-2002-06",
      SLACcitation   = "%%CITATION = HEP-PH/0206293;%%",
}

@Article{Sjostrand:2006za,
     author    = "Sj{\"o}strand, Torbjorn and Mrenna, Stephen and Skands, Peter Z.",
     title     = "{PYTHIA 6.4 physics and manual}",
     journal   = "JHEP",
     volume    = "05",
     year      = "2006",
     pages     = "026",
     eprint    = "hep-ph/0603175",
     archivePrefix = "arXiv",
     doi       = "10.1088/1126-6708/2006/05/026",
     SLACcitation  = "%%CITATION = HEP-PH/0603175;%%"
}

@article{Skands:2010ak,
      author         = "Skands, Peter Zeiler",
      title          = "{Tuning Monte Carlo generators: The Perugia tunes}",
      journal        = "Phys. Rev. D",
      volume         = "82",
      year           = "2010",
      pages          = "074018",
      doi            = "10.1103/PhysRevD.82.074018",
      eprint         = "1005.3457",
      archivePrefix  = "arXiv",
      primaryClass   = "hep-ph",
      reportNumber   = "MCNET-10-08, CERN-PH-TH-2010-113",
      SLACcitation   = "%%CITATION = ARXIV:1005.3457;%%"
}

@Article{Frixione:2003ei,
     author    = "Frixione, Stefano and Nason, Paolo and Webber, Bryan R.",
     title     = "{Matching NLO QCD and parton showers in heavy flavour production}",
     journal   = "JHEP",
     volume    = "08",
     year      = "2003",
     pages     = "007",
     eprint    = "hep-ph/0305252",
     archivePrefix = "arXiv"
}

@Article{Corcella:2000bw,
      author         = "Corcella, G. and others",
      title          = "{HERWIG 6: an event generator for hadron emission
                        reactions with interfering gluons (including supersymmetric processes)}",
      journal        = "JHEP",
      volume         = "01",
      pages          = "010",
      doi            = "10.1088/1126-6708/2001/01/010",
      year           = "2001",
      eprint         = "hep-ph/0011363",
      archivePrefix  = "arXiv",
      reportNumber   = "CAVENDISH-HEP-99-03, CERN-TH-2000-284, RAL-TR-2000-048",
      SLACcitation   = "%%CITATION = HEP-PH/0011363;%%",
}

@Article{Campbell:1999ah,
      author         = "Campbell, John M. and Ellis, R. Keith",
      title          = "{Update on vector boson pair production at hadron colliders}",
      journal        = "Phys. Rev. D",
      volume         = "60",
      pages          = "113006",
      doi            = "10.1103/PhysRevD.60.113006",
      year           = "1999",
      eprint         = "hep-ph/9905386",
      archivePrefix  = "arXiv",
      reportNumber   = "FERMILAB-PUB-99-146-T",
      SLACcitation   = "%%CITATION = HEP-PH/9905386;%%",
}

@article{Aliev:2010zk,
      author         = "Aliev, M. and Lacker, H. and Langenfeld, U. and Moch, S.
                        and Uwer, P. and Wiedermann, M.",
      title          = "{HATHOR -- HAdronic Top and Heavy quarks crOss section
                        calculatoR}",
      journal        = "Comput. Phys. Commun.",
      volume         = "182",
      year           = "2011",
      pages          = "1034-1046",
      doi            = "10.1016/j.cpc.2010.12.040",
      eprint         = "1007.1327",
      archivePrefix  = "arXiv",
      primaryClass   = "hep-ph",
      reportNumber   = "DESY-10-091, HU-EP-10-33, SFB-CPP-10-60",
      SLACcitation   = "%%CITATION = ARXIV:1007.1327;%%"
}

@article{Kant:2014oha,
      author         = "Kant, P. and Kind, O. M. and Kintscher, T. and Lohse, T.
                        and Martini, T. and Mölbitz, S. and Rieck, P. and Uwer, P.",
      title          = "{HatHor for single top-quark production: Updated
                        predictions and uncertainty estimates for single top-quark
                        production in hadronic collisions}",
      journal        = "Comput. Phys. Commun.",
      volume         = "191",
      year           = "2015",
      pages          = "74-89",
      doi            = "10.1016/j.cpc.2015.02.001",
      eprint         = "1406.4403",
      archivePrefix  = "arXiv",
      primaryClass   = "hep-ph",
      reportNumber   = "HU-EP-14-22",
      SLACcitation   = "%%CITATION = ARXIV:1406.4403;%%"
}

@article{Anastasiou:2003ds,
      author         = "Anastasiou, Charalampos and Dixon, Lance and Melnikov,
                        Kirill and Petriello, Frank",
      title          = "{High-precision QCD at hadron colliders: Electroweak
                        gauge boson rapidity distributions at next-to-next-to leading order}",
      journal        = "Phys. Rev. D",
      volume         = "69",
      year           = "2004",
      pages          = "094008",
      doi            = "10.1103/PhysRevD.69.094008",
      eprint         = "hep-ph/0312266",
      archivePrefix  = "arXiv",
      reportNumber   = "SLAC-PUB-10288, UH-511-1042-03",
      SLACcitation   = "%%CITATION = HEP-PH/0312266;%%"
}

@article{Czakon:2013goa,
      author         = "Czakon, Michal and Fiedler, Paul and Mitov, Alexander",
      title          = "{Total Top-Quark Pair-Production Cross Section at
                        Hadron Colliders Through \(O(\alpha_S^4)\)}",
      journal        = "Phys. Rev. Lett.",
      volume         = "110",
      pages          = "252004",
      doi            = "10.1103/PhysRevLett.110.252004",
      year           = "2013",
      eprint         = "1303.6254",
      archivePrefix  = "arXiv",
      primaryClass   = "hep-ph",
      reportNumber   = "CERN-PH-TH-2013-056, TTK-13-08",
      SLACcitation   = "%%CITATION = ARXIV:1303.6254;%%",
}

@article{Frixione:2007nw,
      author         = "Frixione, Stefano and Ridolfi, Giovanni and Nason, Paolo",
      title          = "{A positive-weight next-to-leading-order Monte Carlo for
                        heavy flavour hadroproduction}",
      journal        = "JHEP",
      volume         = "09",
      pages          = "126",
      doi            = "10.1088/1126-6708/2007/09/126",
      year           = "2007",
      eprint         = "0707.3088",
      archivePrefix  = "arXiv",
      primaryClass   = "hep-ph",
      SLACcitation   = "%%CITATION = ARXIV:0707.3088;%%",
}

@Article{Gleisberg:2008ta,
      author         = "Gleisberg, T. and H{\"o}che, S. and Krauss, F. and Sch{\"o}nherr, M. and Schumann, S. and Siegert, F. and Winter J.",
      title          = "{Event generation with SHERPA 1.1}",
      journal        = "JHEP",
      volume         = "02",
      pages          = "007",
      doi            = "10.1088/1126-6708/2009/02/007",
      year           = "2009",
      eprint         = "0811.4622",
      archivePrefix  = "arXiv",
      primaryClass   = "hep-ph",
}

@article{Czakon:2011xx,
  author         = "Czakon, Michal and Mitov, Alexander",
  title          = "{Top++: A program for the calculation of the top-pair cross-section at hadron colliders}",
  journal        = "Comput. Phys. Commun.",
  volume         = "185",
  year           = "2014",
  pages          = "2930",
  doi            = "10.1016/j.cpc.2014.06.021",
  eprint         = "1112.5675",
  archivePrefix  = "arXiv",
  primaryClass   = "hep-ph",
  reportNumber   = "CERN-PH-TH-2011-303, TTK-11-58",
}

@article{Agostinelli:2002hh,
      author         = "Agostinelli, S. and others",
      title          = "{\textsc{Geant4} -- a simulation toolkit}",
      journal        = "Nucl. Instrum. Meth. A",
      volume         = "506",
      year           = "2003",
      pages          = "250",
      doi            = "10.1016/S0168-9002(03)01368-8",
      reportNumber   = "SLAC-PUB-9350, FERMILAB-PUB-03-339",
      SLACcitation   = "%%CITATION = NUIMA,A506,250;%%"
}

@Article{PERF-2007-01,
    author         = "{ATLAS Collaboration}",
    title          = "{The ATLAS Experiment at the CERN Large Hadron Collider}",
    journal        = "JINST",
    volume         = "3",
    year           = "2008",
    pages          = "S08003",
    doi            = "10.1088/1748-0221/3/08/S08003",
    primaryClass   = "hep-ex",
}

@Article{PERF-2010-04,
    author         = "{ATLAS Collaboration}",
    title          = "{Electron performance measurements with the ATLAS detector using the 2010 LHC proton--proton collision data}",
    journal        = "Eur. Phys. J. C",
    volume         = "72",
    year           = "2012",
    pages          = "1909",
    doi            = "10.1140/epjc/s10052-012-1909-1",
    reportNumber   = "CERN-PH-EP-2011-117",
    eprint         = "1110.3174",
    archivePrefix  = "arXiv",
    primaryClass   = "hep-ex",
}

@Article{SOFT-2010-01,
    author         = "{ATLAS Collaboration}",
    title          = "{The ATLAS Simulation Infrastructure}",
    journal        = "Eur. Phys. J. C",
    volume         = "70",
    year           = "2010",
    pages          = "823",
    doi            = "10.1140/epjc/s10052-010-1429-9",
    eprint         = "1005.4568",
    archivePrefix  = "arXiv",
    primaryClass   = "physics.ins-det",
}

@Article{TOPQ-2010-01,
    author         = "{ATLAS Collaboration}",
    title          = "{Measurement of the top quark-pair production cross section with ATLAS in \(pp\) collisions at \(\sqrt{s} = 7\,\text{TeV}\)}",
    journal        = "Eur. Phys. J. C",
    volume         = "71",
    year           = "2011",
    pages          = "1577",
    doi            = "10.1140/epjc/s10052-011-1577-6",
    reportNumber   = "CERN-PH-EP-2010-064",
    eprint         = "1012.1792",
    archivePrefix  = "arXiv",
    primaryClass   = "hep-ex",
}

@Article{DAPR-2011-01,
    author         = "{ATLAS Collaboration}",
    title          = "{Improved luminosity determination in \(pp\) collisions at \(\sqrt{s} = 7\,\text{TeV}\) using the ATLAS detector at the LHC}",
    journal        = "Eur. Phys. J. C",
    volume         = "73",
    year           = "2013",
    pages          = "2518",
    doi            = "10.1140/epjc/s10052-013-2518-3",
    reportNumber   = "CERN-PH-EP-2013-026",
    eprint         = "1302.4393",
    archivePrefix  = "arXiv",
    primaryClass   = "hep-ex",
}

@Article{PERF-2011-02,
    author         = "{ATLAS Collaboration}",
    title          = "{Performance of the ATLAS Trigger System in 2010}",
    journal        = "Eur. Phys. J. C",
    volume         = "72",
    year           = "2012",
    pages          = "1849",
    doi            = "10.1140/epjc/s10052-011-1849-1",
    reportNumber   = "CERN-PH-EP-2011-078",
    eprint         = "1110.1530",
    archivePrefix  = "arXiv",
    primaryClass   = "hep-ex",
}

@Article{PERF-2011-07,
    author         = "{ATLAS Collaboration}",
    title          = "{Performance of missing transverse momentum reconstruction in proton--proton collisions at  \(\sqrt{s} = 7\,\text{TeV}\) with ATLAS}",
    journal        = "Eur. Phys. J. C",
    volume         = "72",
    year           = "2012",
    pages          = "1844",
    doi            = "10.1140/epjc/s10052-011-1844-6",
    reportNumber   = "CERN-PH-EP-2011-114",
    eprint         = "1108.5602",
    archivePrefix  = "arXiv",
    primaryClass   = "hep-ex",
}

@Article{TOPQ-2011-01,
    author         = "{ATLAS Collaboration}",
    title          = "{Measurement of the top quark pair production cross section in \(pp\) collisions at \(\sqrt{s} = 7\,\text{TeV}\) in dilepton final states with ATLAS}",
    journal        = "Phys. Lett. B",
    volume         = "707",
    year           = "2012",
    pages          = "459",
    doi            = "10.1016/j.physletb.2011.12.055",
    reportNumber   = "CERN-PH-EP-2011-103",
    eprint         = "1108.3699",
    archivePrefix  = "arXiv",
    primaryClass   = "hep-ex",
}

@Article{TOPQ-2011-02,
    author         = "{ATLAS Collaboration}",
    title          = "{Measurement of the top quark pair production cross-section with ATLAS in the single lepton channel}",
    journal        = "Phys. Lett. B",
    volume         = "711",
    year           = "2012",
    pages          = "244",
    doi            = "10.1016/j.physletb.2012.03.083",
    reportNumber   = "CERN-PH-EP-2011-201",
    eprint         = "1201.1889",
    archivePrefix  = "arXiv",
    primaryClass   = "hep-ex",
}

@Article{PERF-2012-01,
    author         = "{ATLAS Collaboration}",
    title          = "{Jet energy measurement and its systematic uncertainty in proton--proton collisions at \(\sqrt{s} = 7\,\text{TeV}\) with the ATLAS detector}",
    journal        = "Eur. Phys. J. C",
    volume         = "75",
    year           = "2015",
    pages          = "17",
    doi            = "10.1140/epjc/s10052-014-3190-y",
    reportNumber   = "CERN-PH-EP-2013-222",
    eprint         = "1406.0076",
    archivePrefix  = "arXiv",
    primaryClass   = "hep-ex",
}

@Article{PERF-2012-04,
    author         = "{ATLAS Collaboration}",
    title          = "{Performance of \(b\)-jet identification in the ATLAS experiment}",
    journal        = "JINST",
    volume         = "11",
    year           = "2016",
    pages          = "P04008",
    doi            = "10.1088/1748-0221/11/04/P04008",
    reportNumber   = "CERN-PH-EP-2015-216",
    eprint         = "1512.01094",
    archivePrefix  = "arXiv",
    primaryClass   = "hep-ex",
}

@Article{TOPQ-2012-05,
    author         = "{ATLAS Collaboration}",
    title          = "{Simultaneous measurements of the \(t\bar{t}\), \(W^+W^-\), and \(Z/\gamma^{*}\rightarrow\tau\tau\) production cross-sections in \(pp\) collisions at \(\sqrt{s} = 7\,\text{TeV}\) with the ATLAS detector}",
    journal        = "Phys. Rev. D",
    volume         = "91",
    year           = "2015",
    pages          = "052005",
    doi            = "10.1103/PhysRevD.91.052005",
    reportNumber   = "CERN-PH-EP-2014-119",
    eprint         = "1407.0573",
    archivePrefix  = "arXiv",
    primaryClass   = "hep-ex",
}

@Article{TOPQ-2012-21,
    author         = "{ATLAS Collaboration}",
    title          = "{Comprehensive measurements of \(t\)-channel single top-quark production cross sections at \(\sqrt{s} = 7\,\text{TeV}\) with the ATLAS detector}",
    journal        = "Phys. Rev. D",
    volume         = "90",
    year           = "2014",
    pages          = "112006",
    doi            = "10.1103/PhysRevD.90.112006",
    reportNumber   = "CERN-PH-EP-2014-133",
    eprint         = "1406.7844",
    archivePrefix  = "arXiv",
    primaryClass   = "hep-ex",
}

@Article{PERF-2013-03,
    author         = "{ATLAS Collaboration}",
    title          = "{Electron reconstruction and identification efficiency measurements with the ATLAS detector using the 2011 LHC proton--proton collision data}",
    journal        = "Eur. Phys. J. C",
    volume         = "74",
    year           = "2014",
    pages          = "2941",
    doi            = "10.1140/epjc/s10052-014-2941-0",
    reportNumber   = "CERN-PH-EP-2014-040",
    eprint         = "1404.2240",
    archivePrefix  = "arXiv",
    primaryClass   = "hep-ex",
}

@Article{TOPQ-2013-04,
    author         = "{ATLAS Collaboration}",
    title          = "{Measurement of the \(t\bar{t}\) production cross-section using \(e\mu\) events with \(b\)-tagged jets in \(pp\) collisions at \(\sqrt{s} = 7\) and \(8\,\text{TeV}\) with the ATLAS detector}",
    journal        = "Eur. Phys. J. C",
    volume         = "74",
    year           = "2014",
    pages          = "3109",
    doi            = "10.1140/epjc/s10052-014-3109-7",
    reportNumber   = "CERN-PH-EP-2014-124",
    eprint         = "1406.5375",
    archivePrefix  = "arXiv",
    primaryClass   = "hep-ex",
    related        = "TOPQ-2013-04-addm",
    relatedstring  = "Addendum:",
}

@Article{PERF-2014-05,
    author         = "{ATLAS Collaboration}",
    title          = "{Measurement of the muon reconstruction performance of the ATLAS detector using 2011 and 2012 LHC proton--proton collision data}",
    journal        = "Eur. Phys. J. C",
    volume         = "74",
    year           = "2014",
    pages          = "3130",
    doi            = "10.1140/epjc/s10052-014-3130-x",
    reportNumber   = "CERN-PH-EP-2014-151",
    eprint         = "1407.3935",
    archivePrefix  = "arXiv",
    primaryClass   = "hep-ex",
}

@Article{PERF-2014-07,
    author         = "{ATLAS Collaboration}",
    title          = "{Topological cell clustering in the ATLAS calorimeters and its performance in LHC Run~1}",
    journal        = "Eur. Phys. J. C",
    volume         = "77",
    year           = "2017",
    pages          = "490",
    doi            = "10.1140/epjc/s10052-017-5004-5",
    reportNumber   = "CERN-PH-EP-2015-304",
    eprint         = "1603.02934",
    archivePrefix  = "arXiv",
    primaryClass   = "hep-ex",
}

@Article{TOPQ-2015-02,
    author         = "{ATLAS Collaboration}",
    title          = "{Measurement of lepton differential distributions and the top quark mass in \(t\bar{t}\) production in \(pp\) collisions at \(\sqrt{s} = 8\,\text{TeV}\) with the ATLAS detector}",
    journal        = "Eur. Phys. J. C",
    volume         = "77",
    year           = "2017",
    pages          = "804",
    doi            = "10.1140/epjc/s10052-017-5349-9",
    reportNumber   = "CERN-EP-2017-200",
    eprint         = "1709.09407",
    archivePrefix  = "arXiv",
    primaryClass   = "hep-ex",
}

@Article{TOPQ-2017-03,
    author         = "{ATLAS Collaboration}",
    title          = "{Measurement of the top quark mass in the \(t\bar{t} \to\) lepton+jets channel from \(\sqrt{s} = 8\,\text{TeV}\) ATLAS data and combination with previous results}",
    journal        = "Eur. Phys. J. C",
    volume         = "79",
    year           = "2019",
    pages          = "290",
    doi            = "10.1140/epjc/s10052-019-6757-9",
    reportNumber   = "CERN-EP-2018-238",
    eprint         = "1810.01772",
    archivePrefix  = "arXiv",
    primaryClass   = "hep-ex",
}

@Article{TOPQ-2017-08,
    author         = "{ATLAS Collaboration}",
    title          = "{Measurement of the inclusive \(t\bar{t}\) production cross section in the lepton+jets channel in \(pp\) collisions at \(\sqrt{s} = 7\,\text{TeV}\) with the ATLAS detector using support vector machines}",
    year           = "2022",
    reportNumber   = "CERN-EP-2022-191",
    eprint         = "2212.00571",
    archivePrefix  = "arXiv",
    primaryClass   = "hep-ex",
}

@Article{TOPQ-2018-39,
    author         = "{ATLAS and CMS Collaborations}",
    title          = "{Combination of inclusive top-quark pair production cross-section measurements using ATLAS and CMS data at \(\sqrt{s}= 7\) and \(8\,\text{TeV}\)}",
    year           = "2022",
    reportNumber   = "CERN-EP-2021-222, CMS-TOP-18-014, CERN-LPCC-2022-03",
    eprint         = "2205.13830",
    archivePrefix  = "arXiv",
    primaryClass   = "hep-ex",
}

@Booklet{ATL-SOFT-PUB-2021-001,
    author         = "{ATLAS Collaboration}",
    title          = "{The ATLAS Collaboration Software and Firmware}",
    howpublished   = "{ATL-SOFT-PUB-2021-001}",
    url            = "https://cds.cern.ch/record/2767187",
    year           = "2021",
}

@Booklet{ATL-SOFT-PUB-2021-003,
    author         = "{ATLAS Collaboration}",
    title          = "{ATLAS Computing Acknowledgements}",
    howpublished   = "{ATL-SOFT-PUB-2021-003}",
    url            = "https://cds.cern.ch/record/2776662",
    year           = "2021",
}

@Article{CMS-TOP-11-003,
    author         = "{CMS Collaboration}",
    title          = "{Measurement of the \(t\bar{t}\) production cross section in \(pp\) collisions at \(\sqrt{s} = 7\,\text{TeV}\) with lepton+jets final states}",
    journal        = "Phys. Lett. B",
    volume         = "720",
    year           = "2013",
    pages          = "83",
    doi            = "10.1016/j.physletb.2013.02.021",
    reportNumber   = "CERN-PH-EP-2012-285",
    eprint         = "1212.6682",
    archivePrefix  = "arXiv",
    primaryClass   = "hep-ex",
}

@Article{CMS-TOP-13-004,
    author         = "{CMS Collaboration}",
    title          = "{Measurement of the \(t\bar{t}\) production cross section in the \(e\mu\) channel in proton--proton collisions at \(\sqrt{s} = 7\) and \(8\,\text{TeV}\)}",
    journal        = "JHEP",
    volume         = "08",
    year           = "2016",
    pages          = "029",
    doi            = "10.1007/JHEP08(2016)029",
    reportNumber   = "CERN-EP-2016-044",
    eprint         = "1603.02303",
    archivePrefix  = "arXiv",
    primaryClass   = "hep-ex",
}

@Booklet{ATLAS-CONF-2014-058,
    author         = "{ATLAS Collaboration}",
    title          = "{Estimation of non-prompt and fake lepton backgrounds in final states with top quarks produced in proton--proton collisions at \(\sqrt{s} = 8~\text{TeV}\) with the ATLAS Detector}",
    howpublished   = "{ATLAS-CONF-2014-058}",
    url            = "https://cds.cern.ch/record/1951336",
    year           = "2014",
}

@Booklet{ATL-PHYS-PUB-2011-008,
    author         = "{ATLAS Collaboration}",
    title          = "{New ATLAS event generator tunes to 2010 data}",
    howpublished   = "{ATL-PHYS-PUB-2011-008}",
    url            = "https://cds.cern.ch/record/1345343",
    year           = "2011",
}

\clearpage
 
 
\begin{flushleft}
\hypersetup{urlcolor=black}
{\Large The ATLAS Collaboration}

\bigskip

\AtlasOrcid{M.~Aaboud}$^\textrm{\scriptsize 35c}$,
\AtlasOrcid[0000-0002-6665-4934]{G.~Aad}$^\textrm{\scriptsize 103}$,
\AtlasOrcid[0000-0002-5888-2734]{B.~Abbott}$^\textrm{\scriptsize 122}$,
\AtlasOrcid[0000-0002-7248-3203]{D.C.~Abbott}$^\textrm{\scriptsize 104}$,
\AtlasOrcid[0000-0002-2987-4006]{D.K.~Abhayasinghe}$^\textrm{\scriptsize 95}$,
\AtlasOrcid[0000-0002-8496-9294]{S.H.~Abidi}$^\textrm{\scriptsize 159}$,
\AtlasOrcid[0000-0002-8279-9324]{O.S.~AbouZeid}$^\textrm{\scriptsize 42}$,
\AtlasOrcid{N.L.~Abraham}$^\textrm{\scriptsize 149}$,
\AtlasOrcid[0000-0001-5329-6640]{H.~Abramowicz}$^\textrm{\scriptsize 154}$,
\AtlasOrcid[0000-0002-1599-2896]{H.~Abreu}$^\textrm{\scriptsize 153}$
% ...
\bigskip
\\

% ...
$^{35}$$^{(a)}$Facult\'e des Sciences Ain Chock, R\'eseau Universitaire de Physique des Hautes Energies - Universit\'e Hassan II, Casablanca;$^{(b)}$Facult\'e des Sciences Semlalia, Universit\'e Cadi Ayyad, LPHEA-Marrakech;$^{(c)}$LPMR, Facult\'e des Sciences, Universit\'e Mohamed Premier, Oujda;$^{(d)}$Facult\'e des sciences, Universit\'e Mohammed V, Rabat; Morocco.\\
% ...
$^{42}$Niels Bohr Institute, University of Copenhagen, Copenhagen; Denmark.\\
% ...
$^{95}$Department of Physics, Royal Holloway University of London, Egham; United Kingdom.\\
% ...
$^{103}$CPPM, Aix-Marseille Universit\'e, CNRS/IN2P3, Marseille; France.\\
$^{104}$Department of Physics, University of Massachusetts, Amherst MA; United States of America.\\
% ...
$^{122}$Homer L. Dodge Department of Physics and Astronomy, University of Oklahoma, Norman OK; United States of America.\\
% ...
$^{149}$Department of Physics and Astronomy, University of Sussex, Brighton; United Kingdom.\\
% ...
$^{153}$Department of Physics, Technion, Israel Institute of Technology, Haifa; Israel.\\
$^{154}$Raymond and Beverly Sackler School of Physics and Astronomy, Tel Aviv University, Tel Aviv; Israel.\\
% ...
$^{159}$Department of Physics, University of Toronto, Toronto ON; Canada.\\
% ...
$^{a}$ Also Affiliated with an institute covered by a cooperation agreement with CERN.\\
$^{b}$ Also at Borough of Manhattan Community College, City University of New York, New York NY; United States of America.\\
$^{c}$ Also at Centre for High Performance Computing, CSIR Campus, Rosebank, Cape Town; South Africa.\\
$^{d}$ Also at CERN, Geneva; Switzerland.\\
% ...

\end{flushleft}


\end{document}